\documentclass[twocolumn,times]{aastex631} 
\pdfoutput=1 

\usepackage{graphicx}	
\usepackage{amsmath}	
\usepackage{bm}
\usepackage{float}
\usepackage{xspace}
\usepackage{soul}

\newcommand{\Jy}{\ensuremath{\,{\rm Jy}}}
\newcommand{\K}{\ensuremath{\,{\rm K}}}

\newcommand{\HI}{\textsc{Hi}\xspace}

\newcommand{\add}[1]{\textcolor{black}{#1}}
\newcommand{\revise}[1]{\textcolor{black}{#1}}

\newcommand{\reffg}[1]{Fig.~\ref{#1}}
\newcommand{\reftb}[1]{Table~\ref{#1}}
\newcommand{\refeq}[1]{Eq.~(\ref{#1})}
\newcommand{\refsc}[1]{Sec.~\ref{#1}}

\begin{document}

\title{CRAFTS for \HI cosmology: I. \revise{data processing pipeline and validation tests}}

\author[0009-0006-2521-025X]{Wenxiu Yang}
\affiliation{National Astronomical Observatories, Chinese Academy of Sciences, Beijing 100101, China}
\affiliation{School of Astronomy and Space Science, University of Chinese Academy of Sciences, Beijing 100049, China}
\affiliation{Jodrell Bank Centre for Astrophysics, Department of Physics \& Astronomy, The University of Manchester, Manchester M13 9PL, UK}

\correspondingauthor{Laura Wolz}
\author[0000-0003-3334-3037]{Laura Wolz}
\email{laura.wolz@manchester.ac.uk}
\affiliation{Jodrell Bank Centre for Astrophysics, Department of Physics \& Astronomy, The University of Manchester, Manchester M13 9PL, UK}

\author[0000-0003-1962-2013]{Yichao Li}
\affiliation{Liaoning Key Laboratory of Cosmology and Astrophysics,  College of Sciences, Northeastern University, Shenyang 110819, China}

\author[0000-0002-3108-5591]{Wenkai Hu}
\affiliation{Department of Physics and Astronomy, University of the Western Cape, Robert Sobukhwe Road, Bellville, 7535, South Africa}
\affiliation{ARC Centre of Excellence for All Sky Astrophysics in 3 Dimensions (ASTRO 3D), Australia}

\author[0000-0001-6594-107X]{Steven Cunnington}
\affiliation{Jodrell Bank Centre for Astrophysics, Department of Physics \& Astronomy, The University of Manchester, Manchester M13 9PL, UK}

\author[0000-0002-6780-1406]{Keith Grainge}
\affiliation{Jodrell Bank Centre for Astrophysics, Department of Physics \& Astronomy, The University of Manchester, Manchester M13 9PL, UK}

\author[0000-0001-8075-0909]{Furen Deng}
\affiliation{National Astronomical Observatories, Chinese Academy of Sciences, Beijing 100101, China}
\affiliation{School of Astronomy and Space Science, University of Chinese Academy of Sciences, Beijing 100049, China}
\affiliation{Institute of Astronomy, University of Cambridge, Madingley Road, Cambridge, CB3 0HA, UK}

\author[0000-0003-3858-6361]{Shifan Zuo}
\affiliation{National Astronomical Observatories, Chinese Academy of Sciences, Beijing 100101, China}
\affiliation{Key Laboratory of Radio Astronomy and Technology, Chinese Academy of Sciences, A20 Datun Road, Chaoyang District, Beijing 100101, China}

\author[0009-0004-8919-7088]{Shuanghao Shu}
\affiliation{National Astronomical Observatories, Chinese Academy of Sciences, Beijing 100101, China}
\affiliation{School of Astronomy and Space Science, University of Chinese Academy of Sciences, Beijing 100049, China}

\author[0009-0008-2564-9398]{Xinyang Zhao}
\affiliation{Liaoning Key Laboratory of Cosmology and Astrophysics,  College of Sciences, Northeastern University, Shenyang 110819, China}

\author[0000-0003-3010-7661]{Di Li}
\affiliation{Department of Astronomy, Tsinghua University, Beijing 100084, China}
\affiliation{National Astronomical Observatories, Chinese Academy of Sciences, Beijing 100101, China}

\author[0009-0005-9546-4573]{Zheng Zheng}
\affiliation{National Astronomical Observatories, Chinese Academy of Sciences, Beijing 100101, China}

\author[0000-0001-9659-0292]{Marko Kr\v{c}o}
\affiliation{National Astronomical Observatories, Chinese Academy of Sciences, Beijing 100101, China}

\author[0009-0009-2303-395X]{Yinghui Zheng}
\affiliation{National Astronomical Observatories, Chinese Academy of Sciences, Beijing 100101, China}
\affiliation{School of Astronomy and Space Science, University of Chinese Academy of Sciences, Beijing 100049, China}

\author[0009-0003-4821-5502]{Linjing Feng}
\affiliation{National Astronomical Observatories, Chinese Academy of Sciences, Beijing 100101, China}
\affiliation{School of Astronomy and Space Science, University of Chinese Academy of Sciences, Beijing 100049, China}

\author[0000-0003-3948-9192]{Pei Zuo}
\affiliation{National Astronomical Observatories, Chinese Academy of Sciences, Beijing 100101, China}

\author[0000-0002-3612-9258]{Hao Chen}
\affiliation{Research Center for Astronomical Computing, Zhejiang Laboratory, Hangzhou 311100, China}

\author[0000-0002-8899-4673]{Xue-Jian Jiang}
\affiliation{Research Center for Astronomical Computing, Zhejiang Laboratory, Hangzhou 311100, China}

\author[0000-0001-8923-7757]{Chen Wang}
\affiliation{National Astronomical Observatories, Chinese Academy of Sciences, Beijing 100101, China}

\author{Pei Wang}
\affiliation{National Astronomical Observatories, Chinese Academy of Sciences, Beijing 100101, China}

\author{Chen-Chen Miao}
\affiliation{Research Center for Astronomical Computing, Zhejiang Laboratory, Hangzhou 311100, China}

\correspondingauthor{Yougang Wang}
\author[0000-0003-0631-568X]{Yougang Wang}
\email{wangyg@bao.ac.cn}
\affiliation{National Astronomical Observatories, Chinese Academy of Sciences, Beijing 100101, China}
\affiliation{State Key Laboratory of Radio Astronomy and Technology, Beijing 100101, China}
\affiliation{School of Astronomy and Space Science, University of Chinese Academy of Sciences, Beijing 100049, China}
\affiliation{Liaoning Key Laboratory of Cosmology and Astrophysics,  College of Sciences, Northeastern University, Shenyang 110819, China}

\correspondingauthor{Xuelei Chen}
\author[0000-0001-6475-8863]{Xuelei Chen}
\email{xuelei@cosmology.bao.ac.cn}
\affiliation{National Astronomical Observatories, Chinese Academy of Sciences, Beijing 100101, China}
\affiliation{Liaoning Key Laboratory of Cosmology and Astrophysics,  College of Sciences, Northeastern University, Shenyang 110819, China}
\affiliation{State Key Laboratory of Radio Astronomy and Technology, Beijing 100101, China}
\affiliation{School of Astronomy and Space Science, University of Chinese Academy of Sciences, Beijing 100049, China}



\begin{abstract}

We present the \revise{calibration procedures and validation of source measurement with}
the data of the Commensal Radio Astronomy FAST Survey (CRAFTS) for \HI intensity mapping by the Five-hundred-meter Aperture Spherical Radio Telescope (FAST). 
Using 70-hour drift-scan observation with the L-band (1.05-1.45GHz) 19-beam receiver, we obtain the data covering $270\,\rm deg^2$ sky area. 
We employ both the pulsar backend and the spectrum backend to calibrate the spectral time-ordered-data (TOD) before projecting them onto HEALPix maps. 
We produce calibrated TOD with frequency resolution of 30kHz and time resolution of 1s and the map data-cube with frequency resolution of 30kHz and spatial resolution of $2.95\,\rm arcmin^2$. 
We examine the pointing errors, noise overflow, RFI contamination and their effect on the data quality. The resulting noise level is $\sim$ 5.7mJy for the calibrated TOD and 1.6mJy for the map, consistent with the theoretical predictions within 5\% at RFI-free channels. 
We also validate the data by Principal Components Analysis (PCA) and find \revise{the residual map looks thermal noise dominated after removing 30 modes.}
We identify 447 isolated bright continuum sources in our data matching the NRAO-VLA Sky Survey (NVSS) catalog, with relative flux error of 8.3\% for TOD and \add{6.6\%} for the map-level. We also measure the \HI emission of 90 galaxies with redshift $z<0.07$ and compare with \HI-MaNGA spectra, yielding an overall relative \HI integral flux error of 16.7\%. 
\add{These results provide an important first step in assessing} the feasibility of conducting cosmological \HI detection with CRAFTS.

\end{abstract}



\section{Introduction} \label{sec:intro}

Hydrogen is the most abundant element in the \add{Universe}. The hyperfine spin-flip transition in the ground state of neutral hydrogen (\HI) emits radiation with the wavelength of $\sim$ 21cm (corresponding to frequency $\sim $ 1420MHz). Through observing the large scale distribution of \HI in galaxies, the 21cm emission can provide us with a useful probe to explore the structure, formation, and evolution of the \add{Universe}. 

Many \HI surveys have been carried out by large radio telescopes in recent decades. For example, the Arecibo Legacy Fast ALFA Survey (ALFALFA, \citealt{2018ApJ...861...49H}) and the HI Parkes All Sky Survey (HIPASS, \citealt{2004MNRAS.350.1195M}) are both large surveys for \HI galaxies in the local Universe. However, limited by the spatial resolution of radio telescopes and the faintness of \HI signals, it is difficult to detect individual galaxies at high redshift. Fortunately, the lack of information on small-scale structures associated with individual galaxies will not influence the studies of the large-scale structure of the \add{Universe}. Therefore, a technique called intensity mapping (IM) is proposed \citep{10.1111/j.1365-2966.2004.08416.x, 2008PhRvL.100i1303C, 2008PhRvL.100p1301L,2009astro2010S.234P}. According to the IM method, we can directly record the collective intensity of \HI emission from many unresolved galaxies to obtain a sky map of brightness temperature at different positions and redshift. This strategy is more efficient than traditional galaxy redshift surveys for observing large volumes of sky. Moreover, the intensity mapping technique can be applied not only to \HI signals, but also to other emission lines (e.g. see \citealt{2018MNRAS.479.3490F} for H$_{\alpha}$,  \citealt{2011ApJ...730L..30C,Breysse_2014} for CO, \citealt{2019ApJ...872..126M} for CII, etc.), which is very useful for tracing the distribution of matter in the \add{Universe} and studying features of cosmological Large-Scale Structure (LSS) such as Baryon Acoustic Oscillations (BAO, \citealt{10.1111/j.1365-2966.2004.08416.x, 2008PhRvL.100i1303C, 2015ApJ...803...21B, 2018ApJ...866..135V, 2022MNRAS.516.5454R}). 

At present, there are many advanced instruments used for \HI intensity mapping experiments in the world, and have provided excellent results. The \HI IM has been proved feasible by measuring the cross-correlation between \HI brightness temperature observed by radio telescopes such as Green Bank Telescope(GBT), Parkes, MeerKAT and the Canadian Hydrogen Intensity Mapping Experiment (CHIME) and optical samples \citep{2010Natur.466..463C, 2013ApJ...763L..20M, 2018MNRAS.476.3382A, 2022MNRAS.510.3495W, 2023MNRAS.518.6262C, 2023ApJ...947...16A, 2024arXiv240721626M}. In particular, \cite{2023MNRAS.518.6262C} and \cite{2024arXiv240721626M} measure the \HI cross-correlation power spectrum with the single dish mode of MeerKAT 64-dish array and WiggleZ \citep{2010MNRAS.401.1429D} or the Galaxy And Mass Assembly (GAMA, \citealt{2022MNRAS.513..439D}) galaxies at redshift $z \sim 0.4$. However, the \HI auto power spectrum detection with single dish mode experiments was hindered by the low level of the signal and some instrumental limitations such as the sensitivity, radio frequency interferences (RFIs) contamination \citep{2024arXiv240417908E}, the time-correlated noise like 1/f noise \citep{2021MNRAS.501.4344L, 2021MNRAS.508.2897H,  2024MNRAS.527.4717I}, and the complicated foreground \citep{2015MNRAS.447..400A, 2017MNRAS.464.4938W, 2020MNRAS.499..304C,2021MNRAS.504..208C, 2022MNRAS.509.2048S,2022ApJ...934...83N,2024arXiv241206750C}. Furthermore, \cite{2023arXiv230111943P} reports the first detection of \HI auto power spectrum with the interferometric mode of MeerKAT, which is a milestone progress for \HI IM. However, the field of view of the interferometer restricted this \HI auto-correlation detection to smaller scales, thus detection on larger, cosmological scales ($\gtrsim$ 1 Mpc) still awaits. Additional dedicated \HI IM and general purpose telescopes will be constructed in the near future. For example, the Square Kilometre Array (SKA, \revise{\citealt{2009IEEEP..97.1482D}}) is under construction, which is designed to be the world's most sensitive radio telescope and may make transformative discoveries about the \add{Universe} \citep{2015aska.confE..19S, 2020PASA...37....7S}.

Currently, the Five-hundred-meter Aperture Spherical radio Telescope (FAST) is the most sensitive single-dish radio telescope in the world (\citealt{2019SCPMA..6259502J, 2020RAA....20...64J}). It is located in Guizhou province in southwestern China ($\rm E106.^\circ86, N25,^\circ65$) with a zenith angle that can reach up to $40^{\circ}$, enabling it to cover the sky area in the declination range of $-15^\circ \sim 65^\circ$. FAST is now equipped with the L-band(1-1.5GHz) 19-beam receiver \citep{2020RAA....20...64J} with a field of view (FOV) $\sim$ 2.95 arcmin at 1420MHz for each beam. The illuminated aperture of FAST has a diameter of 300 meters during observation, resulting in a large effective area of $\sim 70000 \rm m^2$. The excellent properties of FAST make it a powerful tool for multiple fields of astronomical research like the study of pulsars \citep{2019SCPMA..6259508Q, 2020ApJ...892L...6P, 2021NatAs...5..788Y, 2023MNRAS.518.1672M}, Fast Radio Bursts (FRBs, \citealt{2021Natur.598..267L, 2022Sci...375.1266F, 2022Natur.606..873N}), interstellar medium (ISM, \citealt{2020RAA....20...77T, 2020MNRAS.499.3085Z, 2021RAA....21..282S, 2022A&A...658A.140L}), HI galaxy and cosmology \citep{2023ApJ...954..139L, 2023A&A...675A..40H, 2024SCPMA..6719511Z}, indirect detection of dark matter particles (\citealt{2023PhRvL.130r1001A, 2023PhRvD.107j3011G}, Yang et al. in preparation), etc. There are many ongoing key projects with FAST for different scientific goals. For example, the Commensal Radio Astronomy FAST Survey (CRAFTS, \citealt{2018IMMag..19..112L}) is a drift scan survey project aimed at simultaneously conducting observation for transients like pulsars \citep{2023MNRAS.518.1672M, 2023RAA....23h5022C} and FRBs \citep{2021ApJ...909L...8N}, as well as a spectral survey like the \HI survey \citep{2019SCPMA..6259506Z, 2023ApJ...954..139L}. 
This project plans to scan the whole sky accessible to FAST twice in the next few decades. Starting in 2020 until 2024, it has covered an area of $\sim$ 7000 $\mathrm{deg^2}$. By exploiting the large amount of data from the CRAFTS project and the high sensitivity of FAST, we can realistically conduct \HI cosmology research with the intensity mapping technique. In addition, because of the large aperture, the high resolution of FAST also enables it to carry out galaxy surveys, especially at low redshift. 

The forecast in \cite{2020MNRAS.493.5854H} indicates that a single scan of a $\sim 20000\,{\rm deg}^2$ sky area by FAST will enable a good detection of the \HI power spectrum, achieving a signal-to-noise ratio (S/N) $> 5$ at redshift $0.05 < z < 0.35$ at the BAO scale ($\sim 0.1 h/{\rm Mpc}$) through both \HI IM and \add{\HI} galaxy survey. Generally, the \HI IM technique will perform much better than the \add{\HI} galaxy survey in terms of \HI power spectrum precision, especially at high redshift. A pilot drift scan survey for \HI IM \citep{2023ApJ...954..139L} at the frequency band of 1050 - 1450MHz ($0 < z < 0.35$ in redshift) is currently ongoing to check the data quality and systematic performance for future \HI cosmology and galaxy studies (\citealt{2021MNRAS.508.2897H}, Shu et al. in preparation). We have now published the data processing pipeline and some preliminary results about point sources for the sky area of $\sim 60\,{\rm deg}^2$ at RA $\sim$ 9-13h, Dec $\sim 25.8^{\circ}-27.1^{\circ}$ in \cite{2023ApJ...954..139L}. However, the dataset of this pilot survey is not large enough for the study of LSS at present and some systematic issues such as the effect of the beam, standing wave features, and irregular temporal fluctuations are still unclear. Therefore, we are currently using the CRAFTS data to conduct more tests for \HI IM and to learn more about the FAST system performance.

In this work, we present the data processing procedures and some preliminary results from the CRAFTS project, based on 
\revise{the observations} covering a sky area of $\sim 270 {\rm deg}^2$. The results include continuum point source measurement and HI emission line detection.
Some systematic issues specific to CRAFTS and more general observations by FAST are also carefully discussed \add{as an initial assessment of data quality}. \add{This study serves as an important first step toward future HI intensity mapping experiments and galaxy surveys with CRAFTS, providing key insights for further improvements in calibration, systematic mitigation, and data validation.}

This paper is organized as follows. In \refsc{sec:data}, we describe the CRAFTS data we use in this work. Then \refsc{sec:process} is about how we process these data. We discuss the data validation including some preliminary foreground removal tests for intensity mapping in \refsc{sec:system}. Then in \refsc{sec:result}, we present results of source measurement from both calibrated data and maps. Finally, we summarize our work in \refsc{sec:summary}.

\section{Data} \label{sec:data}

\subsection{CRAFTS} \label{subsec:data_crafts}

\begin{table} 
    \footnotesize
    \centering
    \begin{tabular}{ccc}
        \hline
        Parameters &  Pulsar Backend & Spectrum Backend (W) \\
         \hline
        Bandwidth & 500MHz & 500MHz \\
        Channels & 4096 & 65536 \\
        Time resolution ($\Delta t$) & 98.304$\mu$s & $\sim$0.2s \\
        Frequency resolution ($\Delta \nu$) & $\sim$122kHz & $\sim$7.6kHz \\
         \hline
    \end{tabular}
    \caption{Parameters of raw data from the pulsar backend and the spectrum backend used in CRAFTS observation}
    \label{tb:backend}
\end{table}

\begin{figure}
    \centering
    \includegraphics[width=0.45\textwidth]{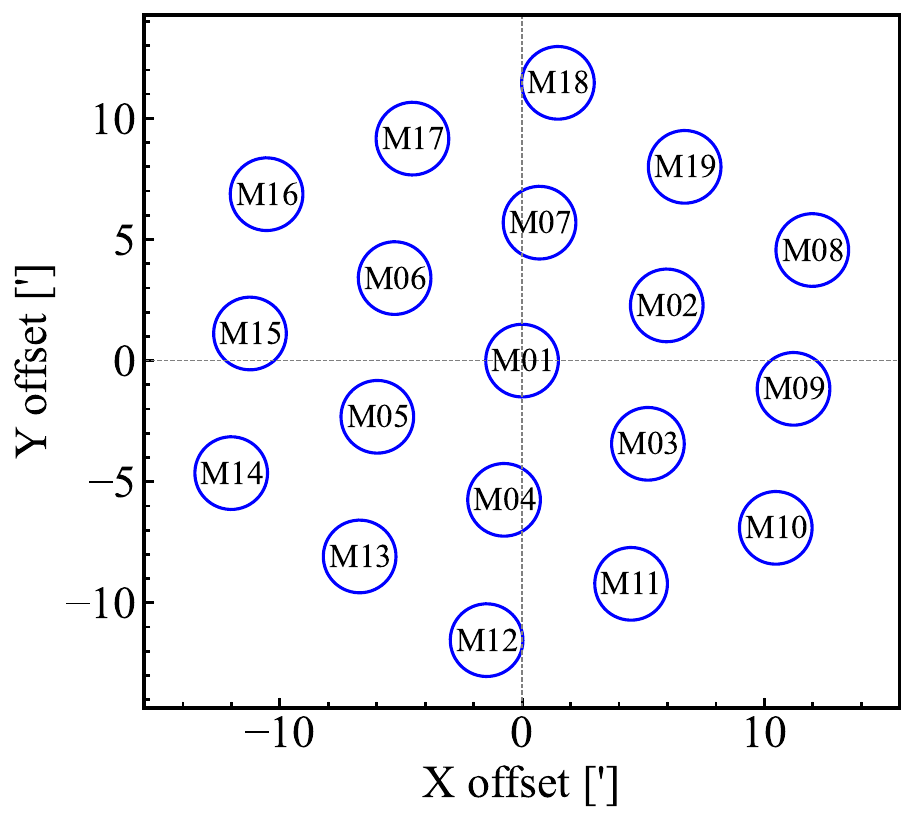}
    \caption{Relative position of the pointing of 19 beams in CRAFTS observation. The 19 beams are rotated by 23.4$^\circ$ for best sampling. The blue circles mark the beam size of $\sim$ 3 arc minutes. In this paper, we add ``M'' before the beam number to indicate that the data is obtained from the corresponding beam. }
    \label{fig:beam_pos}
\end{figure}

\begin{figure}
    \centering
    \includegraphics[width=0.45\textwidth]{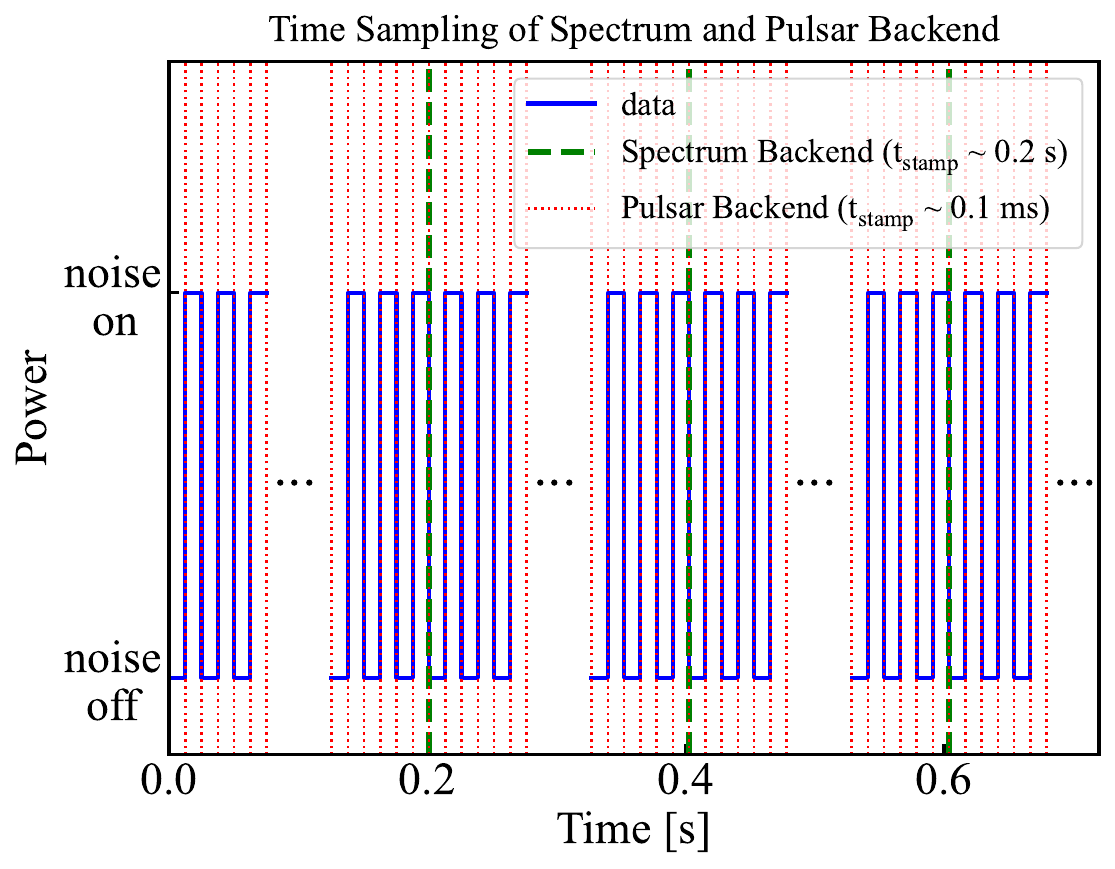}
    \caption{\add{Time sampling for the spectrum backend (green dashed vertical line) and the pulsar backend (red dotted vertical line). The blue solid line represents the variation of data with periodic noise injection. Note that there are actually over 1000 noise on/off periods within 0.2 s. However, for clarity, we display only a few of them and use three dots to indicate the omission.} }
    \label{fig:tbin}
\end{figure}

\begin{figure*}
    \centering
    \includegraphics[width=0.9\textwidth]{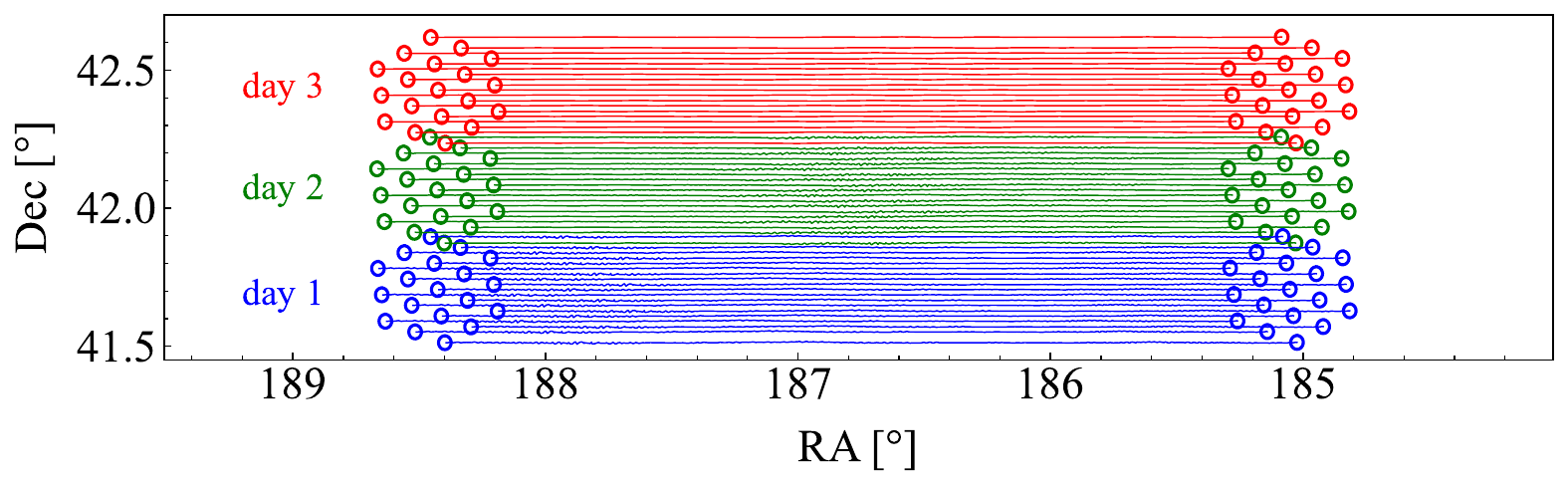}
    \caption{A schematic plot of CRAFTS drift scan survey. Different colors represent different days' observations. The circles mark the position of 19 beams.}
    \label{fig:drift_scan}
\end{figure*}

CRAFTS\footnote{\url{http://groups.bao.ac.cn/ism/english/CRAFTS/CRAFTS/}} \citep{2018IMMag..19..112L} is one of the key projects of FAST designed for commensal observations with high spectral and time resolution. For the purpose of carrying out the spectral survey and transients survey at the same time, the spectrum backend and the pulsar backend are used to record data simultaneously. For the wide-band spectrum backend, the time resolution is $\rm \Delta t \sim 0.2s$ and the frequency resolution is $\rm \Delta\nu \sim 7.6kHz$, while the pulsar backend is $\rm \Delta t \sim 96.304\mu s$ and $\rm \Delta\nu \sim 122kHz$. \add{For illustration, \reffg{fig:tbin} shows the time sampling for the spectrum backend with the green dashed lines and the pulsar backend with the red dotted lines. The time bin of the spectrum backend is 2048 times longer than that of the pulsar backend.} Therefore, we can take advantage of the high time resolution of the pulsar backend and the high-frequency resolution of the spectral backend within our pipeline for different needs. \add{The two} backends record data of four polarization channels related to the four Stokes parameters. In this work, we only focus on the two dual linear polarization data represented with $XX$ and $YY$ to get the intensity $I = XX + YY$. A summary of the parameters of the two backends in CRAFTS observation is presented in \reftb{tb:backend}. A narrow-band spectrum backend with $\rm \Delta t \sim 0.2s$, $\rm \Delta\nu \sim 0.5kHz$, and the total bandwidth of 31.25MHz is also used in CRAFTS observation. However, due to its narrow bandwidth, it is not suitable for our \HI IM and galaxy survey research and is therefore not discussed in this paper. In later data processing sections, we use the subscript ``psr'' to represent data from the pulsar backend and ``spec'' for data from the wide-band spectrum backend. 

The L-band 19-beam receiver of FAST covers the frequency range of 1.05-1.45GHz, corresponding to the \HI redshift range of $0 < z < 0.35$. For optimal sky coverage, these beams are rotated with the angle of $\sim 23.4^{\circ}$ in CRAFTS observations as shown in \reffg{fig:beam_pos}. In each drift scan observation, the 19 beams can cover $\sim 25'$ in declination, see \reffg{fig:drift_scan} for a schematic plot of CRAFTS drift scan mode. A long narrow stripe along the RA direction will be covered in one scan. Note that the distance between the central declination of two adjacent stripes is $\sim 21'$, resulting in a $\sim 4'$ overlap between two days of observation to ensure uniform coverage.

A noise diode signal is injected periodically for data calibration. To avoid data contamination in Fourier space for pulsar search, a novel high-cadence noise injection mode is applied in the CRAFTS project. Generally in other \HI observations, the noise diode signal is injected with a period of several seconds, like 1 s noise injection every 8 s in the FAST \HI IM pilot survey \citep{2023ApJ...954..139L}, 1s noise injection every 600 s in ALFALFA \citep{2018ApJ...861...49H}, or 1.8 s noise injection every 20 s for the MeerKAT single-dish survey \citep{2021MNRAS.505.3698W}. In our case, the noise injection period for CRAFTS, $t_{\rm inj} = \rm 198.608\,\mu s$, is much shorter, which is at the level of the pulsar sampling time scale. The noise diode temperature is $\sim$ 1 K, which is relatively low compared with the system temperature of about 20 K. The time allocation for noise-on and noise-off is $t_{\rm on} = {\rm 81.92\,\mu s}, t_{\rm off} = {\rm 114.688\,\mu s}$ in each injection period. According to the parameters $\Delta t$ in \reftb{tb:backend}, it is obvious that the noise diode signal can not be identified using the data from the spectrum backend only due to its low time resolution. Therefore, the spectrum backend data would contain not only the sky signal but also part of the noise diode signal at the level of $\frac{t_{\rm on}}{t_{\rm on}+t_{\rm off}}{\rm T_{ND}} \sim \frac{1}{2}{\rm T_{ND}} \sim 0.5\, {\rm K}$. Although the sky information we need for \HI intensity mapping is included in the data from the spectrum backend, we also have to rely on the data with and without noise from the pulsar backend for calibration.

\subsection{Data selection}\label{subsec:data_sel}

\begin{figure*}
    \centering
    \includegraphics[width=0.9\textwidth]{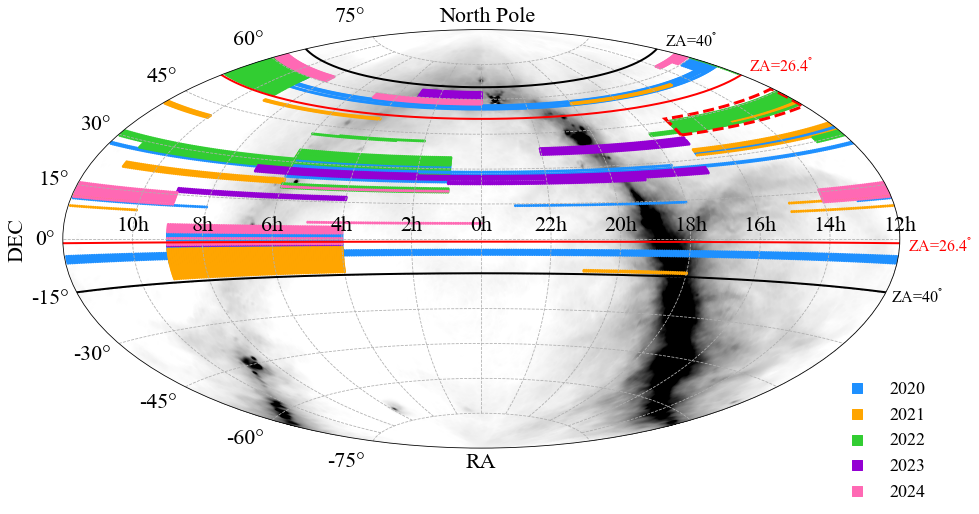}
    \caption{Sky coverage of CRAFTS from February 2020 until July 31, 2024. The gray shadow represents the Haslam radio continuum all-sky map at 408MHz \citep{2015MNRAS.451.4311R} with darker areas indicating stronger foreground radiation. Stripes of different colors represent different observation periods. The red dashed box marks the data used in this work.}
    \label{fig:skycover}
\end{figure*}

\begin{table*} 
    \centering
    \begin{tabular}{ccccc}
        \hline
         \quad Data key \quad & \quad Date \quad & \quad Observation time \quad & \quad RA start \quad & \quad Dec start \quad \\
         \hline
         \quad Dec+4015\_12\_05 \quad & \quad 20221022 \quad & \quad 5h \quad & \quad 12h \quad & \quad $40^\circ15'$ \quad \\
         \quad Dec+4037\_10\_05 \quad & \quad \add{20211001} \quad & \quad 5h \quad & \quad 10h \quad & \quad $40^\circ37'$ \quad \\
         \quad Dec+4037\_15\_03 \quad & \quad 20221005 \quad & \quad 3h \quad & \quad 15h \quad & \quad $40^\circ37'$ \quad \\
         \quad Dec+4059\_10\_05 \quad & \quad \add{20210930} \quad & \quad 5h \quad & \quad 10h \quad & \quad $40^\circ59'$ \quad \\
         \quad Dec+4059\_15\_03 \quad & \quad 20221229 \quad & \quad 3h \quad & \quad 15h \quad & \quad $40^\circ59'$ \quad \\
         \quad Dec+4120\_10\_05 \quad & \quad \add{20210929} \quad & \quad 5h \quad & \quad 10h \quad & \quad $41^\circ20'$ \quad \\
         \quad Dec+4120\_15\_03 \quad & \quad 20221225 \quad & \quad 3h \quad & \quad 15h \quad & \quad $41^\circ20'$ \quad \\
         \quad Dec+4142\_12\_05 \quad & \quad 20221025 \quad & \quad 5h \quad & \quad 12h \quad & \quad $41^\circ42'$ \quad \\
         \quad Dec+4204\_12\_05 \quad & \quad 20221110 \quad & \quad 5h \quad & \quad 12h \quad & \quad $42^\circ04'$ \quad \\
         \quad Dec+4225\_12\_05 \quad & \quad 20221117 \quad & \quad 5h \quad & \quad 12h \quad & \quad $42^\circ25'$ \quad \\
         \quad Dec+4247\_12\_05 \quad & \quad 20221105 \quad & \quad 5h \quad & \quad 12h \quad & \quad $42^\circ47'$ \quad \\
         \quad Dec+4309\_12\_05 \quad & \quad 20221120 \quad & \quad 5h \quad & \quad 12h \quad & \quad $43^\circ09'$ \quad \\
         \quad Dec+4330\_12\_05 \quad & \quad 20221128 \quad & \quad 5h \quad & \quad 12h \quad & \quad $43^\circ30'$ \quad \\
         \quad Dec+4352\_12\_05 \quad & \quad 20221202 \quad & \quad 5h \quad & \quad 12h \quad & \quad $43^\circ52'$ \quad \\
         \quad Dec+4414\_12\_05 \quad & \quad 20221207 \quad & \quad 5h \quad & \quad 12h \quad & \quad $44^\circ14'$ \quad \\
         \quad Dec+4436\_12\_05 \quad & \quad 20221222 \quad & \quad 5h \quad & \quad 12h \quad & \quad $44^\circ36'$ \quad \\
         \quad Dec+4457\_12\_05 \quad & \quad 20221013 \quad & \quad 5h \quad & \quad 12h \quad & \quad $44^\circ57'$ \quad \\
         \hline
    \end{tabular}
    \caption{Data list of the 17 observations for the sky region analyzed in this study.}
    \label{tb:obs_list}
\end{table*}

\begin{figure*}
    \centering
    \includegraphics[width=0.95\textwidth]{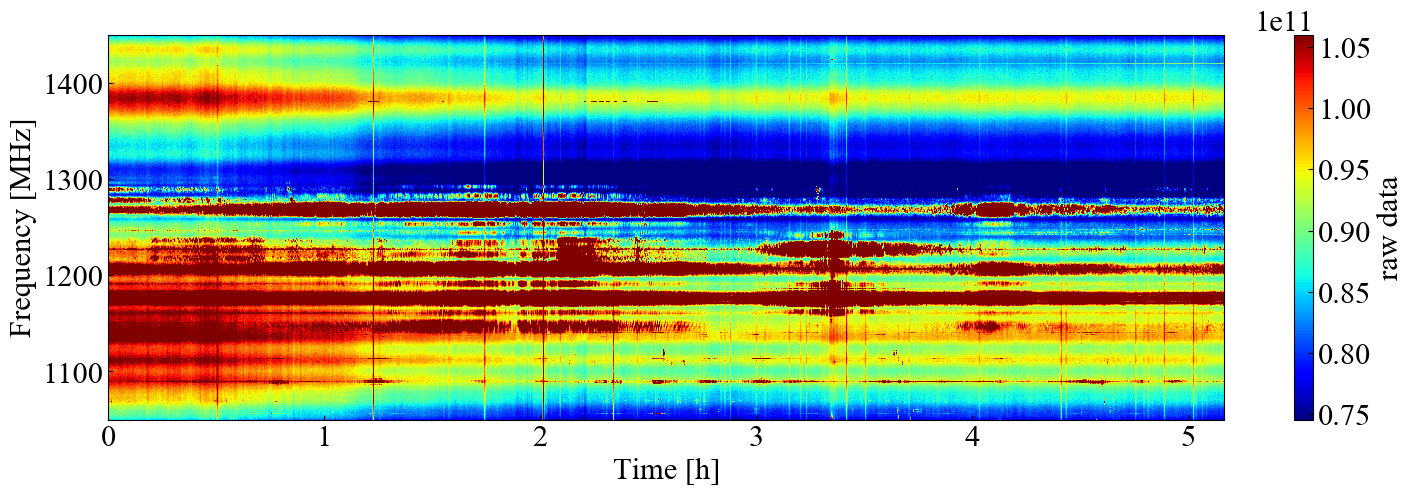}
    \caption{Waterfall plot of the uncalibrated time-ordered data from one 5-hr scan taken by the spectrum backend of one beam and one polarization (M01, XX polarization). The color represents the value of data with no units because they are raw receiver readings with no direct physical meaning. The horizontal stripes are mainly caused by RFI contamination and bandpass shapes.
    }
    \label{fig:raw_wfp}
\end{figure*}

\begin{figure*}
    \centering
    \includegraphics[width=0.9\textwidth]{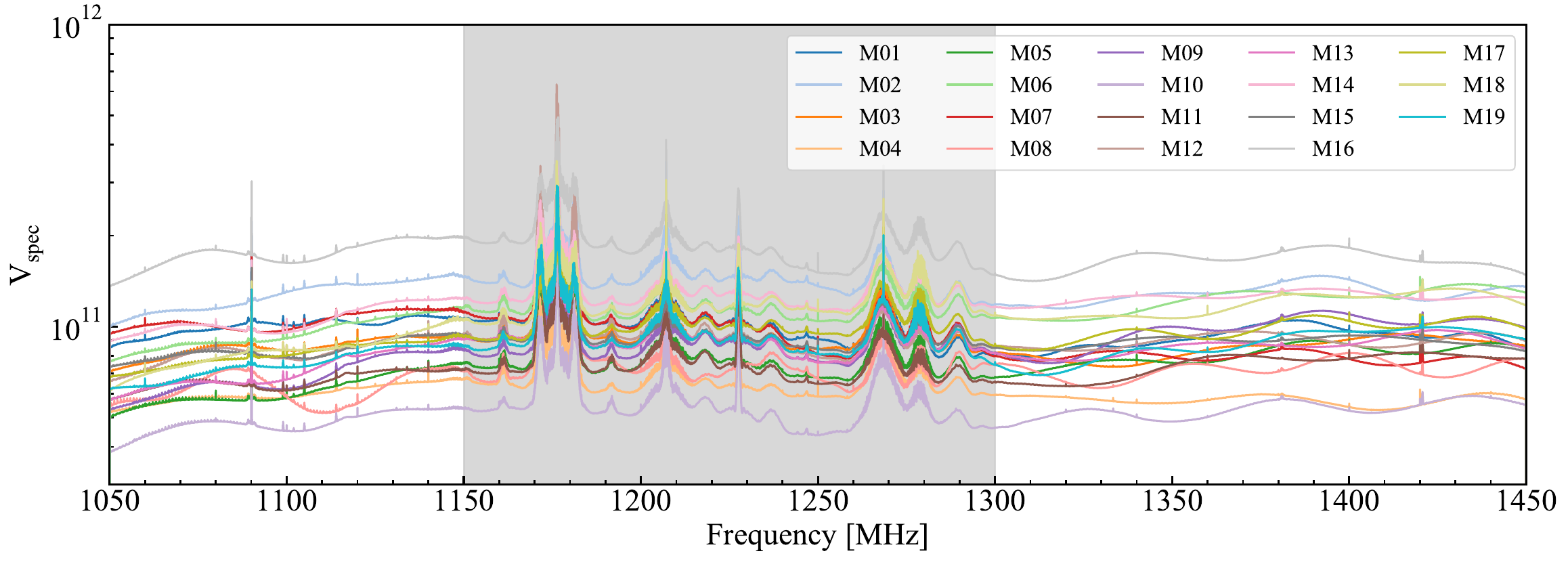}
    \caption{Time averaged spectra of uncalibrated time-ordered-data of all 19 beams (XX polarization, $\sim$ 0.5 hour averaged). The x-axis shows frequency bands and the y-axis ($\rm V_{spec}$) shows the value of raw receiver readings with no unit. Different colors represent different beams. The grey shadow marks the range of RFI contaminated frequency band (1150-1300MHz). }
    \label{fig:raw_spec}
\end{figure*}

\begin{figure*}
    \centering
    \includegraphics[width=0.9\textwidth]{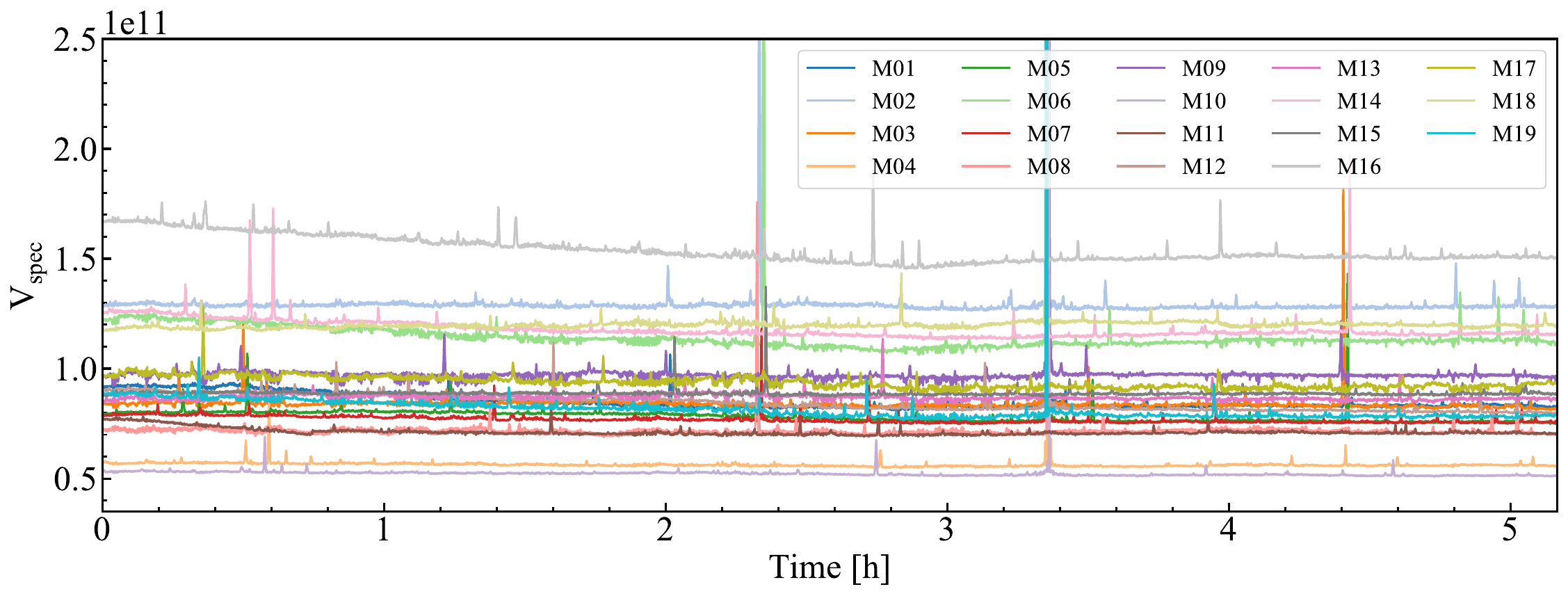}
    \caption{\revise{Frequency averaged time stream of uncalibrated data of all 19 beams (XX polarization, 1300-1450MHz averaged). The x-axis shows the time range and the y-axis ($\rm V_{spec}$) shows the value of raw receiver readings with no unit. Different colors represent different beams.}}
    \label{fig:raw_tod}
\end{figure*}

\add{The current sky coverage of CRAFTS from February 2020 until July 31st, 2024 is shown in \reffg{fig:skycover}. The gray shadow shows the position of galactic plane radiation given by the Haslam map at 408 MHz \citep{2015MNRAS.451.4311R}. The colored regions are scanned areas and different colors represent data in different years. }
\add{Considering the requirement of a continuous sky area for \HI IM} and the data quality of FAST, we use the data from the sky area at $ \rm 12\, h < RA < 17\, h, 40^{\circ} < Dec. < 45^{\circ}$ marked by the red dashed box in \reffg{fig:skycover} in this work. These data cover $\rm \sim 270 deg^2$ observed via 14 stripes obtained from 70 hours of observations distributed over 17 days in 2021 and 2022. \add{All information is listed in detail in \reftb{tb:obs_list}, including five columns for the data key, the observation date in the format of YYYYMMDD, the observation time length, the set start RA and the start Dec of the central beam.}. 
The usable redshift range is $0 < z < 0.07$ and $0.23 < z < 0.35$, limited by the seriously RFI-contaminated frequency bands (1150-1300 MHz) for the FAST L-band 19-beam receiver. The observation time in each day, denoted as $t_{\rm obs}$, is 3 or 5 hours. There are 11 days of scanning that cover 11 stripes within the RA range from 12 h to 17 h, and each of the left three stripes (for Dec $\sim 40^{\circ}37', 40^{\circ}59'$ and $41^{\circ}20'$) is obtained through two days of observation covering RA range from 10h to 15h and from 15h to 18h respectively. As a result, there would be a few re-scanned areas at RA $\sim$ 15 h for these three stripes. Areas outside of the range $\rm 12\,h < RA < 17\,h$ are excluded to simplify the shape of our selected area, which helps in future power spectrum estimation.

\reffg{fig:raw_wfp}, \reffg{fig:raw_spec} \revise{and \reffg{fig:raw_tod}} show the waterfall plot, frequency spectra \revise{and time stream} of the raw data from the spectrum backend, respectively, during a 5-hour scan of Dec $\sim$ $41^\circ42'$, RA from 12 h to 17 h in 2022 as an example. From these \revise{three} figures, we can see variations across both frequency and time axis, which are different for different beams. These structures are mainly caused by the fluctuations of the instrument response and would be suppressed after calibration. The dark red horizontal stripes in the waterfall plot (\reffg{fig:raw_wfp}) and strong peaks in the spectra (\reffg{fig:raw_spec}) indicate the frequency band 1150-1300MHz (highlighted by the grey shadow) is severely contaminated by strong RFIs. At the two nearly RFI-free bands (1050-1150 MHz and 1300-1450 MHz), there are also some small peaks corresponding to weak RFIs, which are similar for all 19 beams and will be further flagged in later RFI flagging process. \revise{The narrow peaks in the time stream (\reffg{fig:raw_tod}) are mainly correspond to the continuum point sources or transient RFIs.}

\section{Data processing} \label{sec:process}

\begin{figure}
    \centering
    \includegraphics[width=0.49\textwidth]{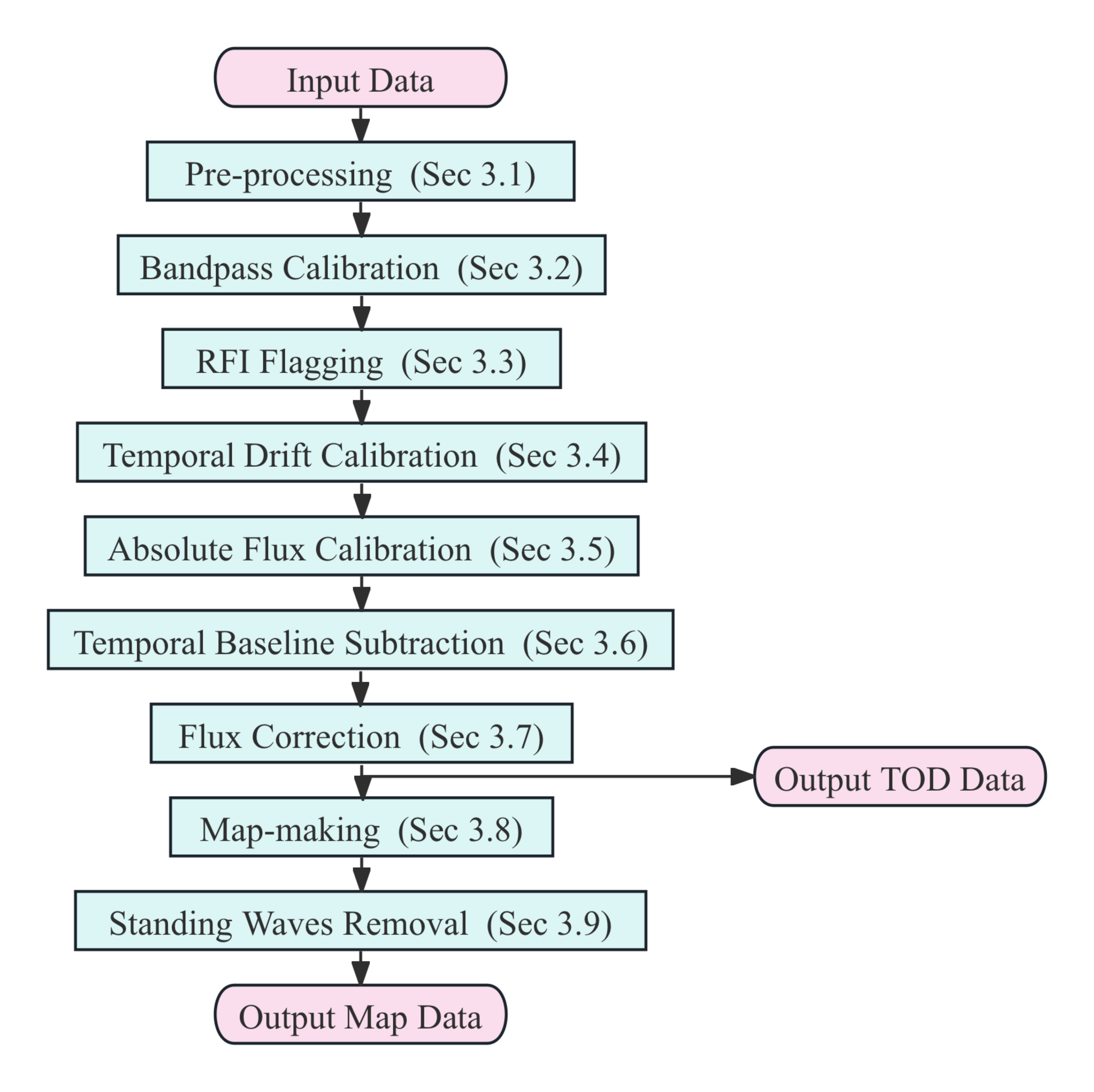}
    \caption{Flow chart of CRAFTS data processing procedures.}
    \label{fig:flowchart}
\end{figure}

The data processing of CRAFTS mainly follows the commonly used procedures for single-dish radio telescope calibration (\citealt{2002ASPC..278..293O}) and the pipelines we have developed for the pilot survey of FAST \HI intensity mapping ({\tt fpipe}, \citealt{2023ApJ...954..139L}) and Tianlai project ({\tt tlpipe}, \citealt{2012IJMPS..12..256C, 2021A&C....3400439Z}). Some adjustments are made for the special characteristics of CRAFTS like the high-cadence calibration mode. The main procedures are shown in the flow chart in \reffg{fig:flowchart}. 
Compared with {\tt fpipe}, the main difference for our CRAFTS pipeline is in the calibration process, in which we employ the pulsar backend in ``bandpass calibration'' (see \refsc{subsec:bpcal}) and ``temporal drift calibration'' (see \refsc{subsec:gtcal}). Besides, due to the lack of a specific sky calibrator in CRAFTS observation, we use the antenna efficiency parameters given in \cite{2020RAA....20...64J} in our ``absolute flux calibration'' (see \refsc{subsec:fluxcal}) instead of measuring the antenna efficiency every time. Furthermore, to reduce systematic errors, we add the ``flux correction'' (see \refsc{subsec:flux_corr}) and ``standing waves removal'' (see \refsc{subsec:rmsw}) in our processes. Some detailed parameters, such as the window size for bandpass smoothing and the kernel size used in the RFIs flagging algorithm, also differ slightly due to the different observation parameters between CRAFTS and the FAST \HI IM pilot survey. Details and tests at each step are described below. In this section, we usually show the calibration results of data from about 5 hours of observation on October 25, 2022, for the stripe at Dec $\sim 41^{\circ}42'$, RA from 12h to 17h (same as \reffg{fig:raw_wfp} and \reffg{fig:raw_spec}) as examples.

\subsection{Pre-processing}\label{subsec:pre_process}

\begin{figure}
    \centering
    \includegraphics[width=0.46\textwidth]{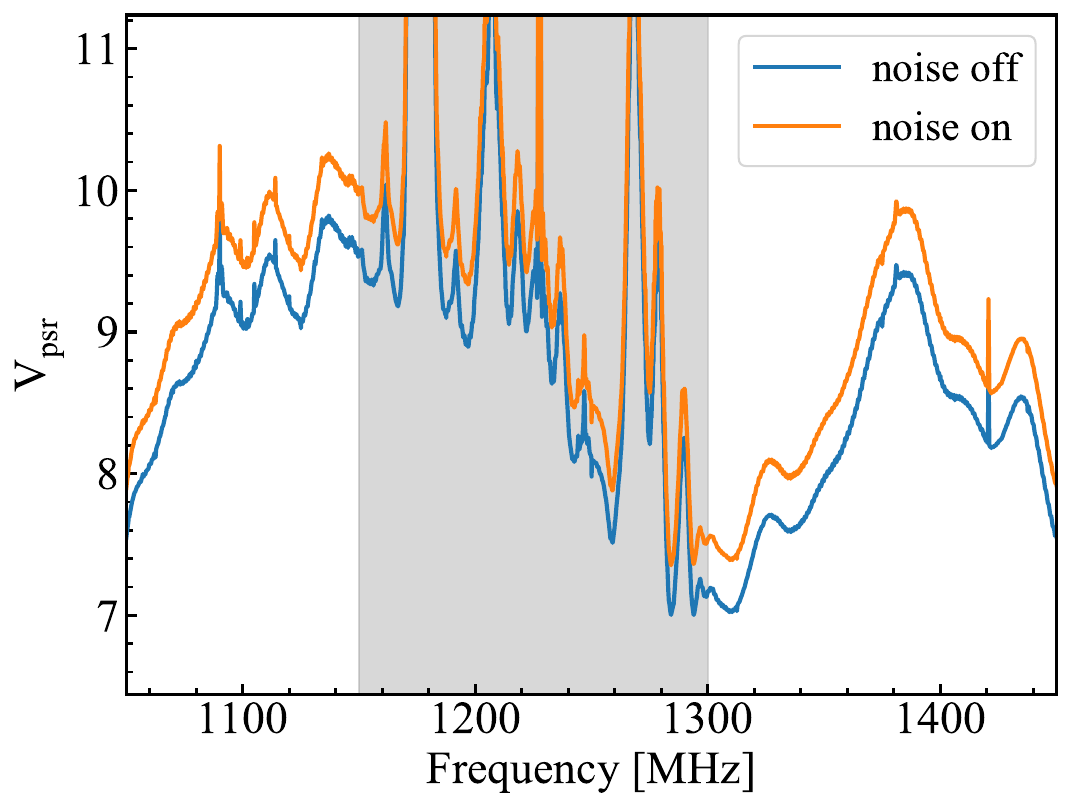}
    \caption{Noise-on (orange line) and noise-off (blue line) spectra from the pulsar backend for M01, XX polarization, $\sim$ 0.5 hour averaged. The values represent the raw receiver reading with no unit. The grey shadow marks the RFI-contaminated frequency band 1150-1300MHz. }
    \label{fig:raw_psr}
\end{figure}


As mentioned in \refsc{sec:data}, the high-cadence injected noise cannot be distinguished by the spectrum backend only, so we need to rely on the pulsar backend for data calibration. The original data from the pulsar backend follow the {\tt psrfits}\footnote{\url{https://www.atnf.csiro.au/research/pulsar/psrfits_definition/Psrfits.html}} format with $\rm \Delta t = 96.304\mu s$. However, the high time resolution of the pulsar backend, which is necessary for transient observation, is not needed for the \HI spectrum survey. For data storage and computation efficiency, a compression process is carried out by the FAST data center to rebin the noise-on and noise-off data in every 0.2s \add{(as shown in \reffg{fig:tbin}) respectively. Specifically, the average value of the noise-on (or noise-off) data from the pulsar backend over 1024 time stamps within each 0.2 s is taken as the new $V_{\rm psr, on}(t,\nu)$ (or $V_{\rm psr, off}(t,\nu)$), where $t$ represents the rebinned time with a resolution of $\Delta t = 0.2$ s.} The results are stored in calibration files for later processing, which contain noise-on and -off data at every time point with interval of $\sim$ 0.2s, consistent with the spectrum backend output file for each frequency channel, beam, and polarization. \reffg{fig:raw_psr} shows an example of the spectrum of noise-on and -off data from the pulsar backend, \add{computed by averaging the raw pulsar backend data from all noise-on time stamps and all noise-off time stamps respectively during half an hour’s observation. Theoretically, the raw noise-on and noise-off data can be written by 
\begin{equation}
    V_{\rm psr, on}(t,\nu) = g_{\rm psr}(t,\nu)(T_{\rm off}(t,\nu)+T_{\rm ND}(\nu)+n(t,\nu))
\end{equation}
and 
\begin{equation}
    V_{\rm psr, off}(t,\nu) = g_{\rm psr}(t,\nu)(T_{\rm off}(t,\nu)+n(t,\nu))\,,
\end{equation}
Where $V_{\rm psr,on}$ and $V_{\rm psr,off}$ are the raw data from pulsar backend for noise-on and noise-off time stamps respectively, $g_{\rm psr}(t,\nu)$ is the gain of the pulsar backend, \revise{$T_{\rm off}(t,\nu)$ is the input temperature without noise diode injection}, $T_{\rm ND}(\nu)$ is the spectrum of noise diode and $n(t,\nu)$ is the noise. Since the gain and input temperature are stable during one on-off period ($\sim 200\mu s$), the noise-on spectrum is supposed to be similar to the noise-off spectrum in the overall shape but has a slightly higher amplitude caused by the noise diode signal injection, which is to the features in \reffg{fig:raw_psr}.} 

\subsection{Bandpass calibration}\label{subsec:bpcal}

\begin{figure}
    \centering
    \includegraphics[width=0.49\textwidth]{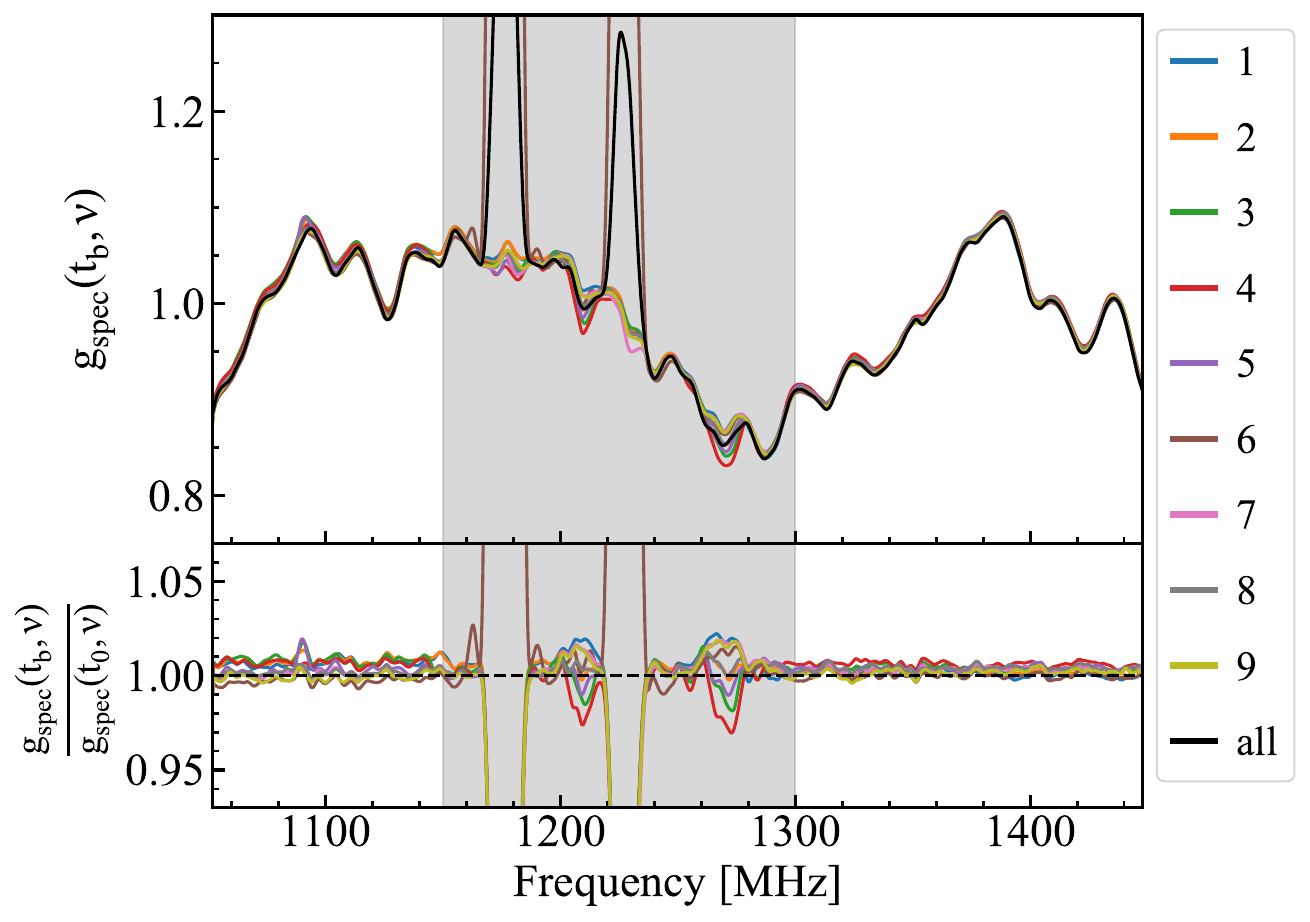}
    \caption{ Comparison of the bandpass shape of nine 0.5-hour time blocks taken during one 5-hour observation. Upper panel: bandpass normalized by its median value. Lower panel: bandpass at each time block relative to the all-time-block averaged bandpass. Each colored line represents one time block of $\sim$ 0.5 hour. The black thick line is the average of all time blocks. The grey shadow between 1150 MHz and 1300 MHz is the severely RFI-contaminated band, which is not used in our work. The significant outliers in the grey area are influenced by very strong RFIs, which means data in these frequency bands are not reliable. }
    \label{fig:bp_shape}
\end{figure}

\begin{figure*}
    \centering
    \includegraphics[width=0.49\textwidth]{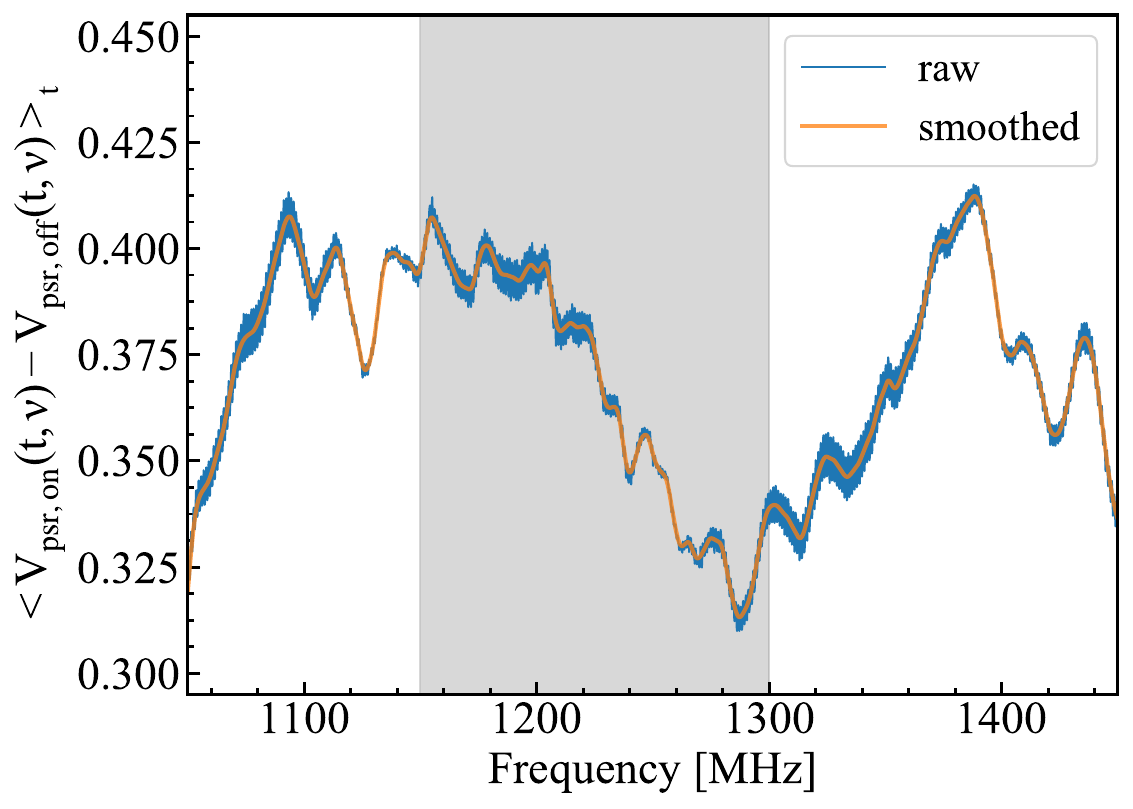}
    \includegraphics[width=0.495\textwidth]{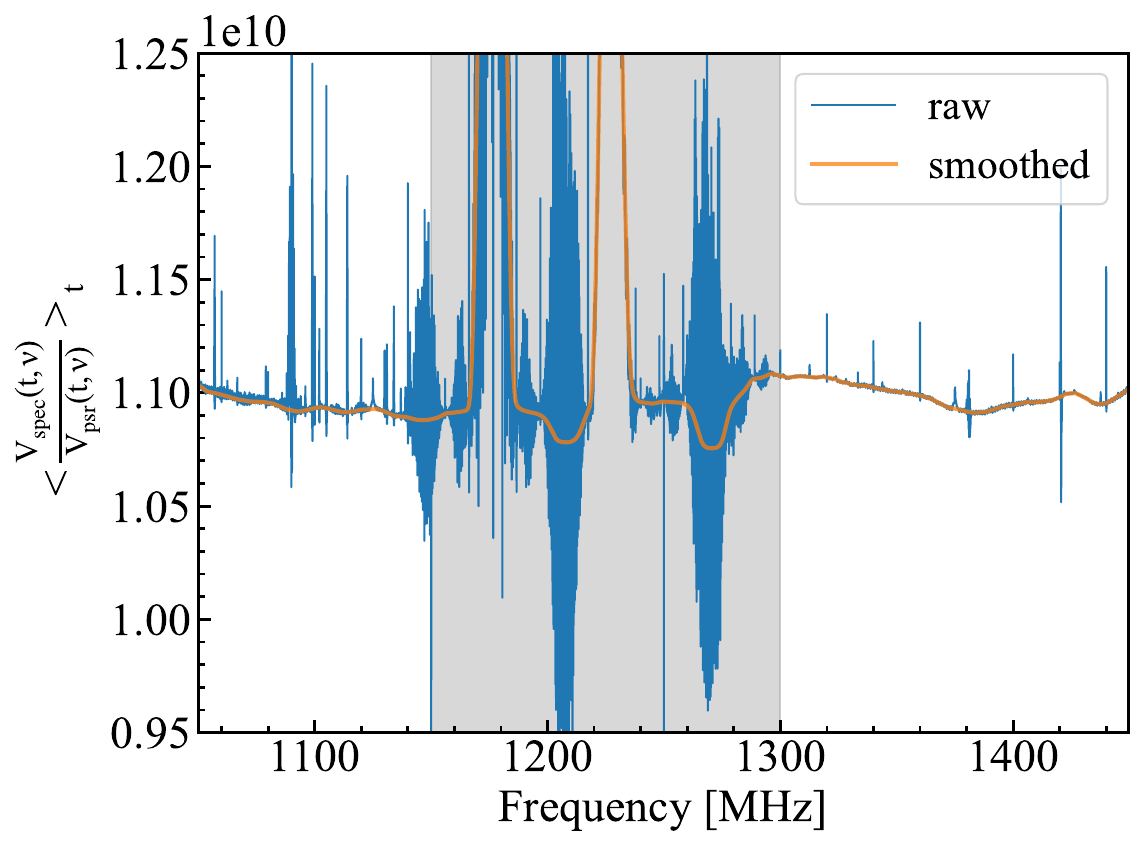}
    \caption{Left: $<V_{\rm psr, on}(t, \nu) - V_{\rm psr, off}(t, \nu)>_t$ before (blue line) and after (orange line) smoothing. Right: $<\frac{V_{\rm spec}(t, \nu)}{V_{\rm psr}(t, \nu)}>_t$ before (blue line) and after (orange line) smoothing. The grey shadows in the two plots are the 1150MHz-1300MHz RFI contaminated band.
    }
    \label{fig:bp+Cv}
\end{figure*}

The raw data from the spectrum backend can be written as
\begin{equation}\label{eq:V_spec}
    \begin{split}
    V_{\rm spec}(t, \nu) 
    &= g_{\rm spec}(t, \nu) ( T(t, \nu) + n(t,\nu) ) \\
    &= g_{t, \rm spec}(t) g_{\nu, \rm spec}(\nu) ( T(t, \nu) + n(t,\nu) ) \,,
    \end{split}
\end{equation}
where $T(t,\nu)$ is the input signal including both the sky signal and all other components such as the receiver noise, atmospheric emission, and the radiation spilling over from the surroundings, $g_{\rm spec}(t,\nu)$ is the gain of the spectrum backend and $n(t,\nu)$ is the noise with $\langle n(t,\nu)\rangle_t = 0$. 
Assuming the gain fluctuations with time and frequency are independent, we have
\begin{equation}\label{eq:gtv_spec}
    g_{\rm spec}(t,\nu) = g_{t, \rm spec}(t) \cdot g_{\nu, \rm spec}(\nu) \,,
\end{equation}
where $g_{t, \rm spec}(t)$ is the temporal drift of gain and $g_{\nu, \rm spec}(\nu)$ is the bandpass. This decomposition is based on the assumption that the shape of the bandpass would not have significant change during our observation time $t_{\rm obs}$. This assumption is proved to be reasonable by comparing the bandpass shape of different time blocks($t_{\rm b} \sim$ 30 minutes) during one day's observation as in \reffg{fig:bp_shape}, from which we can see the bandpass shape at each time block is very similar, with the relative error of $\lesssim 1\%$ at clean frequency bands. Data in the severe RFI contaminated band (marked by the grey shadow) is unreliable and excluded in our later astrophysical analysis.

Similarly, the raw data from the pulsar backend can be written in the same way 
\begin{equation}\label{eq:V_psr}
    \begin{split}
    V_{\rm psr}(t, \nu) 
    &= g_{\rm psr}(t, \nu) ( T(t, \nu) + n(t,\nu) ) \\
    &= g_{t, \rm psr}(t) g_{\nu, \rm psr}(\nu) ( T(t, \nu) + n(t,\nu) ) \,,
    \end{split}
\end{equation}
where $g_{\rm psr}(t,\nu)$ is the gain of the pulsar backend and also decomposed into $g_{t, \rm psr}(t) \cdot g_{\nu, \rm psr}(\nu)$. Data from the pulsar backend can be divided to $V_{\rm psr,on}(t,\nu)$ and $V_{\rm psr,off}(t,\nu)$, representing data with and without noise diode injection, respectively. Therefore, the total power $V_{\rm psr}(t,\nu)$, which includes both noise-on and noise-off data, can also be written as
\begin{equation}\label{eq:V_psr_onoff}
    V_{\rm psr}(t, \nu) = \frac{V_{\rm psr,on}(t,\nu) + V_{\rm psr,off}(t,\nu)}{2} \,.
\end{equation}

To connect the data from the two backends together, we introduce a coefficient $C(t,\nu)$ between them as in
\begin{equation}\label{eq:V_psr2spec}
    V_{\rm spec}(t, \nu) 
    = C(t, \nu) \cdot V_{\rm psr}(t, \nu) 
    = C_t(t) \cdot C_{\nu}(\nu) \cdot V_{\rm psr}(t, \nu) \,,
\end{equation}
in which we apply the decomposition of time and frequency variation of $C(t,\nu)$ again. By introducing \refeq{eq:V_psr} into \refeq{eq:V_psr2spec}, we have
\begin{equation}\label{eq:V_spec2}
    V_{\rm spec}(t, \nu) = C_t(t)g_{t,\rm psr}(t) \cdot C_{\nu}(\nu)g_{\nu, \rm psr}(\nu) \cdot ( T(t, \nu) + n(t,\nu) ) \,,
\end{equation}
in which we regard 
\begin{equation}\label{eq:gv_spec}
    g_{\nu,\rm spec}(\nu) = C_{\nu}(\nu)g_{\nu,\rm psr}(\nu) \,,
\end{equation}
as the bandpass of the spectrum backend.

The coefficient for frequency variation is derived by taking the transposition and time average of \refeq{eq:V_psr2spec},
\begin{equation}\label{eq:Cv_psr2spec}
    C_{\nu}(\nu) 
    = \frac{1}{\overline{C}_t}\langle\frac{V_{\rm spec}(t, \nu)}{V_{\rm psr}(t, \nu)}\rangle_t 
    = \frac{1}{\overline{C}_t}\langle\frac{2 \cdot V_{\rm spec}(t, \nu)}{V_{\rm psr,on}(t, \nu)+V_{\rm psr,off}(t, \nu)}\rangle_t \,,
\end{equation}
where $\overline{C}_t$ is the mean value of $C_t(t)$ over $t_{\rm obs}$ in one day's observation, $V_{\rm psr,on}(t, \nu)$ and $V_{\rm psr,off}(t, \nu)$ refer to the pulsar backend data with and without noise diode signal in each time stamp of $\sim$ 0.2s, respectively, consistent with the time resolution of the spectrum backend. 

As noted in \refsc{subsec:pre_process}, the calibration files contain separately the noise-on and -off data from the pulsar backend. According to \refeq{eq:V_psr} and relying on the injected noise diode spectrum, the bandpass $g_{\nu,\rm psr}(\nu)$ and temporal drift $g_{t,\rm psr}(t)$ of the pulsar backend can be expressed as
\begin{equation}\label{eq:V_psr_diff}
    V_{\rm psr, on}(t, \nu) - V_{\rm psr, off}(t, \nu) = g_{t, \rm psr}(t)g_{\nu, \rm psr}(\nu)( T_{\rm ND}(\nu) + n(t,\nu) ) \,,
\end{equation}
where $T_{\rm ND}(\nu)$ is the temperature spectrum of the noise diode measured using the hot-load method every  few months \footnote{See noise diode calibration reports at FAST website:\,\url{https://fast.bao.ac.cn/cms/category/telescope_performence_en/noise_diode_calibration_report_en/}}. Averaging \refeq{eq:V_psr_diff} over the observation time $t_{\rm obs}$ in one day, 
\begin{equation}\label{eq:gv_psr}
    g_{\nu, \rm psr}(\nu) = \frac{\langle V_{\rm psr, on}(t, \nu) - V_{\rm psr, off}(t, \nu)\rangle_t}{\overline{g}_{t,\rm psr}T_{\rm ND}(\nu)} \,,
\end{equation}
where $\overline{g}_{t,\rm psr}$ is the mean value of $g_{t,\rm psr}(t)$ over $t_{\rm obs}$.
Combining \refeq{eq:V_spec2}, \refeq{eq:Cv_psr2spec} and \refeq{eq:gv_psr}, the bandpass calibrated data is
\begin{equation}\label{eq:bpcal}
    \begin{split}
        V_1(t,\nu) 
        &\equiv \frac{V_{\rm spec}(t,\nu)}{\langle\frac{V_{\rm spec}(t, \nu)}{V_{\rm psr}(t, \nu)}\rangle_t \cdot \langle V_{\rm psr, on}(t, \nu) - V_{\rm psr, off}(t, \nu)\rangle_t} \\ \\
        &= \frac{C_t(t)g_{t, \rm psr}(t)}{\overline{C}_t \, \overline{g}_{t, \rm psr}} \cdot \frac{T(t,\nu)+ n(t,\nu)}{T_{\rm ND}(\nu)}  \,.
    \end{split}
\end{equation}
Note that since the frequency resolution of the spectrum backend ($\sim$ 7.6kHz) is 16 times higher than that of the pulsar backend ($\sim$ 122kHz), we interpolate data from the pulsar backend by linear interpolation to make them consistent.

Since the RFI-contaminated time stamps may affect the bandpass shape, we exclude the data points that deviate over 3$\sigma$ from the temporal variation before performing the time average of $\frac{V_{\rm spec}(t, \nu)}{V_{\rm psr}(t, \nu)}$ and $V_{\rm psr, on}(t, \nu) - V_{\rm psr, off}(t, \nu)$. \add{Here the temporal variation is computed by averaging the spectrum backend data for each beam and polarization across the RFI-free bands (1050-1150 MHz and 1300-1450 MHz excluding known strong RFIs as listed in \cite{2021RAA....21...18W}).} There are also thermal noise and standing waves in the bandpass. In order to remove them, we smooth $\langle\frac{V_{\rm spec}(t, \nu)}{V_{\rm psr}(t, \nu)}\rangle_t$ and $\langle V_{\rm psr, on}(t, \nu) - V_{\rm psr, off}(t, \nu)\rangle_t$ by a median filter with window size of $\sim$ 1MHz and a Hanning filter with window size of $\sim$ 5MHz before performing the bandpass calibration as in \refeq{eq:bpcal}. The window size is chosen to ensure the complete removal of thermal noise and standing waves with the period of $\sim$ 1.1MHz, preserve the bandpass shape at the scale of $\sim$ 10MHz and suppress the influence of most RFIs at the scientific bands. Examples of $\langle\frac{V_{\rm spec}(t, \nu)}{V_{\rm psr}(t, \nu)}\rangle_t$ and $\langle V_{\rm psr, on}(t, \nu) - V_{\rm psr, off}(t, \nu)\rangle_t$ before and after smoothing are shown in the left and right plots of \reffg{fig:bp+Cv}, respectively. We can see that the smoothed lines fit the overall shape of the raw noisy lines well, and the influence of spikes are suppressed at the RFI-free bands. \add{One thing to be clarified is that the filters are only applied to the bandpass $g(\nu)$ instead of the spectral survey data. }
After bandpass calibration, the variation of data over frequency is much reduced.  

\subsection{RFI flagging}\label{subsec:RFIflag}

\begin{figure*}
    \centering
    \includegraphics[width=0.49\textwidth]{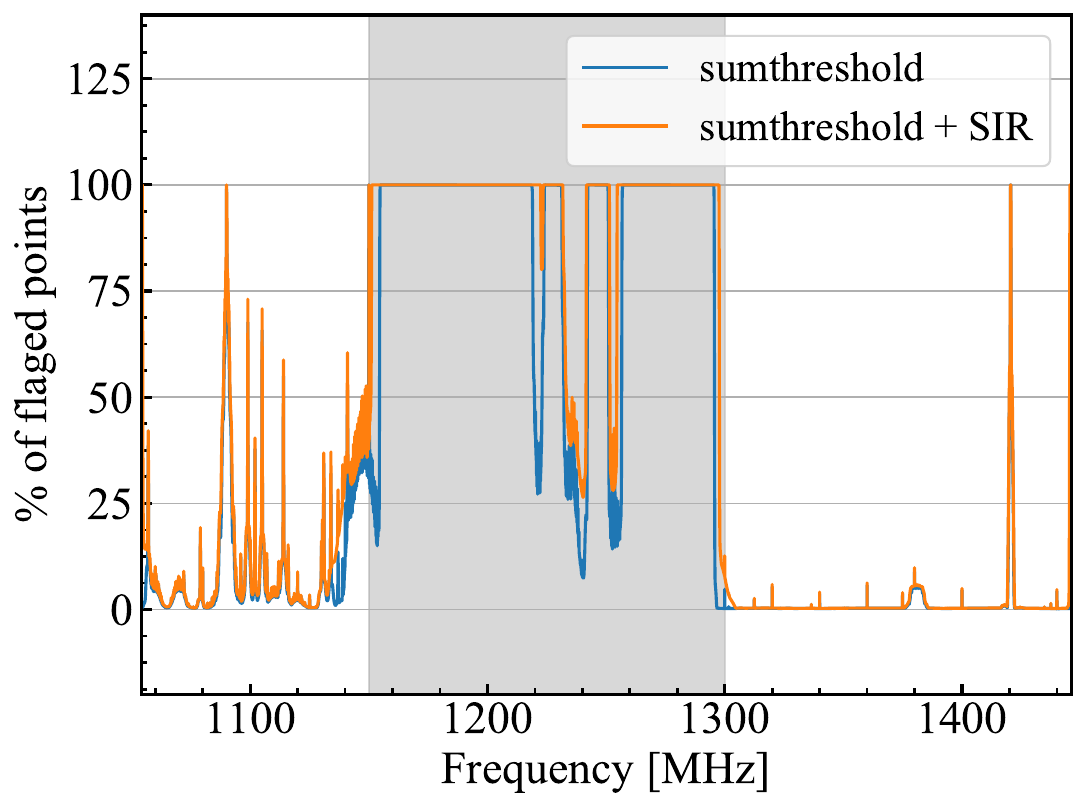}
    \includegraphics[width=0.495\textwidth]{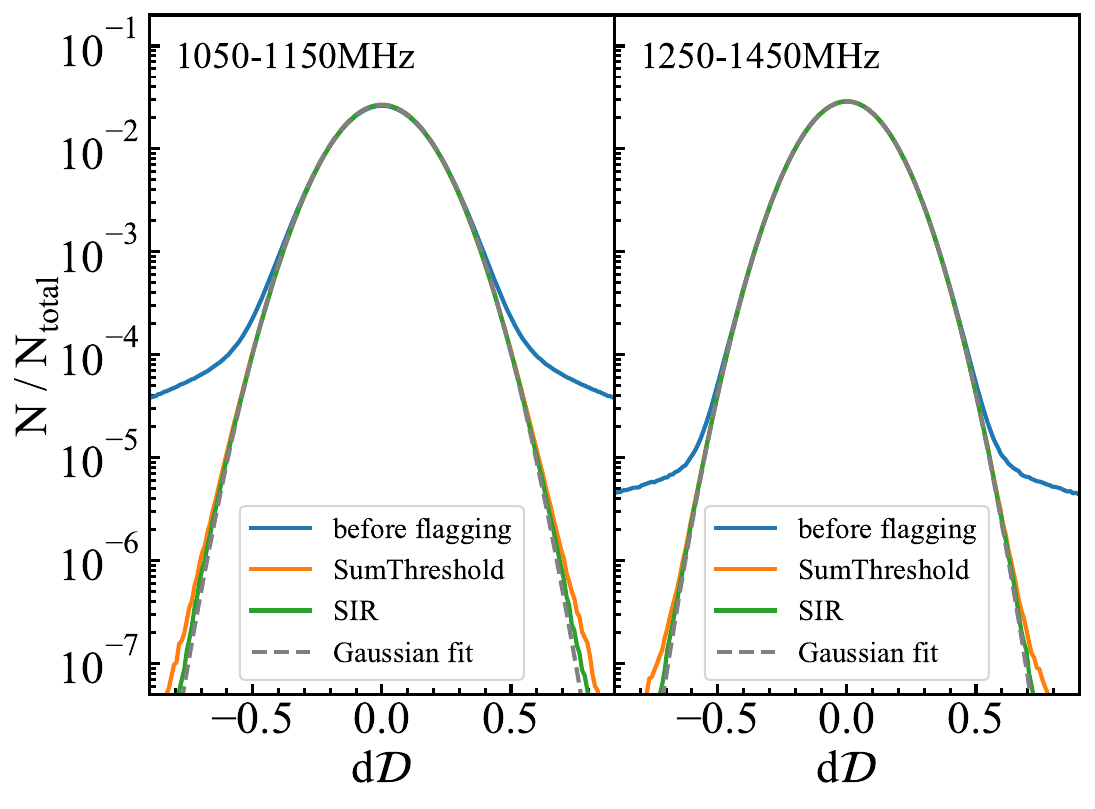}
    \caption{Left: Percentage of flagged points at each frequency channel for 5 hours observation. The blue line shows the result by {\tt SumThreshold} and the orange line shows result from {\tt SumThreshold} and {\tt SIR} algorithm. The grey shadow marks the frequency range (1150-1300MHz) severely contaminated by RFIs. \add{Right: Histogram of the difference (${\rm d}\mathcal{D}$) between adjacent channels after RFI flagging. The left and right panel show the results for 1050-1150 MHz and 1250-1450 MHz respectively. The solid lines with different colors represent the histograms of data before RFI flagging (blue), after {\tt SumThreshold} flagging (orange) and after {\tt SumThreshold} $+$ {\tt SIR} flagging (green). The gray dashed lines are Gaussian fit results for RFI-flagged data.}}
    \label{fig:RFIflag_hist}
\end{figure*}

In \reffg{fig:raw_wfp}, it is clear that strong RFIs appear in the range of 1150-1300 MHz, which are mainly caused by satellites \citep{2021RAA....21...18W}. This RFI contaminated band is ignored in later processes. However, even in the frequency bands 1050-1150 MHz and 1300-1450 MHz that are relatively free of RFIs, there are still some low level interferences present. An effective method is needed to distinguish RFIs from real sky signals. Note that the main reason for applying RFI flagging after bandpass calibration is to lower fluctuations in the frequency domain to identify low-level RFIs in the data.

As a first stage, we average the bandpass calibrated time-ordered data from all 19 beams and 2 polarizations. This process lowers the noise level and helps us find weak RFIs based on the assumption that these RFIs would exist in several beams and polarizations. 

We apply the {\tt SumThreshold} and {\tt SIR} (scale-invariant-rank) methods in {\tt tlpipe} described in \cite{2021A&C....3400439Z} to flag RFI in the beam and polarization averaged bandpass calibrated data. Specifically, the {\tt SumThreshold} algorithm flags data deviating from the smoothed baseline over the threshold $\chi_n = \chi_1/1.5^{{\rm log}2^n}$, in which $\chi_1 = 10$ is the first threshold and $n$ is the number of the data samples considered. This flagging is only applied in the frequency domain to avoid masking sky continuum sources\add{, meaning it is performed on the frequency spectra at each time stamp.}. After that, the {\tt SIR} is performed to further flag weak interferences near the RFIs found by {\tt SumThreshold} if more than $1-\eta_{\rm SIR} (\%) = 95\%$ samples in the sub-sequence are flagged. 
The left plot of \reffg{fig:RFIflag_hist} presents the percentage of flagged time points for each frequency channel. We can see even for the ``RFI-free'' bands, there is still some RFI contamination, especially for 1050-1150MHz. The {\tt SIR} helps to extend the masking array at the weak tails of strong RFIs identified by {\tt SumThreshold}. The flagging results are then applied on each beam and polarization.

To validate the flagging results, we perform some statistical checks to see if the noise in unmasked data is mainly thermal noise. As an example in the right plot of \reffg{fig:RFIflag_hist}, we present the histograms of the difference $d\mathcal{D}$ between adjacent frequency channels of the data from M01, XX polarization, for two frequency bands. We confirmed that these results are representative of other polarizations and beams. \add{From the histograms, clear tails can be observed in the blue lines, particularly in the lower frequency band. Before applying {\tt SIR}, most non-Gaussian components have already been flagged by {\tt SumThreshold}, though small deviations remain in the tails. After completing the two-step RFI flagging process, as shown by the green histograms,} the non-Gaussian tails at high $d\mathcal{D}$ almost disappear and the histogram of residual data follows the Gaussian curve well, which conforms to the feature of thermal noise. A detailed comparison between the noise of RFI-free data and the theoretical variance, discussed in \refsc{subsec:detection_limit}, shows a good agreement between the data noise and the theoretical predictions.

We also perform Jackknife tests to check the influence of individual beams. We try 19 different input datasets for RFI flagging with each of them containing the 18 beams averaged data excluding one beam. We find that there is usually 1-3\% more data flagged when 19 beams of data are averaged compared with 18 beams averaged, which is consistent with the $\sim 2.7\%$ lower noise level when we include all beams. The masking results for the 19 datasets are very similar, indicating the robustness of our RFI flagging process. Details about the Jackknife test results are shown in Appendix~\ref{appx:jackknife}.

\subsection{Temporal drift calibration}\label{subsec:gtcal}

After RFI-flagging, we rebin data for computation efficiency while retaining the scientifically required resolution like the high requirement of spatial resolution for point sources detection and the high requirement of frequency resolution for galaxy emission detection. The time resolution of $\Delta t \sim 0.2{\rm s}$ is reset to be $\Delta t \sim 1{\rm s}$ corresponding to the spatial size of $\sim$ 0.25 arcmin. The frequency resolution of $\Delta \nu \sim 7.6{\rm kHz}$ is reset to be $\Delta \nu \sim 30{\rm kHz}$ driven by the typical width of neutral hydrogen emission lines of $\sim$ 1MHz and the requirement of redshift resolution for intensity mapping.  

Since all receiver systems suffer from 1/f noise \citep{2021MNRAS.508.2897H} which will introduce time-correlated fluctuations, the temporal drift calibration is required. Similar to \refeq{eq:Cv_psr2spec} and 
\refeq{eq:gv_psr}, we use 
\begin{eqnarray}\label{eq:Ct_psr2spec}
    C_t(t) 
    &=& \frac{1}{\overline{C}_{\nu}}\langle\frac{V_{\rm spec}(t, \nu)}{V_{\rm psr}(t, \nu)}\rangle_{\nu} \nonumber \\
    &=& \frac{1}{\overline{C}_{\nu}}\langle\frac{2 \cdot V_{\rm spec}(t, \nu)}{V_{\rm psr,on}(t, \nu)+V_{\rm psr,off}(t, \nu)}\rangle_{\nu} \,,
\end{eqnarray}
and
\begin{equation}\label{eq:gt_psr}
    g_{t,\rm psr}(t) = \frac{\langle V_{\rm psr, on}(t, \nu) - V_{\rm psr, off}(t, \nu)\rangle_{\nu}}{\overline{g}_{\nu,\rm psr}(\langle T_{\rm ND}(\nu)\rangle_{\nu}+n(t))} \,,
\end{equation}
to calculate the temporal coefficient $C_t(t)$ and the temporal drift of the pulsar backend $g_{t,\rm psr}(t)$.

\begin{figure}
    \centering
    \includegraphics[width=0.47\textwidth]{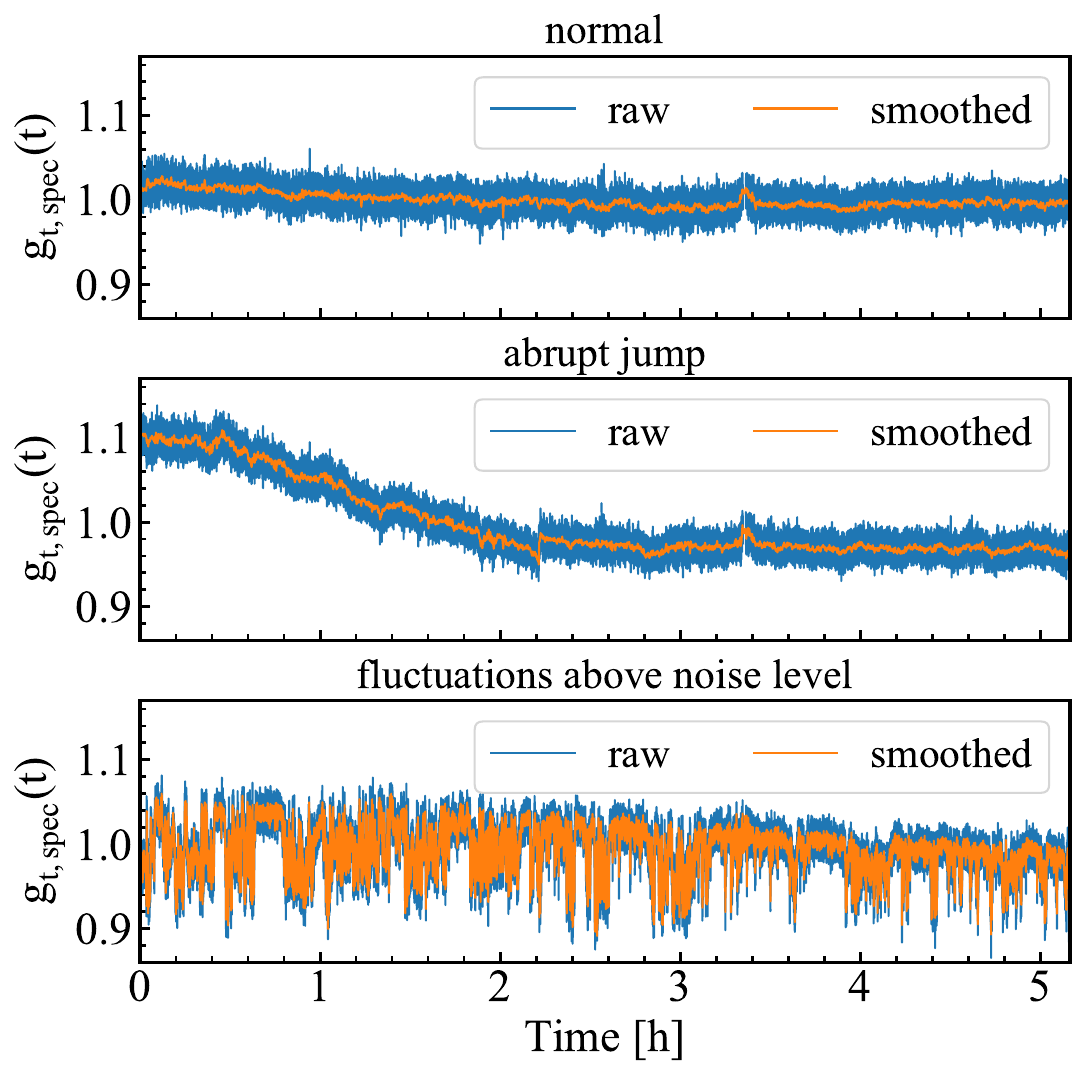}
    \caption{Examples of $g_{t,\rm spec}(t)$ before and after smoothing. The blue line and orange line in each sub-plot show the $g_{t,\rm spec}$ before and after smoothing, respectively. Top: a normal $g_{t,\rm spec}$. Middle: $g_{t,\rm spec}$ with an abrupt jump at $\sim$ 2.2h. Bottom: $g_{t,\rm spec}$ with violent fluctuations. 
    }
    \label{fig:gt}
\end{figure}

By normalizing the two equations above by their time-averaged values over $t_{\rm obs}$, we obtain
\begin{equation}\label{eq:Ct_norm}
    \frac{C_t(t)}{\overline{C}_t} 
    = \frac{\langle V_{\rm spec}(t, \nu)/V_{\rm psr}(t, \nu)\rangle_{\nu}}{\langle V_{\rm spec}(t, \nu)/V_{\rm psr}(t, \nu)\rangle_{\nu,t}} \,,
\end{equation}
and
\begin{equation}\label{eq:gt_psr_norm}
    \frac{g_{t,\rm psr}(t)}{\overline{g}_{t,\rm psr}} 
    = \frac{\langle V_{\rm psr, on}(t, \nu) - V_{\rm psr, off}(t, \nu)\rangle_{\nu}}{\langle V_{\rm psr, on}(t, \nu) - V_{\rm psr, off}(t, \nu)\rangle_{\nu,t}}  \,.
\end{equation}
Furthermore, we apply a Hanning filter with window size $\sim$ 15s for $g_{t,\rm psr}(t)$ and $C_t(t)$ to suppress noise fluctuations. \add{This window size is smaller than the time it takes for a source to pass through a single beam ($ t_{\rm beam} = \rm {\frac{3\, arcmin}{0.25\, arcsec/s \times cos(Dec)} \sim 16 s}$ where $\rm 3\, arcmin$ is the beam size, $\rm 0.25 arcsec/s$ is the Earth rotation speed and Dec$\sim 42^{\circ}$ is the declination). 
} The top figure in \reffg{fig:gt} presents an example of the $g_{t,\rm spec}(t)$ before and after smoothing. We can see the temporal drift varies slowly over time with thermal noise which could be further removed by smoothing.

According to \refeq{eq:bpcal}, we perform the temporal drift calibration as
\begin{equation}\label{eq:gtcal}
    V_2(t,\nu) = 
    \frac{V_1(t,\nu)}{\frac{C_t(t)}{\overline{C}_t} \cdot \frac{g_{t,\rm psr}(t)}{\overline{g}_{t,\rm psr}}}  \,.
\end{equation}
where the denominator $\frac{C_t(t)}{\overline{C}_t} \cdot \frac{g_{t,\rm psr}(t)}{\overline{g}_{t,\rm psr}}$ can be regarded as $g_{t,\rm spec}(t) = g_{t,\rm psr}(t) \cdot C_t(t)$ normalized by its time average over $t_{\rm obs}$.

During the temporal drift calibration process, we found some irregularities in the data that require correction so that they would not bias the calibration process. For example, we find some abrupt changes in amplitude in $g_{t,\rm spec}(t)$, as shown in the middle plot of \reffg{fig:gt} at $t \sim$2.2h\add{, which exists at all frequency channels. The jumps appear 23 times in our 17 days observation with 19 beams and 2 polarizations, which is about 3.7\% of total data and there is no significant regularity of their appearance}. Our current understanding is that this feature could be a transient change in instrument response because it also appears in the uncalibrated data and can be removed after the temporal drift calibration if we keep this structure in $g_{t,\rm spec}(t)$. Therefore, to avoid removing the abrupt jumps in $g_{t,\rm spec}(t)$, before we smooth $g_{t,\rm spec}(t)$, we visually inspect the data to identify the time index of the temporal irregularities. We then divide the dataset such that the abrupt change is maintained in the smoothed $g_{t,\rm spec}(t)$. In this way, we can avoid introducing this kind of systematic structure in our calibrated data. 

Additionally, in some beams and polarizations, there are also some fluctuations significantly above the noise level as shown in the bottom panel in \reffg{fig:gt}. The reasons for these structures are still unknown. These sharp drops severely influence the smoothing results of $g_{t,\rm spec}(t)$ and would introduce more error in the calibrated data. To reduce the influence of these data, we set criteria to find such kind of ``bad data'' and exclude them before map-making. The selection procedures are as follows:\\
(1) Remove the smooth baseline of $g_{t,\rm spec}(t)$ for each beam and polarization by Hanning filter with window size $\sim$ 10 minutes;\\
(2) Plot histogram of each baseline subtracted $g_{t,\rm spec}(t)$ and calculate the kurtosis and skewness values;\\
(3) Calculate the mean value $\mu'$ and standard deviation $\sigma'$ of kurtosis and skewness respectively. Delete the significant outliers (over $3\sigma'$ deviated from $\mu'$) and recalculate $\mu'$ and $\sigma'$ again. Iterate this process until no new outlier is found and get the final $\mu$ and $\sigma$. Record one beam and polarization as bad data if its kurtosis or skewness value $x$ satisfies $|x-\mu| > 3\sigma$. \\

\add{The abnormal fluctuations above noise level are identified in 43 datasets, where each dataset corresponds to the data collected from a single day, a single beam and a single polarization. These data is excluded before making the map in \refsc{subsec:map-making}, resulting in a loss of $\sim$ 6.7\% of the total data.} More details and examples about identification of bad data are shown in Appendix~\ref{appx:bad_data}. We have not been able to identify a pattern in the irregularities of the data, however, we find that abnormal fluctuations in data are more likely to appear in some beams and polarizations (e.g. M10 YY polarization). Therefore, more attention should be paid to data from these beams and polarizations.

\subsection{Absolute flux calibration}\label{subsec:fluxcal}

The absolute flux calibration is performed by introducing the noise diode spectrum $T_{\rm ND}(\nu)$ which is $\sim$ 1K measured by the hot-load method, as well as the aperture efficiency $\eta$ which varies with frequency $\nu$ and zenith angle $\theta_{\rm ZA}$. Following \cite{2023ApJ...954..139L}, the calibrated temperature is
\begin{equation}\label{eq:Tcal}
    T(t,\nu) = V_2(t,\nu) \cdot \frac{2 t_{\rm on}}{t_{\rm on}+t_{\rm off}} T_{\rm ND}(\nu) / \eta(\theta_{\rm ZA}(t),\nu)  \,.
\end{equation}
The $t_{\rm on} = {\rm 81.92\mu s}, t_{\rm off} = {\rm 114.688\mu s}$ are the time allocation for noise injection. The noise diode temperature $T_{\rm ND}(\nu)$ is measured every year by the FAST group. In our process, we choose the $T_{\rm ND}(\nu)$ measured at the time closest to the observation date of our data. The $\eta(\theta_{\rm ZA}(t),\nu)$ used here is a fitting results with the measured $\eta$ at some certain ($\theta_{\rm ZA}$, $\nu$) obtained from \cite{2020RAA....20...64J}, instead of observing a sky calibrator to get a continuous $\eta(\nu)$ at a certain $\theta_{\rm ZA}$ every time. Therefore, there would be flux error in this process because of the instability of the telescope over a long time period. 
\add{According to the measurement of $\eta$ in \cite{2023ApJ...954..139L} and $T_{\rm ND}$ by FAST data center, both $\eta$ and $T_{\rm ND}$ have stable shapes, with their main source of error arising from variations in amplitude. These variations are $\sim$4\% for $\eta$ at similar zenith angle and $\sim$5\% for $T_{\rm ND}$ as illustrated in \refsc{appx:TND}. The flux error introduced by these amplitude variations affects only the overall scaling of the calibrated data and can be corrected through our flux correction process, which is discussed in detail in \refsc{subsec:flux_corr}. Therefore, we believe the variation of $\eta$ and $T_{\rm ND}$ would not introduce a significant systematic error to our calibration.}

The measured temperature is converted to a flux unit by
\begin{equation}\label{eq:T2Jy}
    S(t,\nu) =  \frac{2k_{\rm B}}{A_{\rm geo}} T(t,\nu) \,,
\end{equation}
where $S(t,\nu)$ is the absolute flux density, $k_B$ is the Boltzmann constant, $A_{\rm geo} = 70700{\rm m}^2$ is the geometric illuminated area of FAST. Note that the antenna efficiency $\eta$ has already been included in $T(t,\nu)$ as shown in \refeq{eq:Tcal}.

\subsection{Temporal baseline subtraction}\label{subsec:rmbsl}

The calibrated data contains the signals as well as the system temperature. In \cite{2020RAA....20...64J}, the system temperature $T_{\rm sys}$ is at the level of $\sim 20~ \rm K$ which includes the receiver temperature, continuum brightness from the sky (including CMB and some galactic non-thermal emission), atmosphere emission, and terrain radiation. 
\add{However, the system temperature also varies by several Kelvins depending on the beam, zenith angle and sky background which contributes to the systematic difference between beams at the level of $\sim$ 3K. Another potential source of systematics is the variation in diffuse galactic synchrotron emission. By extrapolating the Haslam map at 408MHz, we find the diffuse synchrotron emission in our observed sky area is about 0.8K, accounting for 4\% of the total system temperature and, within this sky area, the standard deviation of 0.08 K, with a maximum variation of 0.45 K.}
To lower the systematic difference between different beams, polarizations, and observation time, we subtract the temporal baseline of the calibrated data.

\begin{figure}
    \centering
    \includegraphics[width=0.49\textwidth]{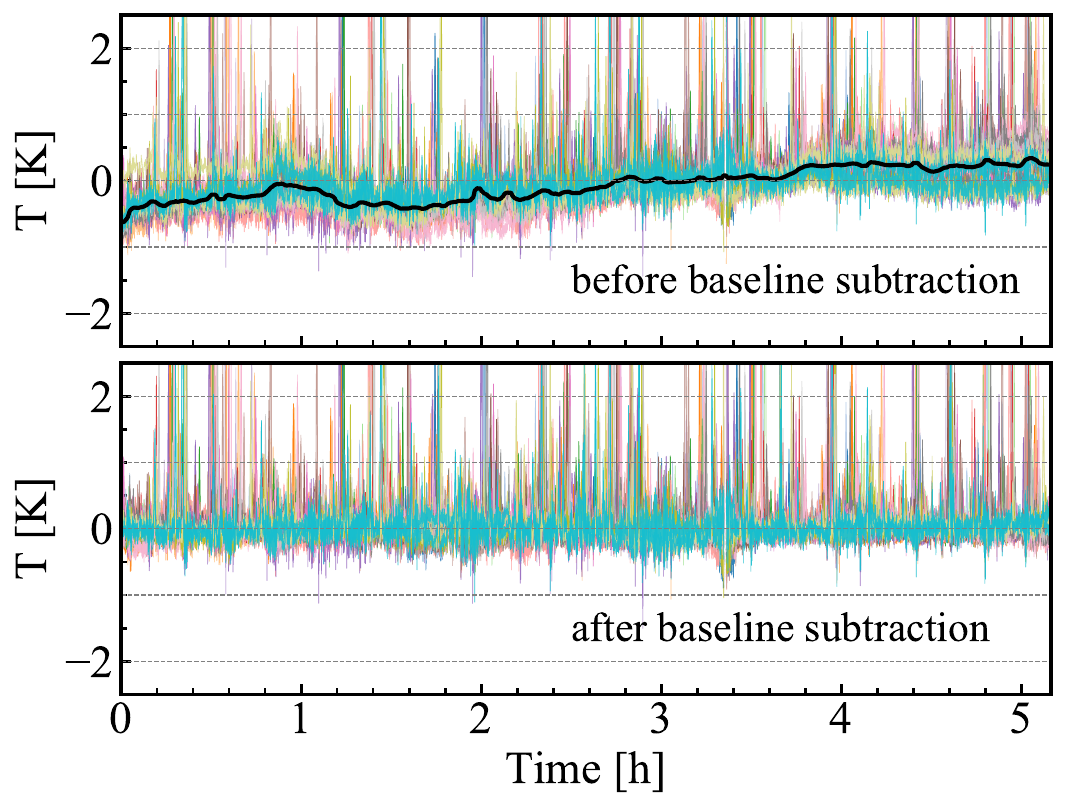}
    \caption{Upper: Temporal fluctuations of 19 beams from one day's observation as an example. The temporal variation for each beam has been centered by subtracting the median value over $t_{\rm obs}$ and unmasked frequency channels. Different colors represent different beams. The black thick line is the template baseline. Lower: Temporal fluctuations of 19 beams after temporal baseline subtraction. Each colored line represents one beam. }
    \label{fig:rmbsl}
\end{figure}

We firstly subtract the median value of $T(t,\nu)$ over $t_{\rm obs}$ and unmasked frequency bands for each beam and polarization to center them at zero. Then we calculate a template baseline by averaging the zero-centered data across all unmasked frequency channels, beams, and polarizations. To remove spikes and thermal noise, the averaged data is then smoothed by a median filter with window size $\sim$ 200s. The template baseline, denoted as an $n_t \times 1$ matrix $\textbf{b}$, is shown by the black line in the upper panel of \reffg{fig:rmbsl} as an example from $\sim$5 hours observation. The temporal fluctuation of all 19 beams and 2 polarizations are also shown with the colored lines in the upper panel of \reffg{fig:rmbsl}. It can be seen that there are some small fluctuations in the baseline for individual beam and polarization, but the overall variations for all lines are similar, which can be generally represented by the template baseline. The spikes in \reffg{fig:rmbsl} are bright continuum sources. We fit the calibrated time-ordered data (TOD) with the template baseline $\textbf{b}$ following \cite{2023ApJ...954..139L} with
\begin{equation}\label{eq:fit_bsl}
    \textbf{A} = \left (\textbf{b}^{\rm T} \textbf{b} \right )^{-1}\textbf{b}^{\rm T} \textbf{T}\,,
\end{equation}
where $\textbf{T}$ is the $n_t \times n_{\nu}$ matrix of the TOD and $\textbf{A}$ is the $1 \times n_{\nu}$ fitting parameters matrix. Then we perform the temporal baseline subtraction as
\begin{equation}\label{eq:rmbsl}
     \textbf{T}^c_1 = \textbf{T} - \textbf{b}\textbf{A} \,,
\end{equation}
where $\textbf{T}^c_1$ is the preliminary baseline subtracted TOD. To remove the temporal baseline better, we do an additional subtraction by
\begin{equation}\label{eq:rmbsl2}
     T_2^c(t, \nu) = T^c_1(t, \nu) - b'(t) \,,
\end{equation}
where $T_2^c(t, \nu)$ is the final baseline centered TOD, $b'(t)$ is the minor baseline obtained by averaging $T^c_1(t, \nu)$ over unmasked frequency bands and then smoothing it along time using a median filter with window size $\sim$ 200s. 

In our approach, we assume that the subtracted components contain the slow systematic variation of all beams and polarizations with similar shapes, as well as some extended foreground synchrotron emission. The frequency averaged time variation of $T_2^c(t, \nu)$ for each beam and polarization is shown in the lower panel of \reffg{fig:rmbsl}. Compared with the upper plot, the temporal baseline is significantly subtracted and the difference between beams can be suppressed from $\sim$ 1K to $\sim$ 0.1K. \add{Notably, the spikes in both upper and lower panel are continuum point sources, which appear as Gaussian-like peaks in the time-stream data. The diversity in the spike structure among different beams is expected, as the beams observe different sky regions. }

\subsection{Flux correction}\label{subsec:flux_corr}

Since there is no specific sky calibrator observation in CRAFTS data, we perform a systematic correction using the measured results of a group of known continuum point sources. These sources are carefully selected from the NVSS (NRAO VLA Sky Survey \footnote{\url{https://heasarc.gsfc.nasa.gov/W3Browse/radio-catalog/nvss.html}}, \citealt{1998AJ....115.1693C}) catalog with the criteria same as \cite{2023ApJ...954..139L}:

\begin{itemize}
    \item \textbf{No bright neighbors}: no neighbors within 9 arcmin with flux larger than 10\% of the central source;
    \item \textbf{Not too faint}: with flux larger than 14 mJy in NVSS catalog;
    \item \textbf{Well scanned}: distance from beam center is smaller than 1.5 arcmin.
\end{itemize}

This strict selection is to avoid complications such as the measurement error induced by nearby sources and sky background fluctuations. After that, over 95\% of sources in the NVSS catalog at our survey area are rejected. 

To measure the flux of a chosen source in FAST TOD data, we follow the procedure below:\\
(1) For each source passing through the above selection, we obtain its position from the NVSS catalog and identify which beam scanned the source at which time point $t_0$.\\
(2) During the time range ($t_0$ - 120 s, $t_0$ + 120 s) when the beam scans sky near this source, we re-flag channels in the frequency spectrum of each time sample that deviate over 3$\sigma$ from its smoothed baseline and interpolate these channels with linear fitting.\\
(3) Average the interpolated data of the 1375-1425MHz frequency bands and 2 polarizations to get the time-varying data of $\sim$4 minutes near the time point when the beam scans the source. The choice of 4 minutes ensure a sufficient time length for baseline fitting and avoids including too many other sources.\\
(4) Assuming the beam pattern as Gaussian profile, we fit the time-varying data with a Gaussian profile plus a 2nd-order polynomial, and record the amplitude in Gaussian function as the measured flux of this source. \add{This is a reasonable assumption as we find the relative difference between a Gaussian beam model and the new Zernike polynomial (ZP) model \citep{2024arXiv241202582Z} is only $< 3\%$ within the main beam size ($\theta_{\rm FWHM} \sim 3$ arcmin).}

\begin{figure}
    \centering
    \includegraphics[width=0.47\textwidth]{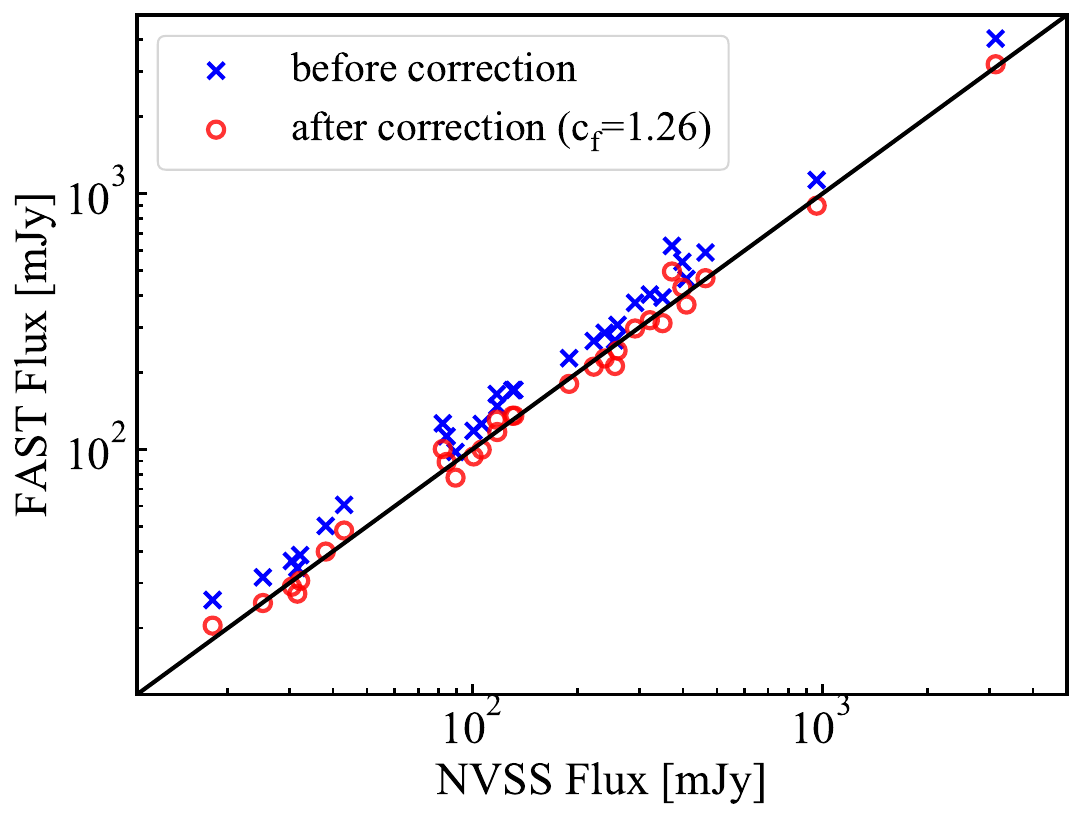}
    \caption{\add{Comparison between flux measured by FAST and NVSS catalog for 30 sources scanned in Dec+4037\_10\_05. Blue crosses are for FAST measurement before flux correction. Red circles are with FAST flux after correction. The black line is the $y=x$ line for comparison.}}
    \label{fig:flux_corr}
\end{figure}

\add{For example, during the $\sim$5 hours observation of the sky area at Dec $\sim 40^\circ37'$, RA from 10h to 15h on October 1, 2021, we measured 30 sources satisfying the selection criteria.} The blue crosses in \reffg{fig:flux_corr} show the comparison of the flux of these sources measured by FAST and given in the NVSS catalog respectively. The error caused by the fitted aperture efficiency $\eta$ as mentioned in \refsc{subsec:fluxcal} would also be mixed in the flux error. We calculate a correction factor $c_f$ for each day's observation by the least square fitting of the blue crosses in \reffg{fig:flux_corr} with the function $y = c_f \cdot x$. The processed TOD is then corrected by
\begin{equation} \label{eq:flux_corr}
    T^c(t, \nu) = T^c_2(t, \nu) / c_f \,,
\end{equation}
where $T^c(t, \nu)$ is the final output TOD after flux correction. \add{For the above example data of Dec+4037\_10\_05, the correction factor is $c_f = 1.26$. The flux comparison for the 30 sources after flux correction is shown with the red circles in \reffg{subsec:fluxcal}.} After this systematic correction, the comparison results are slightly improved and most sources are well matched with the NVSS flux with the acceptable error of $\lesssim$10\%. More results of continuum point sources measurement from TOD are shown in \refsc{subsubsec:tod_continuum}.


\revise{The correction factor arises from multiple reasons. One contributor is the instability of the absolute calibration parameters ($T_{\rm ND}$ and $\eta$), which can introduce an uncertainty of approximately 5\%, according to the measurements of these parameters in Appendix~\ref{appx:TND} and \cite{2023ApJ...954..139L}. Another major contributor is the noise injection synchronisation effect in early CRAFTS data as described in \refsc{subsec:noise_overflow}, which can lead to a flux overestimation of $\sim$ 30\%. Taking into account the combined influence of these factors, the observed correction factor ($c_f=1.26$ in \reffg{fig:flux_corr} for one observation in 2021) is consistent with our expectation. The noise injection issue was corrected in winter 2021, so data taken after that time should be free of this effect and only need a correction factor between 0.95 - 1.05, which also agrees well with our results.}

\subsection{Map-making}\label{subsec:map-making}

\begin{figure*}
    \centering
    \includegraphics[width=0.99\textwidth]{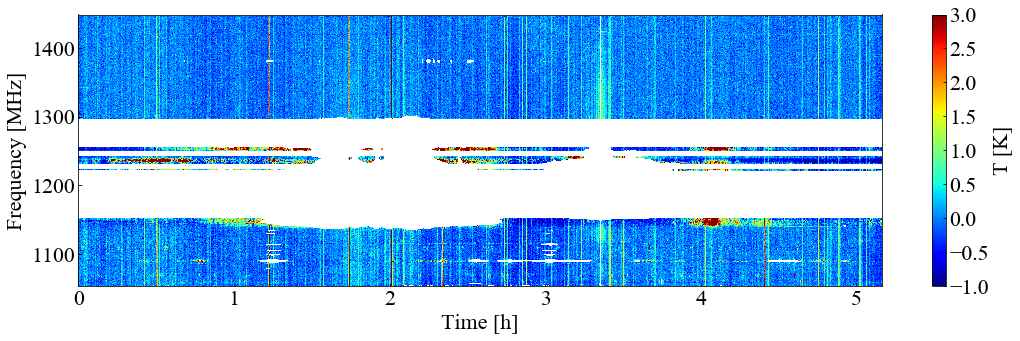}
    \caption{Waterfall plot of TOD after all calibration and correction processes for $\sim$ 5h observation of Dec+4142\_12\_05, M01, XX polarization. The color represents the temperature and the blank areas are flagged RFIs.}
    \label{fig:cal_tod}
\end{figure*}

The TOD after applying all calibration and correction processes above is illustrated in the waterfall plot in \reffg{fig:cal_tod} for 5 hours of observation of Dec+4142\_12\_05 from M01, XX polarization. Compared to the raw TOD shown in \reffg{fig:raw_wfp}, the strong RFIs have been removed and the gain fluctuations over time and frequency have both been suppressed significantly. \add{The vertical stripes in \reffg{fig:cal_tod} correspond to continuum point sources. The blank areas are masked RFIs. However, some RFI residuals remain within the scientific frequency bands, such as the extended emission around 3.3h or the edges of RFIs near 1150MHz. While applying a stricter threshold in our RFI flagging could be helpful to remove these residuals, it would also lead to significant signal loss for bright sources. Therefore, instead of performing a stricter RFI flagging on the TOD, we defer further mitigation of residual RFIs to the map stage in \refsc{subsec:PCA}. }

Before map-making, the XX and YY polarization data are averaged to remove any linear polarization signal and obtain a Stokes I intensity map. We use the map-making code in {\tt fpipe} \citep{2023ApJ...954..139L} to project our TOD onto a {\tt healpix} sky map, based on the idea in \cite{1997ApJ...480L..87T}. The sky map $\hat{\textbf{m}}$ is obtained by
\begin{equation}\label{eq:map}
    \hat{\textbf{m}} = (\textbf{P}^{\rm T}\textbf{N}^{-1}\textbf{P})^{-1}\textbf{P}^{\rm T}\textbf{N}^{-1}\textbf{T}^c \,,
\end{equation}
in which $\textbf{P}$ is the pointing matrix to connect the observation time samples and sky coordinates, $\textbf{N}$ is the noise covariance matrix which is assumed to be diagonal, and $\textbf{T}^c$ is the TOD after all calibration and correction processes. We apply the map-making code in {\tt fpipe} with the parameter {\tt nside} = 2048, which is equivalent to an angular resolution (pixel size) of $\sim$1.7 arcmin. With this method, the intensity at each pixel is only determined by the TOD samples located within it. Therefore, we call it a `center-only' map from this point onwards for simplicity.

\begin{figure*}
    \centering
    \includegraphics[width=0.99\textwidth]{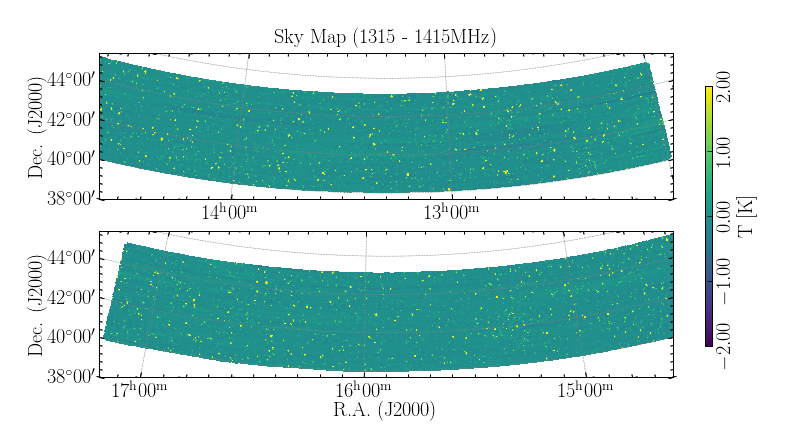}
    \caption{Map-making result for the 5h $\times$ 5deg sky area at RA from 12h to 17h, Dec from $\sim 40^{\circ}$ to $\sim 45^{\circ}$. This map is the averaged intensity of 1315-1415MHz. }
    \label{fig:map_example}
\end{figure*}

We present the 1315-1415 MHz averaged map of $\sim 270~ \rm deg^2$ of the sky from $\sim$ 70 hours observation in 17 days in \reffg{fig:map_example}. Note the conversion between the unit Jy and Kelvin for the map can be made according to \refeq{eq:T2Jy}. Many continuum point sources are visible as the yellow spots on the map. There also appears to be some weak extended background variation, particularly some stripes along the RA direction, which may be caused by the systematic difference between different beams or different days. \add{Before applying temporal baseline subtraction (see \refsc{subsec:rmbsl}), the variations across different beams or observation days can reach several Kelvins. After temporal baseline subtraction, these variations are typically reduced to the $\sim$ 0.1 K level, though for a few beams, the residual differences can be as large as $\sim$ 0.4 K, contributing to the darker stripes in the sky map. We did not apply stricter temporal baseline subtraction for better removal of these structures, considering the risk of over-subtract real signals in the calibration processes. These stripes can be effectively subtracted by PCA as presented in \refsc{subsec:PCA} due to their smooth frequency structures. In the future, a repeated scan may be helpful to reduce systematics and weaken the stripes.}

Similar to \cite{2023ApJ...954..139L}, we also use an alternative map-making algorithm for better point source measurement, which is 
\begin{equation}\label{eq:map_GK}
    \hat{m}_p = \big[(\textbf{PK})^{\rm T}\textbf{T}^c\big]_p\big/\big[(\textbf{PK})^{\rm T}\textbf{I}\big]_p \,,
\end{equation}
where $p$ is the pixel index on the map, $\textbf{I}$ is a uniform column vector, and $\textbf{K}$ is a kernel function, which we set to be a Gaussian kernel as
\begin{equation}\label{eq:GK}
    K_{pq} = {\rm exp}\Big[-\frac{1}{2}\Big(\frac{r_{pq}}{\sigma_K}\Big)^2\Big] \,,
\end{equation}
where $K_{pq}$ is the $(p,q)$ element of $\textbf{K}$ matrix, $r_{pq}$ is the spherical distance between the $p$-th and $q$-th pixel of the map and $\sigma_{K}$ is the kernel size, which is set to be $\sigma_K = {\rm FWHM}/2\sqrt{2{\rm log}2} \sim 1.27\ {\rm arcmin}$ in this work. In the point sources measurement of \refsc{subsubsec:map_continuum}, we use the Gaussian kernel map with pixel size $\sim$ 0.86 arcmin ({\tt nside} = 4096), whose spatial resolution is higher than the ``center-only'' map, to improve the results.

Further results about sources measurement on the map and foreground removal tests for \HI intensity mapping are presented in \refsc{sec:result}. Note the results are primarily based on the center-only map, except for the continuum point sources measurement.

\subsection{Standing waves removal}\label{subsec:rmsw}

\begin{figure*}
    \centering
    \includegraphics[width=0.61\textwidth]{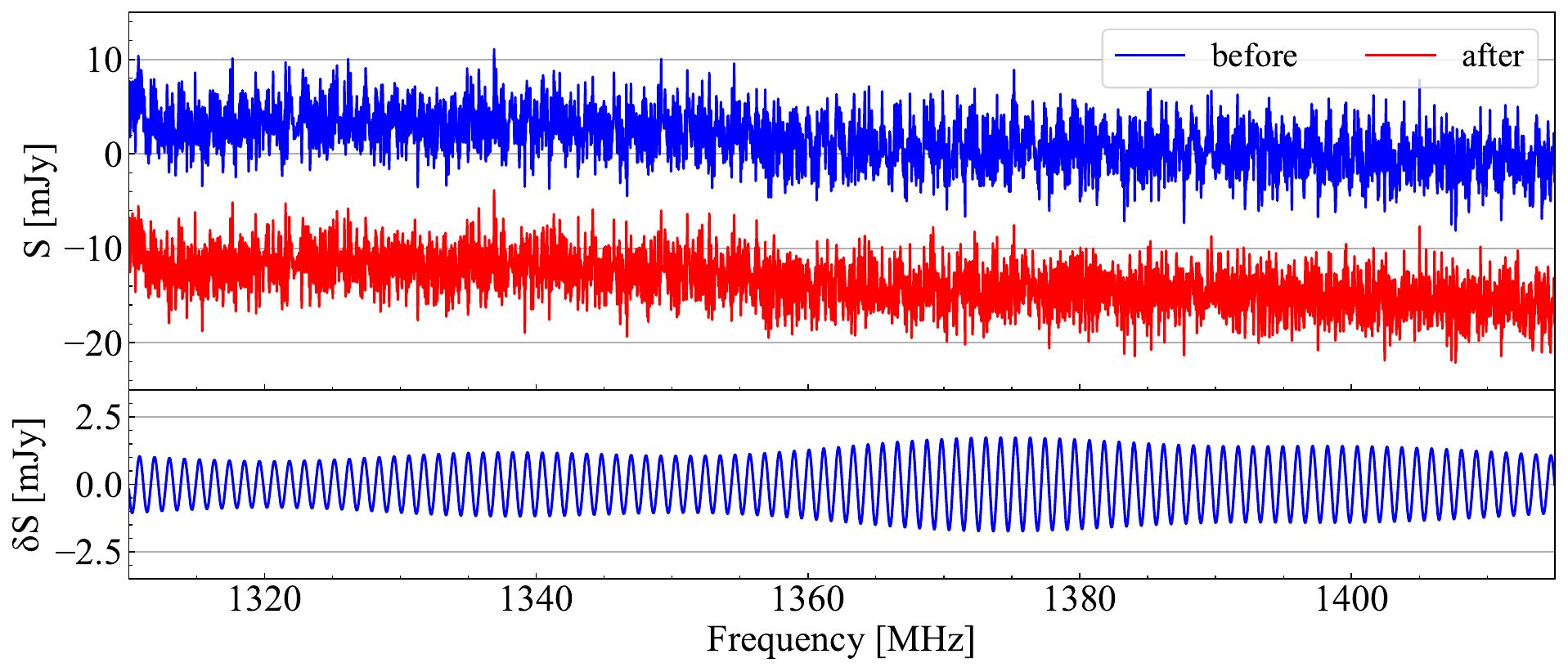}
    \includegraphics[width=0.37\textwidth]{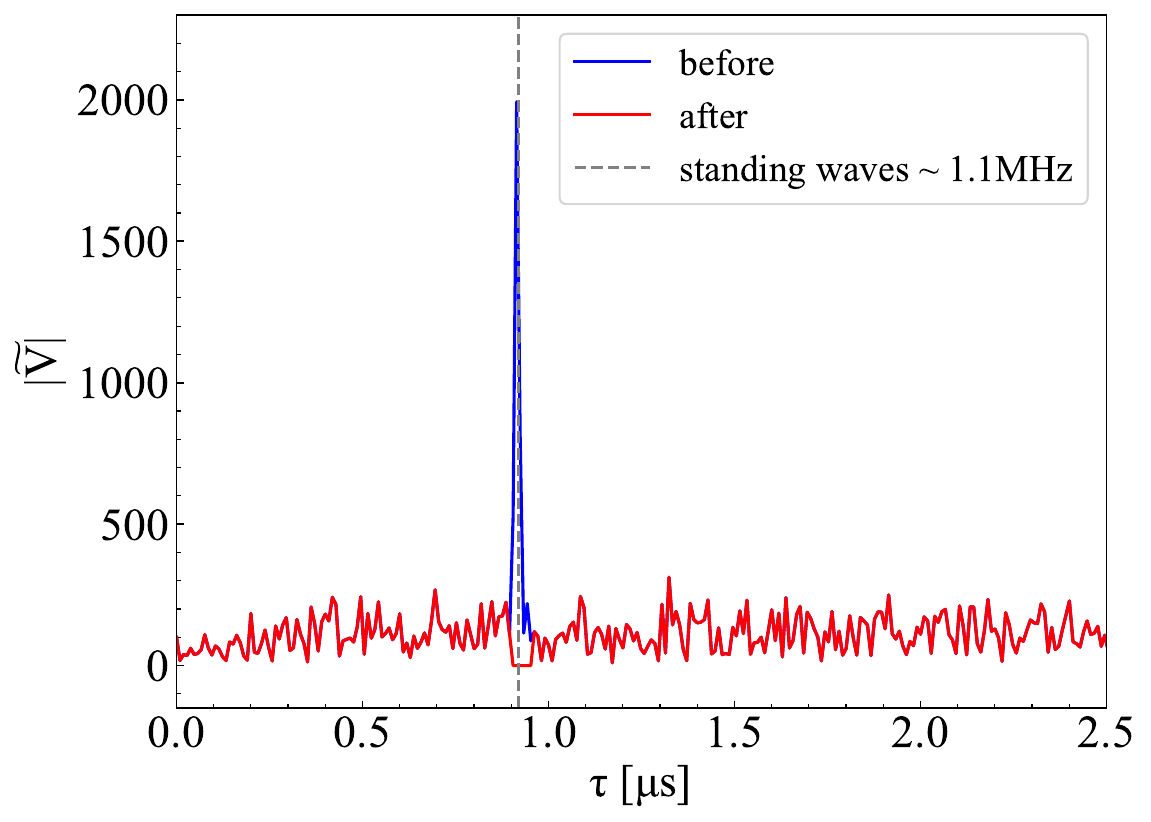}
    \caption{Left: an example of the comparison of the spectrum at one pixel before (blue line in upper panel) and after (red line in upper panel) standing waves removal. Note that the red line has been artificially shifted downward by 15 mJy for clearer comparison. The line in the lower panel shows the difference between the upper two lines. Right: delay spectra of the left upper two spectra. The blue line presents the delay spectrum before standing wave removal, which is largely overlapped with the red line indicating the delay spectrum after standing wave removal. The grey dashed vertical line marks the theoretical position of the peak corresponding to the $\sim$ 1.1MHz standing waves. }
    \label{fig:map_rmsw}
\end{figure*}

Due to the standing waves between the FAST feed cabin and the reflector which has a focal length of 138 m, there are ripples in our data with a period of $\frac{c}{2f} \sim$ 1.1 MHz. These standing waves may bias the analysis of \HI emissions because of their comparable frequency scale. 
These structures have complicated time-varying phase and amplitude which are hard to model \add{\citep{2025RAA....25a5011X}}, so we use a simple method based on the fixed period for mitigation. At each pixel, we compute the Fourier transform of the spectrum, i.e. the delay spectrum. Then we extract the standing waves component $s_{p}$ in the frequency domain by performing the inverse Fourier transform of the peaks at $\sim 0.9 - 0.96 \mu \rm s$ in the delay spectrum. After that, we subtract them from the map by
\begin{equation}\label{eq:rmsw}
    \hat{m}'_p = \hat{m}_p - s_{p} \,,
\end{equation}
where $\hat{m}_p$ is the spectrum at $p$-th pixel, $s_p$ is the standing waves components at this pixel and $\hat{m}'_p$ is the standing waves removed spectrum. The comparisons of the frequency spectrum at one pixel and its delay spectrum before and after standing wave removal are shown in \reffg{fig:map_rmsw}. The standing waves with the amplitude of $\sim$ 2mJy are mixed in the spectrum as indicated by the blue line in the upper panel of the left plot in \reffg{fig:map_rmsw}, and the corresponding peak is very clear in the Fourier space in the right plot. After subtracting the standing wave component, the spectrum is dominated by thermal noise as shown with the red line in the upper panel of the left plot. In the lower panel of the left plot, we can see that the subtracted components contain clear standing wave features with the period of $\sim$ 1.1 MHz enveloped with other periodic structures wider than 20 MHz. 

One concern is that applying this standing wave removal method may lead to information loss in some $k$-modes when we estimate the power spectrum. \add{This impact is expected to be confined to a narrow range around $k_{\parallel} \sim 2\, h\,{\rm Mpc}^{-1}$ in the power spectrum measurement at low redshift. For example, at $z=0.07$, this effect is constrained within $2.35\, h\,{\rm Mpc}^{-1} < k_{\parallel} < 2.5\, h\,{\rm Mpc}^{-1}$. Its effect on large-scale structures ($k< 1 \, h\,{\rm Mpc}^{-1}$) is negligible. We could also address its influence on the power spectrum using a transfer function derived from simulations in future work.} The measurement results of small-scale sources presented in this paper are not affected by this issue as we do not analyze them in Fourier space.

Considering the different standing wave features present in the different beams, one might be tempted to remove the standing waves in the TOD instead of on the map, because the spectrum in each pixel is the mixture of data from two or three beams. However, the noise level for the spectrum at individual time samples is usually higher than the amplitude of standing waves, and processing the data at each time point separately would be time-consuming. Even though the noise level of spectra in map pixels is not low enough either to see the standing waves very clearly by eyes, it is several times better than that of the TOD spectra. Besides, the period of standing waves in different beams is almost the same,so our method here is also suitable to be applied to the beam-combined spectrum of each pixel.

\section{Data validation tests}\label{sec:system}

\subsection{Pointing}\label{subsec:pointing}

For drift scan observation, the telescope is set with a certain altitude (Alt) and azimuth (Az), its pointing on the sky moves with the Earth's rotation. We expect the coordinates of the pointed position to be ${\rm RA} \approx {\rm RA_0} + 15^\circ/{\rm h} \cdot t, \rm Dec = Dec_0$ with observation time $t$ in the unit of hours and the beginning position ($\rm RA_0, Dec_0$).

\begin{figure*}
    \centering
    \includegraphics[width=0.50\textwidth]{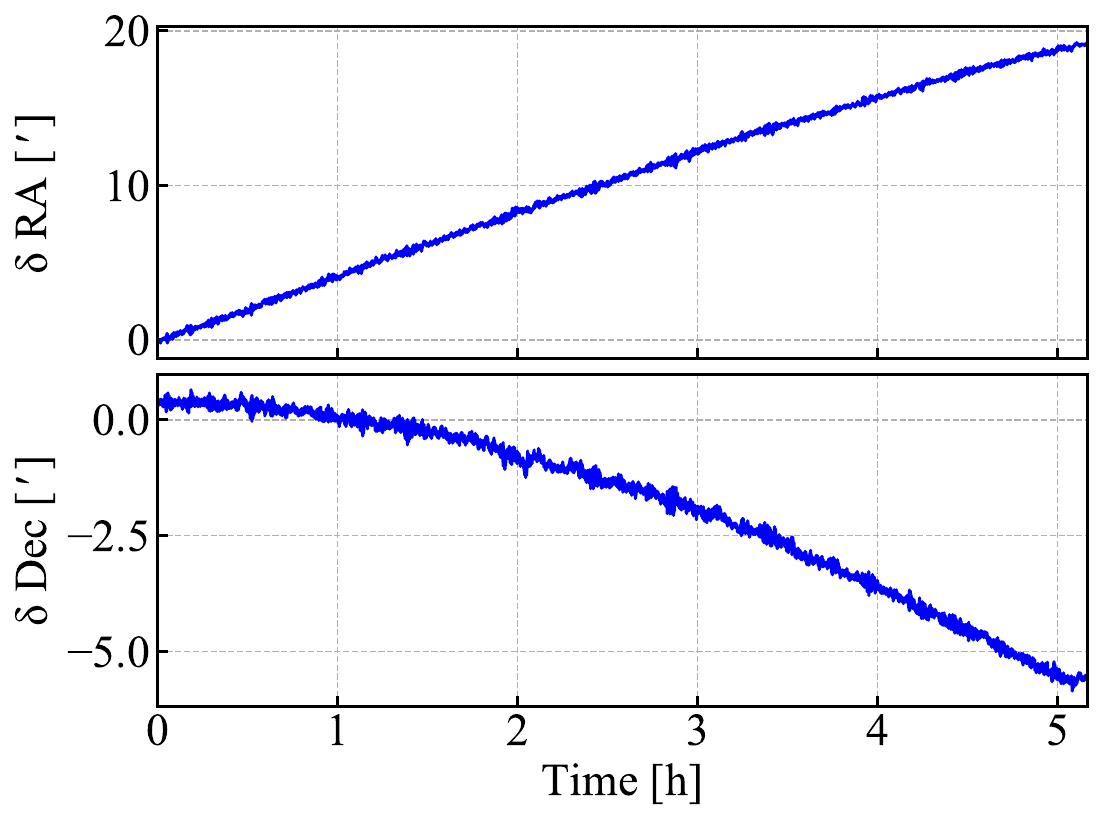}
    \includegraphics[width=0.49\textwidth]{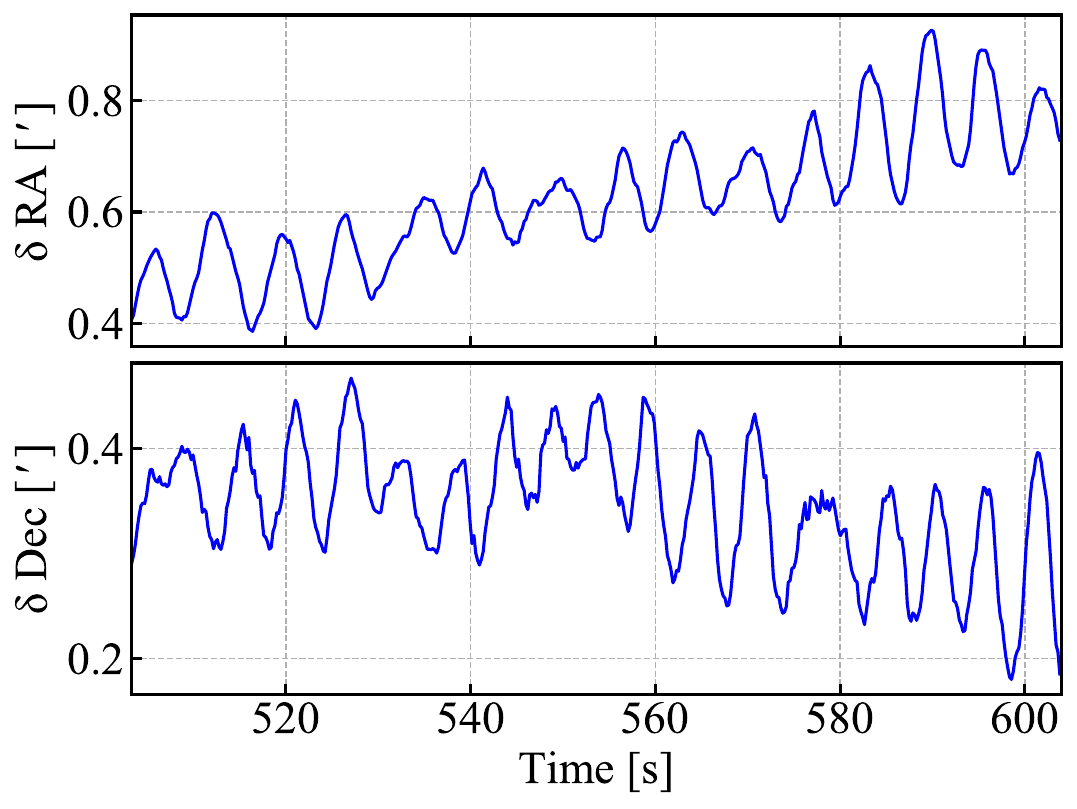}
    \caption{Left figure: pointing deviation during $\sim$ 5 hours observation for Dec+4142\_12\_05 at October 25, 2022. The upper panel shows the error of RA and the lower panel shows the error of Dec. Right figure: enlarged figure of the left figure at time range from $t \sim 500 \rm s$ to $t \sim 600 \rm s$. }
    \label{fig:pointing}
\end{figure*}

The pointing deviation, defined as the difference between the real position ($\rm RA_{obs}, Dec_{obs}$) and the expected position ($\rm RA, Dec$), is given by $\delta {\rm RA} = {\rm RA_{obs}} - {\rm RA}$ and $\delta {\rm Dec} = {\rm Dec_{obs}} - {\rm Dec}$. An example of the pointing deviation for the central beam during 5 hours of observation in one day with initial coordinates set $\rm RA_0 = 12h, Dec_0 = 41^{\circ}42'$ is displayed in the left plot of \reffg{fig:pointing}. The deviation increases steadily with time, primarily because the RA and Dec are defined in the J2000 coordinate system, which differs slightly from the coordinates in 2022 due to precession of the Earth. 
However, besides this steady deviation, the pointing also has a slight periodic fluctuations as shown in the right figure of \reffg{fig:pointing}, which is an enlarged plot of the time range from $t \sim 500 \rm s$ to $t \sim 600 \rm s$ in the left figure of \reffg{fig:pointing}. We checked the observation logs to see if the variations of telescope pointing were related to some environmental parameters such as temperature, humidity, and wind speed. However, no significant correlation is found between them and the pointing fluctuations. Therefore, the reason for these short-period oscillations might be some intrinsic mechanical resonance. This is not a major concern because the amplitude of the periodic fluctuation is just at the level of a few arc seconds, which is very small compared to the 3 arc minutes beam size.

\subsection{Noise injection synchronisation error}\label{subsec:noise_overflow}

The FAST group conducted a test observation in December 2021 by adjusting the time resolution of the pulsar backend to be $\Delta t_{\rm psr} = 8.192\mu \rm s$ (private communication by FAST collaboration), which enables us to identify the noise-on and noise-off data more accurately. 
Through the test observation, we find the noise overflow effect in the early CRAFTS observations -- the noise-on condition continued into the noise-off time when the pulsar backend data are binned to the $\sim 100 \mu$s time stamps. The consequence of this effect is that a smaller $V_{\rm psr,on}$ and larger $V_{\rm psr,off}$ than real conditions are recorded, resulting in an amplified $V_1$ in \refeq{eq:bpcal}.

Nevertheless, during the test observation of $\sim$ 30 minutes that day, we find the overflow effect is stable with a constant time delay during one observation, which will only introduce a constant coefficient to the flux scale. \add{However, the situation might be different for different days. Besides, the error introduced by the overflow effect is entangled with the uncertainties like the flux error introduced by $T_{\rm sys}$ and $\eta$, making it difficult to model precisely. Therefore,} in \refsc{subsec:flux_corr}, we applied the flux correction on each day's observation separately to correct the noise overflow effect and obtained improved results. \add{A rough estimation suggests that the uncertainties from $T_{\rm sys}$ and $\eta$ contribute approximately 5\%, while test observations indicate that the noise overflow effect could introduce an error of around 30\%. Given these contributions, the flux correction factor (e.g. $f_c$=1.26 in \reffg{fig:flux_corr}) falls within the expected range.}

\add{The noise overflow effect was caused by a mistake in early parameter setting of CRAFTS and corrected in the winter of 2021. Most of our data in this work are collected in 2022 except for 3 observations, so the correction factor in most of our data is very close to 1.}

\subsection{RFI contamination}\label{subsec:mask_map}

\begin{figure*}
    \centering
    \includegraphics[width=0.58\textwidth]{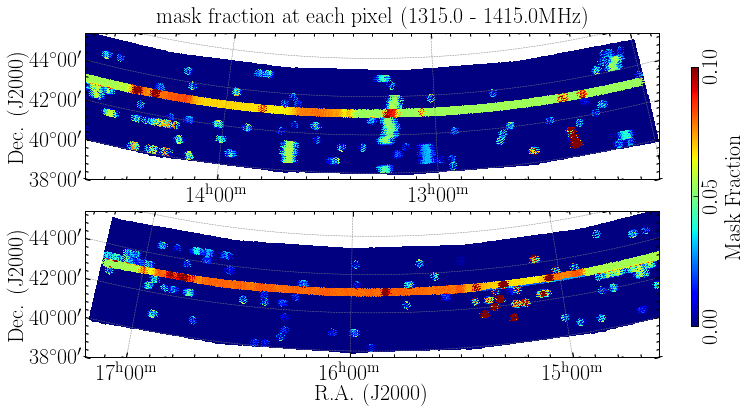}
    \hspace{3mm}
    \includegraphics[width=0.39\textwidth]{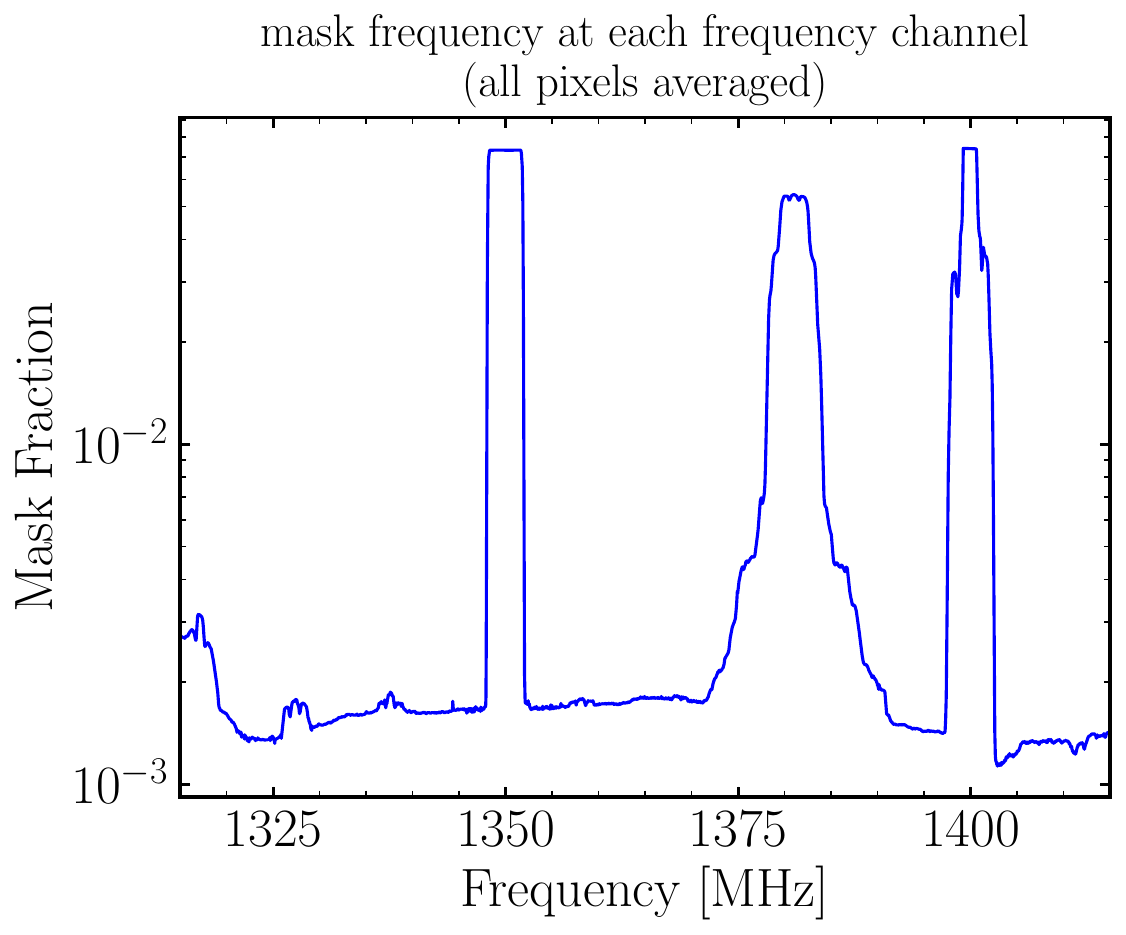}
    \caption{ Left: Sky map of the mask array with the colorbar representing the mask fraction at each pixel, averaged over 1315-1415MHz. Right: mask fraction at different frequency channels in the 1315-1415MHz band.}
    \label{fig:maskmap}
\end{figure*}

In the RFI flagging process as described in \refsc{subsec:RFIflag},  we obtain the mask array with 1 at the flagged data point and 0 at the unflagged data point for the TOD. We convert these flags from the TOD domain into the map-space via the map-making algorithm, which is shown in \reffg{fig:maskmap}. We can see the mask fraction at most pixels is almost zero, which means these areas are RFI-free at the 1315-1415MHz frequency band. \add{In general, the redder regions indicate a higher mask fraction, meaning more frequency channels are contaminated by RFIs. The bright spots correspond to short-term RFIs, revealing the approximate positions of 19 beams. Some of these RFIs, such as the bright green region near RA 13h42m, Dec 41$^{\circ}$), last for a few seconds. Another clear feature is the long stripe at Dec $\sim 43^{\circ} $, which is a long-term RFI that appears at a narrow frequency band in all 19 beams during the whole 5-hour observation.} The all-pixel averaged mask fraction spectrum is shown in the right figure of \reffg{fig:maskmap}. The mask fraction at some frequency channels (e.g. $\sim$ 1350MHz, $\sim$ 1380MHz and $\sim$ 1400MHz) could reach $\sim$ 0.1, indicating these channels are contaminated at $\sim$ 10\% pixels. The mask map can give us valuable guidance to check the data quality for further signal detection.

\subsection{Detection limit} \label{subsec:detection_limit}

\begin{figure*}
    \centering
    \includegraphics[width=0.48\textwidth]{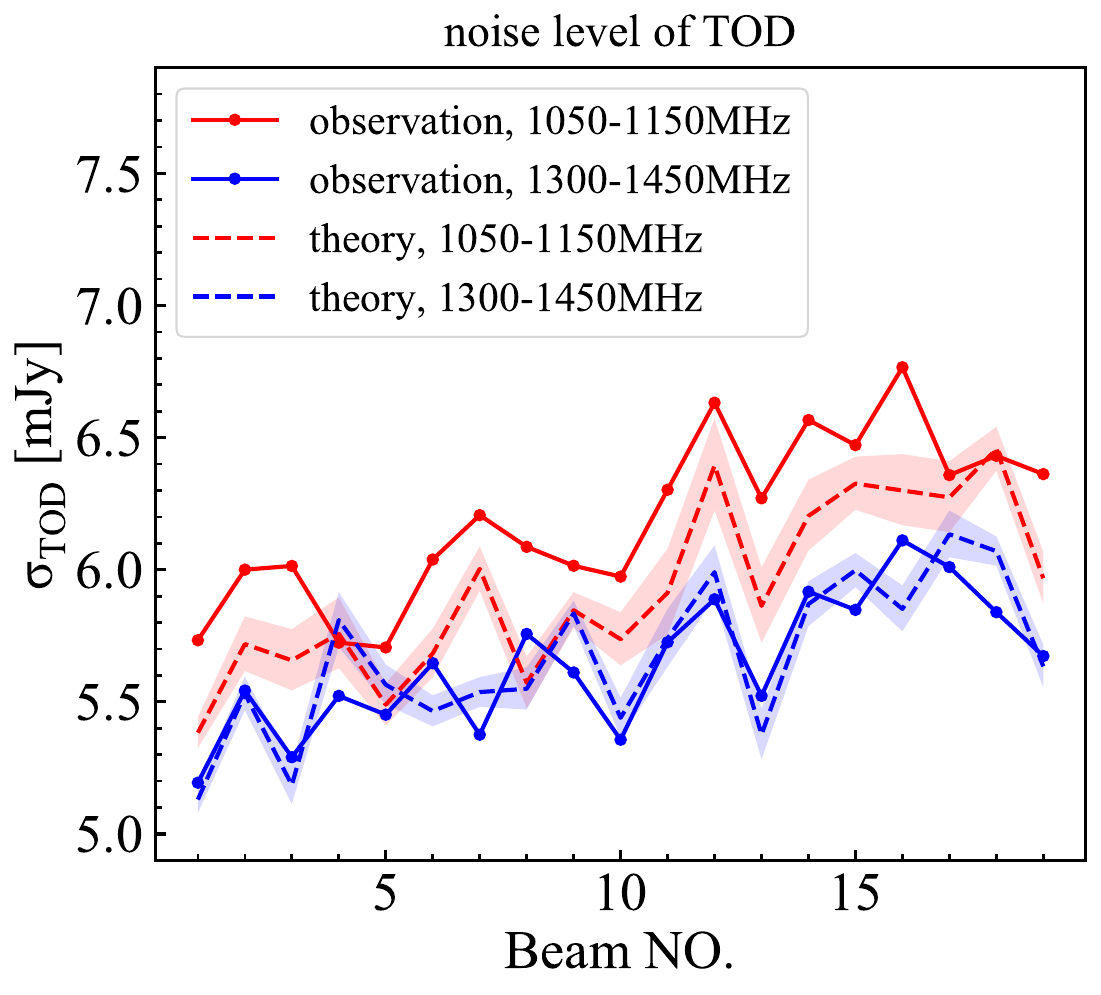}
    \hspace{1mm}
    \includegraphics[width=0.48\textwidth]{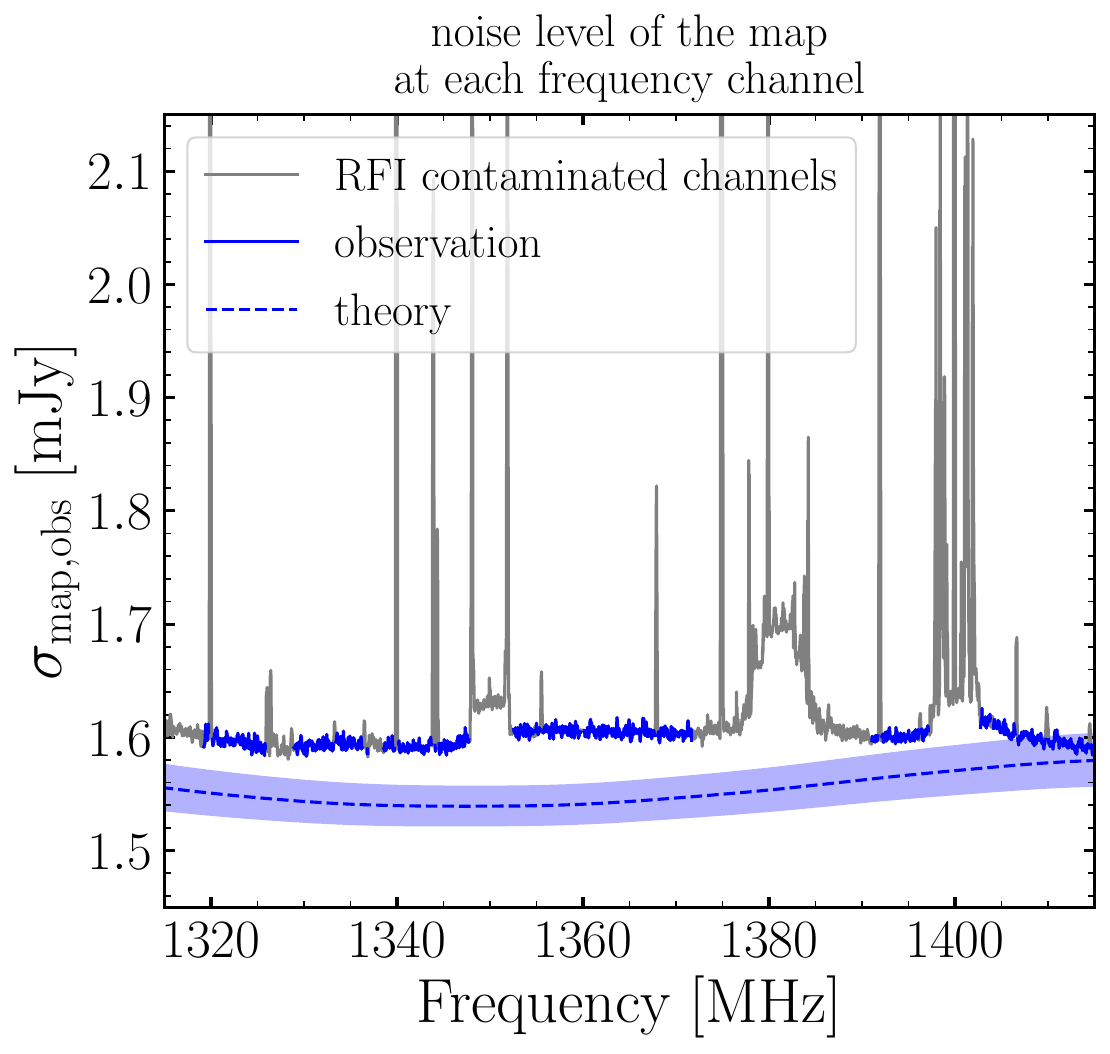}
    \hspace{1mm}

    \caption{ Left: noise level of calibrated TOD from 19 beams. The red line represents the noise level of the frequency band 1050-1150 MHz while the blue line shows the noise level of the 1300-1450 MHz band. The red and blue dashed lines represent the averaged theoretical noise level at 1050-1150 MHz and 1300-1450 MHz respectively. The red and blue shaded areas show the uncertainty in theoretical noise level from the parameters we use to estimate the sensitivity. 
    \add{Right: the blue solid line shows the noise level of the map at each frequency channel at 1315-1415 MHz, with the gray solid line marks the RFI contaminated channels. The blue dashed line represents the theoretical noise level and the blue shadow is its error range caused by the uncertainty of fitting parameters.} 
    }
    \label{fig:noise_TOD_map}
\end{figure*}

Theoretically, the sensitivity of the telescope can be estimated by the radiometer equation
\begin{equation}\label{eq:sensitivity}
    \sigma = \frac{2k_{\rm B}T_{\rm sys}}{\eta A_{\rm illu}\sqrt{N_{\rm beam} N_{\rm pol} \Delta t \Delta \nu}} \,,
\end{equation}
in which $k_{\rm B}$ is the Boltzmann constant, $T_{\rm sys}$ is the system temperature, $\eta$ is the antenna efficiency, $A_{\rm illu}$ is the illuminated area, $N_{\rm beam}$ is the beam number and $N_{\rm pol}$ is the polarization number. The parameters of our calibrated TOD are $A_{\rm illu} = 70700 \rm m^2, N_{\rm beam} = 1, N_{\rm pol} = 2, \Delta t = 1{\rm s},\, \Delta \nu = 30{\rm kHz}$. Additionally, to estimate the theoretical noise level accurately, we use the fitting results of $T_{\rm sys}(\theta_{\rm ZA},\nu)$ and $\eta(\theta_{\rm ZA},\nu)$ for each beam in \cite{2020RAA....20...64J} instead of simply assuming constant values. Based on these parameters, the theoretical sensitivity $\sigma_{\rm TOD, theo}$ is calculated to be at the level of $\sim$ 5.5 mJy and varies across different beams and frequency bands as shown with the red and blue dashed lines in the left plot of \reffg{fig:noise_TOD_map}. It can be seen that the noise level increases for beams in the outer areas of the receiver. The error of $\sigma_{\rm TOD, theo}$ caused by the uncertainty of fitting parameters is also marked by the red and blue shadow.

We also estimate the sensitivity of real data by fitting the noise histogram with a Gaussian function. To subtract the continuum background, instead of getting the histogram of the calibrated flux $S(t,\nu)$ directly, we use the difference between four nearby frequency channels following the method given in \cite{2021MNRAS.505.3698W} and \cite{2023ApJ...954..139L}, i.e. 
\begin{equation}\label{eq:sensitivity_diff}
    \delta S(t,\nu) = \frac{1}{2}(S(t,\nu_{1})+S(t,\nu_{3})) - \frac{1}{2}(S(t,\nu_{2})+S(t,\nu_{4})) \,,
\end{equation}
for noise level estimation. We \add{fit the histogram of $\delta S(t,\nu)$ with Gaussian model and record the standard deviation of the fitted Gaussian profile as the sensitivity of $S(t,\nu)$, which helps to minimize the influence of non-Gaussian tails. The observed noise level of TOD is denoted as $\sigma_{\rm TOD,obs}$.} The observed noise level of 19 beams in 5 hours of observation is $\sigma_{\rm TOD,obs} \sim 6.2 {\rm mJy}$ for 1050-1150 MHz and $\sim 5.5 {\rm mJy}$ for 1300-1450 MHz, as shown in the left plot of \reffg{fig:noise_TOD_map}. The  $\sigma_{\rm TOD,obs}$ at 1300-1450 MHz band match our expectation, while the noise level for the 1050-1150 MHz band is slightly larger than the theoretical sensitivity, which may be due to the presence of RFI and standing wave residuals. \add{It may be noticed that the observed noise level at the high frequency band is lower than theoretical expectation for some beams. One possible reason is that the theoretical noise level is derived from system parameters ($T_{\rm sys}$ and $\eta$) following \cite{2020RAA....20...64J}, but these parameters inherently have uncertainties. While we have included the expected error regions in \reffg{fig:noise_TOD_map} based on the uncertainties provided in \cite{2020RAA....20...64J}, the actual deviation in certain observations may be larger than expected. }

Based on \refeq{eq:sensitivity}, we can also derive the theoretical sensitivity of our maps. Since one pixel on a map is usually scanned by more than one beam over a few seconds, we get the hits maps
by counting the number of time samples by each beam contained in each pixel, which helps to make a precise estimation of the noise level for the map. We find that most pixels contain $\sim$14 time samples, though the pixels on edge are scanned few times than those in the central region. By weighted averaging the theoretical $\sigma_{\rm TOD, theo}(\theta_{\rm ZA},\nu)$ of each beam on each pixel according to the hits map, we obtain the averaging theoretical sensitivity of all pixels with $\sigma_{\rm map,theo} \sim 1.58{\rm mJy/beam}$, and varies across frequency as shown by the \revise{blue} dashed line in the right plot of \reffg{fig:noise_TOD_map}. The gray shadow represents the error from the uncertainty of fitted $\rm T_{sys}$ and $\eta$ given in \cite{2020RAA....20...64J}. 

The sensitivity of the real map is also estimated by fitting the histogram of $\delta S(p,\nu)$ similar to \refeq{eq:sensitivity_diff} but for the spectrum in each pixel $p$. \add{RFI contaminated pixels have been excluded in this noise estimation following the pixel selection in \refsc{subsec:PCA}.} Compared with the theoretical sensitivity, the observed noise level for the clean band 1315-1415 MHz is $\sigma_{\rm map,obs} = 1.6{\rm mJy/beam}$, which is very close to the theoretical value. We plot the noise level of the map at each frequency channel between 1315 MHz and 1415 MHz in the right plot of \reffg{fig:noise_TOD_map}. We can see that the noise levels at most channels are generally consistent with the theoretical values with acceptable excess of $< 5 \%$ at clean channels. Higher noise levels are likely to appear at those RFI-contaminated channels like $\sim$ 1350 MHz, $\sim$ 1380 MHz and $\sim$ 1400 MHz.\add{To avoid the influence of contaminated pixels and frequency channels on cosmological analysis, they would be removed before applying PCA as described in \refsc{subsec:PCA}. } 

\begin{figure*}
    \centering
    \includegraphics[width=0.99\textwidth]{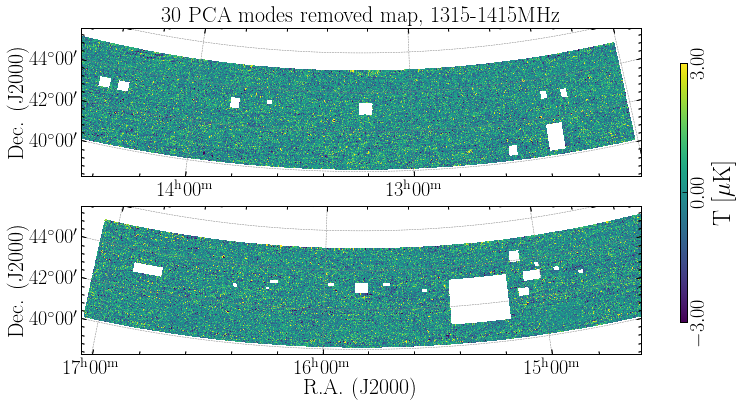}
    \caption{Sky map averaged over RFI-free channels in the 1315-1415 MHz band after 30 foreground modes removed by PCA. The blank areas are RFI-contaminated regions according to the discussion in \refsc{subsec:mask_map}. 
    }
    \label{fig:map_rmfg30}
\end{figure*}

\subsection{PCA tests} \label{subsec:PCA}

The maps we obtained in \refsc{subsec:map-making} would include several components, such as the galactic synchrotron radiation, free-free radiation, and radio sources in addition to the cosmological \HI signal. Relying on the widely accepted assumption that these bright foreground signals are correlated with frequency and dominate the whole data, we use the Principal Components Analysis (PCA, \citealt{2015MNRAS.447..400A, 2021MNRAS.504..208C}) to subtract the bright foreground to learn more about the data quality. The main procedures of PCA are outlined below: 

\noindent 1. Calculate the frequency covariance matrix by 
\begin{equation}\label{eq:covar_matrix}
    \textbf{C} = \textbf{X}_{\rm obs}^{\rm T} \textbf{X}_{\rm obs}/(N_{p}-1) \,.
\end{equation}
where $\textbf{X}_{\rm obs}$ represents the matrix of the map dataset and $N_{p}$ is the number of pixels.

\noindent 2. Apply eigendecomposition of the covariance matrix by
\begin{equation}\label{eq:eigen_decompose}
    \textbf{CV} = \textbf{V}\bm{\Lambda} \,,
\end{equation}
in which $\textbf{V}$ is the eigenvector matrix and $\bm{\Lambda}$ is the diagonal matrix with declining ordered eigenvalues as diagonal elements. 

\noindent 3. Using the first $N_{\rm fg}$ columns of $\textbf{V}$ as the mixing matrix $\hat{\textbf{A}}$, we can reconstruct the map of principal components with
\begin{equation}\label{eq:eigen_map}
    \textbf{S} = \textbf{A}^{\rm T}\textbf{X}_{\rm obs} \,.
\end{equation}
and the foreground map
\begin{equation}\label{eq:fg_map}
    \hat{\textbf{X}}_{\rm fg} = \textbf{A}\textbf{S} \,.
\end{equation}

A large part of the foreground components are expected to be removed by subtracting the foreground map $\hat{\textbf{X}}_{\rm fg}$ from the sky map. However, some systematic issues like the residual RFIs and the uneven scanning would introduce more complexity to the components' decomposition. We make some improvements before applying PCA on the sky map: 
\begin{itemize}
    \item \textbf{Remove RFI contaminated pixels and frequency channels}: According to the mask map described in \refsc{subsec:mask_map}, we remove all pixels with a mask fraction over 0.1 and all frequency channels deviating over 0.01\% from the baseline obtained by fitting the mask fraction spectrum in \reffg{fig:maskmap} by the asymmetrically reweighted penalized least squares (arPLS, \citealt{baek2015baseline}) algorithm. 
    \add{We aim to keep the mask fairly regular and continuous, which is important in power spectrum estimation that is sensitive to ringing.} We therefore further remove the pixels within a guard area around the masked pixels (shown as the blank rectangular spaces in \reffg{fig:map_rmfg30}). 
    \item \textbf{Cut off the map edges}: Since the pixels close to the edges of the map are scanned by fewer beams, we cut off the edges pixels to obtain a sky map with a more uniform noise level.
\end{itemize}

\begin{figure}
\centering
\includegraphics[width=0.49\textwidth]{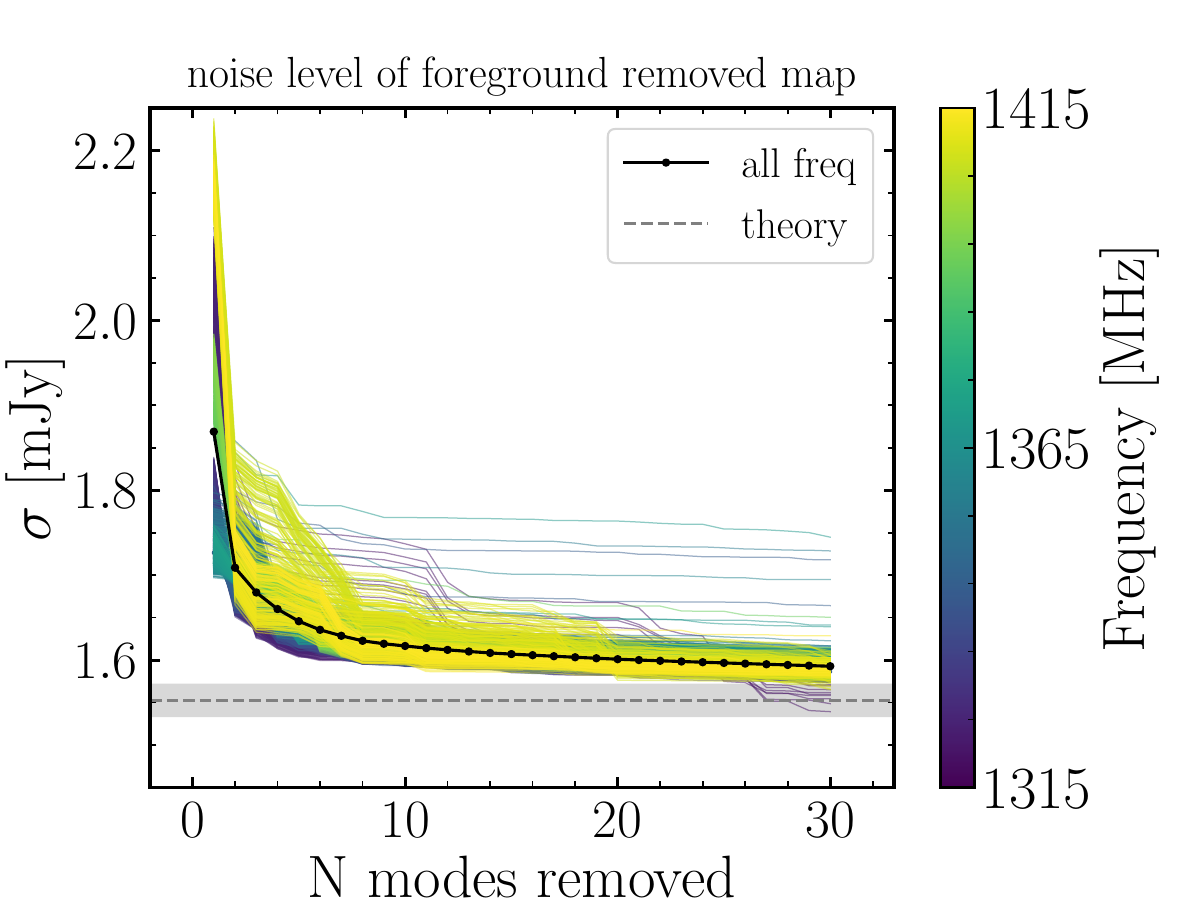}
\caption{Noise level of the PCA modes removed maps. The black thick line is the noise level of the map containing all RFI-free frequency channels between 1315-1415 MHz and each colored line represents an individual frequency channel. The gray dashed line and the gray shadow also show the averaged theoretical sensitivity at 1315-1415 MHz and its error from the fitting parameters uncertainty.}
\label{fig:maskmap_pca}
\end{figure}

After the conservative selection above, 15\% pixels and 44\% frequency channels are removed. We apply PCA on the remaining $\sim$ 320,000 pixels (corresponding to $\sim 230\rm deg^2$) and 1834 frequency channels ($\sim$ 56MHz). The sky map averaged over 1315-1415MHz frequency range after the removal of 30 PCA modes is shown in \reffg{fig:map_rmfg30}. Compared to the original map in \reffg{fig:map_example}, most point sources and sky background have been subtracted, resulting in a new map which appears to be dominated by noise. \add{There are still faint residuals even after 30 modes removed, which may caused by multiple reasons like the primary beam effect and polarization leakage \citep{2015ApJ...815...51S, 2022MNRAS.509.2048S}.} To further characterize the data, we check the noise level of the maps after applying PCA.
As shown with the black thick line in \reffg{fig:maskmap_pca}, the noise level decreases gradually when we remove more and more PCA modes, approaching the theoretical sensitivity when over 10 PCA modes are removed. After removing 30 PCA modes, the noise level of the map is about 1.59 mJy, which is only $\sim 2.4\%$ higher than the expected value of 1.55 mJy marked by the gray shadow. The reduction of noise level can also be seen at each frequency channel as shown with the colored lines in \reffg{fig:maskmap_pca}, which can help us check if there are inconsistent frequency channels \add{and refine the RFI masking. For example, the jumps at some frequency channels (mainly between 1319.7-1320.1 MHz) and the higher noise level for a few channels are likely to be some subdominant RFI structures in small region of the sky map, which will be further removed in future foreground removal improvements}. 

We compare the noise level of our map with the expected fluctuations from cosmological \HI. After removing 30 PCA modes in foreground cleaning, the rms in the cleaned maps is $\sim 13\,$mK and is fairly constant across our clean frequency channels. Note that the maps are rebinned to $\Delta \nu = 0.27\rm MHz$. In comparison, the rms of pure \HI in the simulations used in \cite{2021MNRAS.504..208C} is $\sim 0.142$\,mK. The residue noise level remain significantly higher than the expected \HI fluctuations, even after the 30 PCA foreground modes have been removed. It is about an order of magnitude higher than the noise level of $\sim 1.2$\,mK in \cite{2024arXiv240721626M} for the MeerKAT single-dish mode intensity mapping experiment, which is mainly caused by the different integration time, and note also that the beam size of FAST is a factor of $\sim$ 20 times smaller than that of MeerKAT.

These preliminary analyses with PCA provide guidance for future foreground removal, which is a great challenge for \HI intensity mapping experiments. \add{In future work, we will also carefully investigate how RFI residuals and other systematic effects affect the foreground removal performance. Particularly, primary beam effects and their deviation from the Gaussian approximation need careful considerations \citep{2021MNRAS.506.5075M, 2022MNRAS.509.2048S}. However, most previous single-dish experiments use Gaussian approximations for the beam in their cosmological analysis \citep{2013ApJ...763L..20M, 2018MNRAS.476.3382A, 2022MNRAS.510.3495W, 2023MNRAS.518.6262C}. We will also need to consider the \HI signal loss and compute a transfer function. More observational data and the cross-correlation with optical galaxy observations would also be helpful to detect the \HI signal.}

\section{Results}\label{sec:result}

\subsection{Continuum point sources}\label{subsec:continuum}

\subsubsection{Measurement with TOD}\label{subsubsec:tod_continuum}

\begin{figure*}
   \centering
   \includegraphics[width=0.48\textwidth]{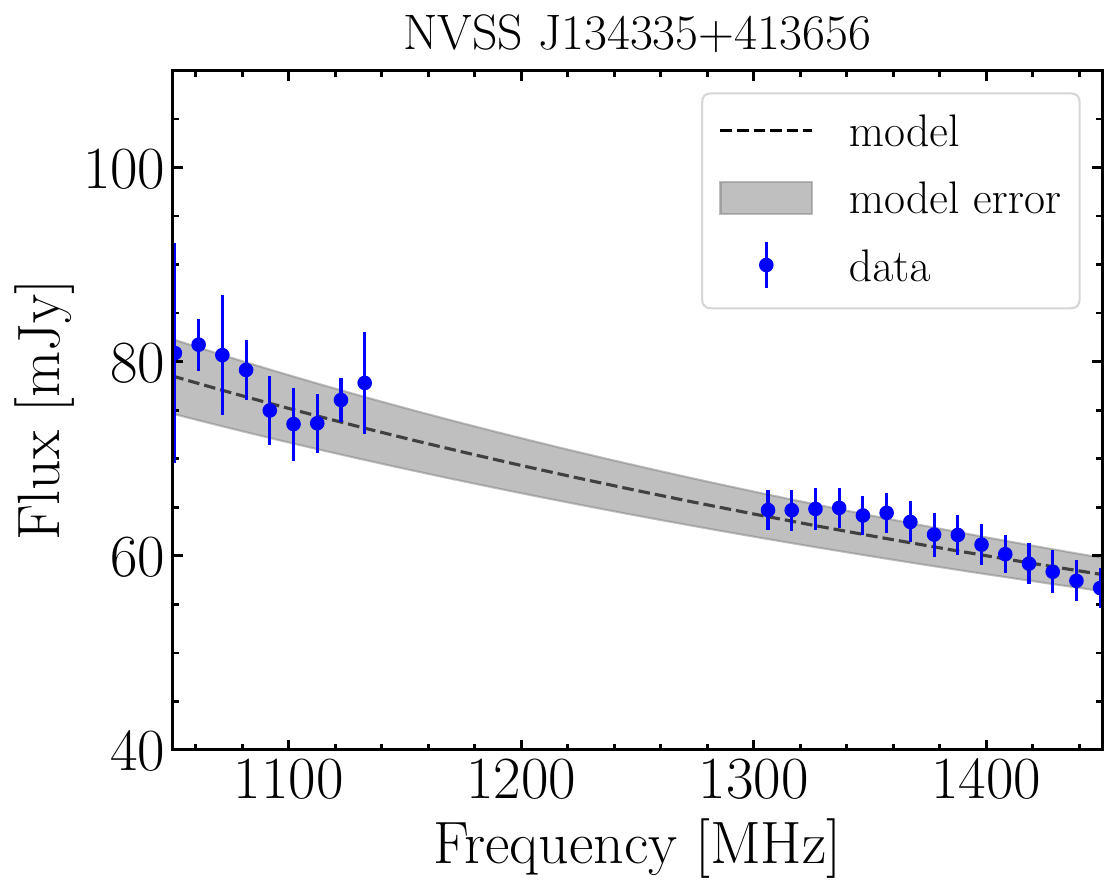}
   \hspace{3mm}
   \includegraphics[width=0.475\textwidth]{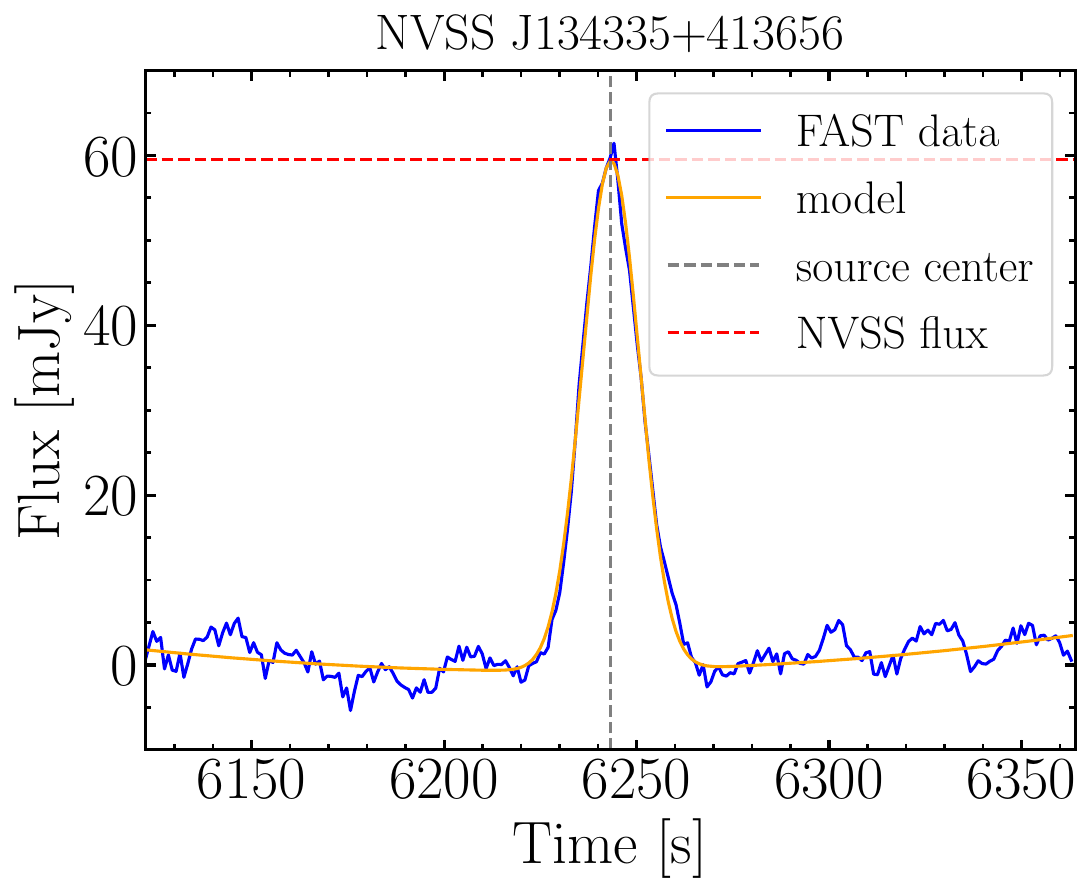}
   \caption{\add{Spectrum (left) and time-stream data (right) of source NVSS J134335+413656 as an example of continuum point sources measurement with CRAFTS TOD. In the left figure, the blue line shows our measured spectrum, with the flux obtained by averaging the data within each 10 MHz frequency bin. The black dashed line shows the spectrum model by fitting the flux density measured by VLA at 1400 MHz and the Northern Cross Radio Telescope at 408 MHz, with the error of model shown with the gray area. In the right plot, the blue line represents our continuum observation data, the orange line is the model with a Gaussian profile plus a 2nd-order polynomial and the red horizontal line marks the NVSS flux. }}
   \label{fig:NVSS_TOD_example}
\end{figure*}

\begin{figure}
    \centering
    \includegraphics[width=0.45\textwidth]{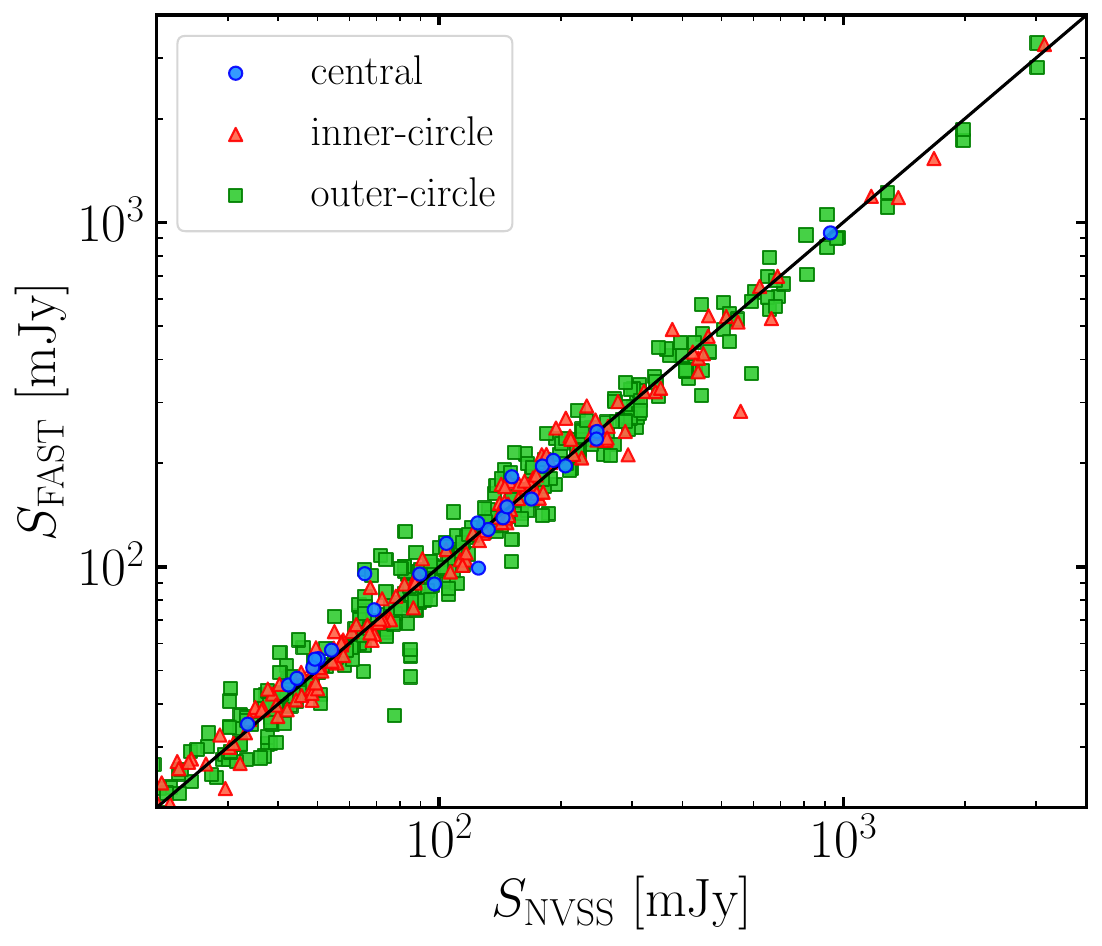}\\
    \includegraphics[width=0.45\textwidth]{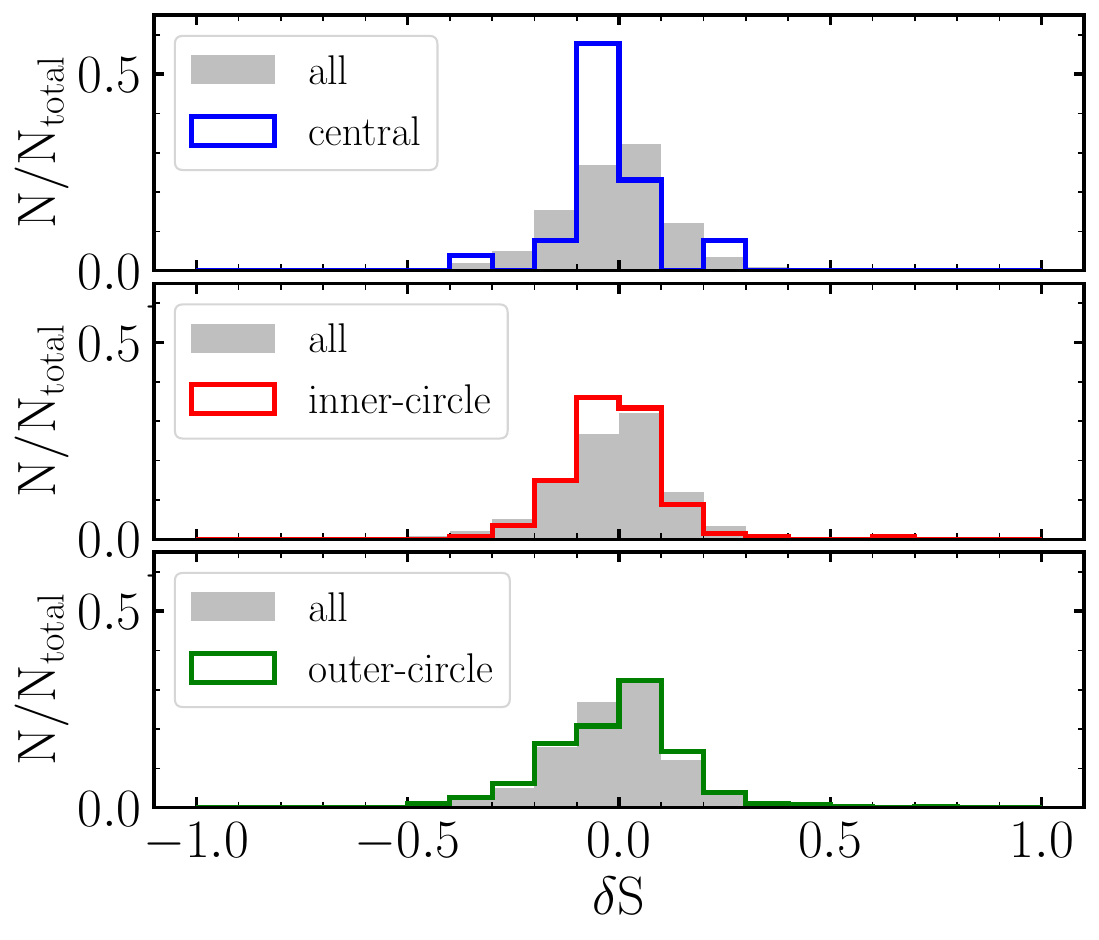}
    \caption{Top: comparison of the flux of \add{447 continuum point sources} measured by CRAFTS TOD to NVSS catalog in the 1375-1425 MHz band. The black line marks equal flux. Different colors distinguish the source measured by beams at different positions, i.e. blue circles for the central beam (M01), red triangles for the inner circle beams (M02 - M07), and green squares for the outer-circle beams (M08 - M19). Bottom: histogram of the relative flux error $\delta S$ for the measurement by beams at the center, inner circle and outer circle positions. }
    \label{fig:NVSS_TOD}
\end{figure}

With the selection criteria and source measurement method mentioned in \refsc{subsec:flux_corr}, we \add{obtain a sample containing 447 continuum point sources} from the NVSS catalog and measure their flux near 1400 MHz with our TOD in the selected area. 
\add{An example of the continuum spectrum ($\sim$ 12 s averaged) and time-stream data (1375-1425 MHz averaged) while scanning source NVSS J134335+413656 is shown in \reffg{fig:NVSS_TOD_example}.} 
Most spectra follow a power-law profile, which confirms the validity of our bandpass calibration process. A comparison of flux measured with FAST and given in the NVSS catalog is in the left-hand side scatter plot of \reffg{fig:NVSS_TOD}. Our results match well with NVSS for the majority of sources, with a small number of outliers. There are many possible reasons for these outliers, like the intrinsic flux variability of some sources, overlap with RFIs, calibration error, etc. To quantify the comparison results, we calculate the relative error of each source between FAST and NVSS by
\begin{equation}\label{eq:err_relative}
    \delta S = (S_{\rm CRAFTS} - S_{\rm NVSS})/\sqrt{S_{\rm CRAFTS} S_{\rm NVSS}} \,,
\end{equation}
in which $S_{\rm NVSS}$ is the flux value in NVSS catalog and $S_{\rm CRAFTS}$ is the flux measured with CRAFTS data. The gray histogram in each subplot of the right figure of \reffg{fig:NVSS_TOD} presents the histogram of $\delta S$ for the 447 sources. The overall relative error of the total samples ($\delta S_{\rm total}$) is then obtained using the median of the absolute values. \add{For the 447 sources}, we have $\delta S_{\rm total} = 8.3\%$, comparable to the error $\sim 6.3\%$ given in \cite{2023ApJ...954..139L} for the FAST \HI IM pilot survey. 

As noted in \cite{2023ApJ...954..139L}, the flux error might be related to which beam scans the source, because the real beam pattern is not perfectly described by a 2D Gaussian profile, especially for beams located in the outer circle.  We divide the 19 beams into three groups as shown in different colors in \reffg{fig:NVSS_TOD}, and find the histograms become wider from top to bottom, corresponding to the increasing relative errors. We calculate the relative error of each group, yielding values of 6.7\%, 7.2\%, and 9.3\% respectively, which are consistent with our expectation. A more accurate beam model \citep{2024arXiv241202582Z} might be helpful for improving the result and will be analyzed in future work.

\subsubsection{Measurement on map}\label{subsubsec:map_continuum}

\begin{figure}
    \centering
    \includegraphics[width=0.485\textwidth]{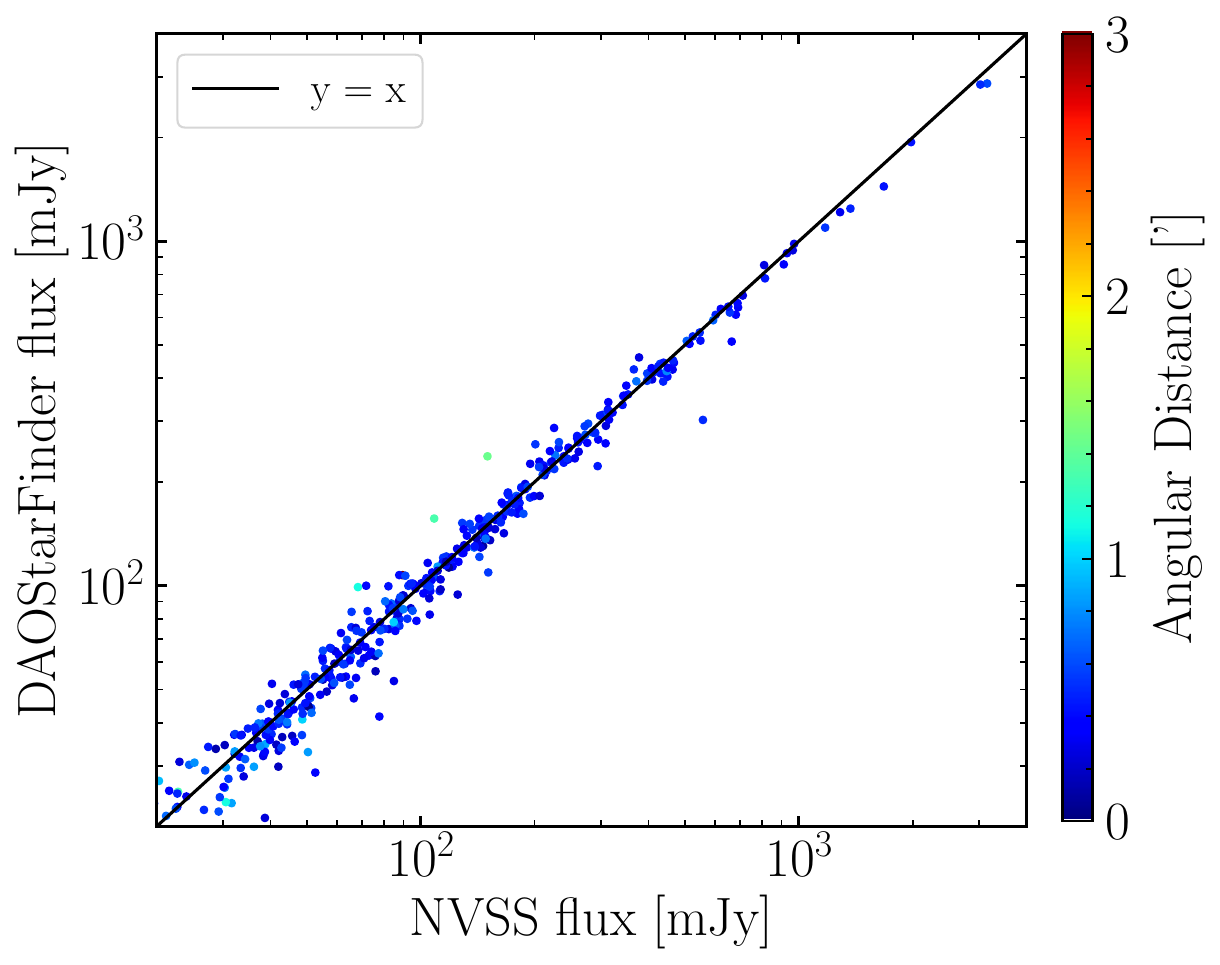}\\
    \includegraphics[width=0.46\textwidth]{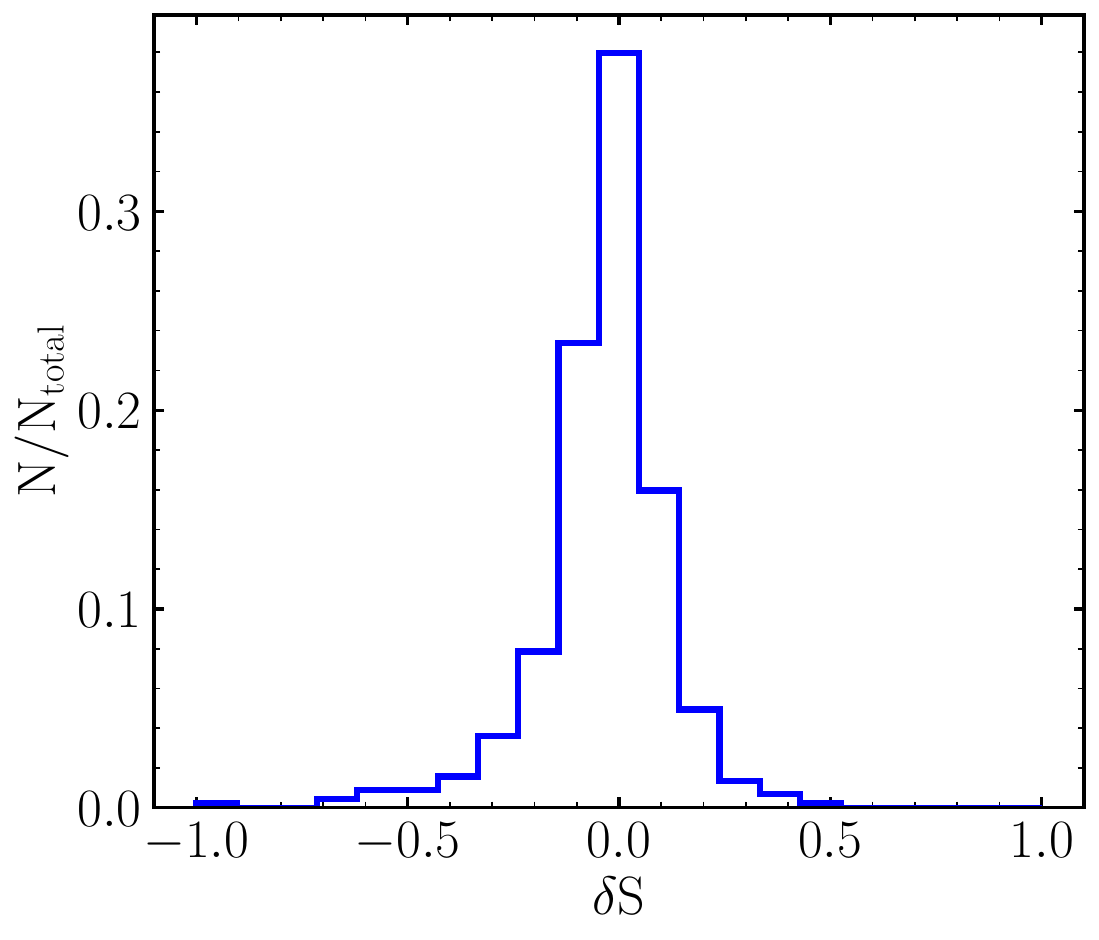}
    \caption{Top: Comparison of the 1375-1425 MHz flux of \add{447 continuum point sources} measured from the CRAFTS averaged map and the NVSS catalog. The colors of these dots represent the angular distance between the position given by {\tt DAOStarFinder} and the position in the NVSS catalog. The black line marks the equal flux relation. Bottom: histogram of the relative flux error $\delta S$ between our measurement on the map and NVSS flux.}
    \label{fig:NVSS_map}
\end{figure}

The continuum point sources are also measured on the Gaussian Kernel convolved map with {\tt nside}=4096, kernal size $\sigma_K = {\rm FWHM}/2\sqrt{2{\rm log}2}\sim 1.27\ {\rm arcmin}$ and averaged across 1375-1425 MHz, which can help us utilize data from more nearby time samples than the center-only map. We apply a source finder {\tt DAOStarFinder} \citep{1987PASP...99..191S} on our map with threshold = 7 mJy (corresponding to the five times of confusion limit of FAST map given in \cite{2023ApJ...954..139L}) and FWHM = add{$\sqrt{\rm FWHM_{beam}^2 + FWHM_{kernel}^2}$}. Over five thousand candidates are detected with our parameters, \add{most of which are largely influenced by their neighboring sources due to the spatial resolution limitation of FAST}. For the ideal sample containing 447 well-scanned bright isolated sources as described in \refsc{subsubsec:tod_continuum}, the comparison of their flux measured by {\tt DAOStarFinder} on the CRAFTS map and from NVSS catalog is shown in \reffg{fig:NVSS_map}. For each NVSS source, the corresponding FAST flux is from the measurement of the closest candidate found by {\tt DAOStarFinder} and the color of the dots in the left figure of \reffg{fig:NVSS_map} represent their angular distance. If the angular distance between the positions given by the star finder and NVSS catalog exceeds 3 arcmin, the corresponding source is likely a misidentification on our map. 
All 447 sources exhibit position errors smaller than 3 arcmin.
The {\tt DAOStarFinder} successfully finds all ideal sources that satisfy the selection criteria we mentioned in \refsc{subsec:flux_corr} and yields well-matched flux shown as the histogram of $\delta S$ in \reffg{fig:NVSS_map} with the overall relative error of \add{$\sim$ 6.6\%}. This error is slightly smaller than the measurement with TOD thanks to the multi-beam combination to lower the random background fluctuations and thermal noise. 
The measurement of continuum point sources on the map complements the source measurement in \refsc{subsubsec:tod_continuum} and validates our map-making process.

\subsection{\HI emission lines}\label{subsec:HIgal}

\begin{figure*}
    \centering
    \includegraphics[width=0.465\textwidth]{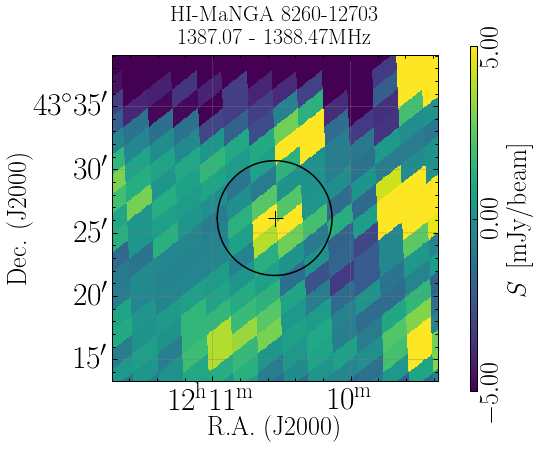}
    \hspace{3mm}
    \includegraphics[width=0.47\textwidth]{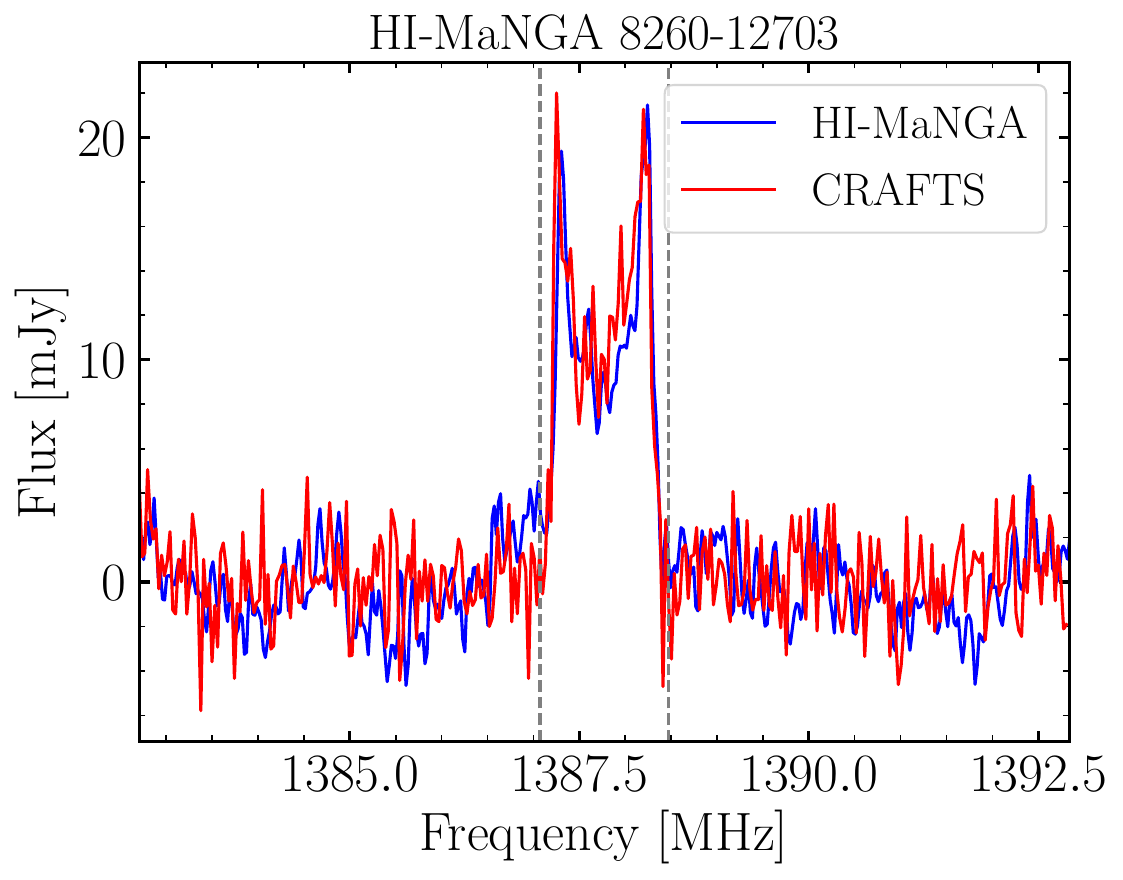}
    \caption{ Sky map (left) and spectra (right) of the galaxy \HI-MaNGA 8260-12703 as an example of \HI galaxies measured by CRAFTS and GBT \HI-MaNGA observation. Left: The sky map averaged over 1.4MHz near this galaxy. The colors represent the intensity at each pixel, the cross marks the central position given by \HI-MaNGA and the black circle shows the aperture size we use for the flux integration. Right: comparison between the spectra from CRAFTS map (red line) and \HI-MaNGA (blue line). }
    \label{fig:HIgal_example}
\end{figure*}

In contrast to the continuum sources, many low-redshift \HI galaxies can not be regarded as point sources compared with FAST beam size. Therefore, it is better to measure these extended sources on the map rather than with time-ordered data. We choose galaxies from \HI-MaNGA \footnote{\url{https://greenbankobservatory.org/science/gbt-surveys/hi-manga/};\,data available at \url{https://data.sdss.org/sas/dr17/env/MANGA_HI/}}\citep{2019MNRAS.488.3396M,2021MNRAS.503.1345S} catalog observed by the Green Bank Telescope (GBT), which is a 21cm follow-up program for the MaNGA (Mapping Nearby Galaxies at Apache Point Observatory) survey of SDSS-IV (the Sloan Digital Sky Survey - IV), to check our data. We select galaxies in the \HI-MaNGA catalog with S/N $>$ 6 located in the overlapped sky area and obtain a sample containing 90 sources. According to \cite{1980A&A....81..167S}, we obtain the \HI spectrum by
\begin{equation}\label{eq:HI_sumpix}
    S_{\nu} = \frac{\sum_{i} s_{\nu}(p_i)}{\sum_{i} B_{\nu}(p_i)} \,,
\end{equation}
in which $p_i$ is the $i$-th pixel, $s_{\nu}(p_i)$ is the flux at this pixel and $B_{\nu}(p_i)$ is the beam pattern at the angular distance between $p_i$ to the galaxy center given by \HI-MaNGA. To compare with GBT results, we use the nearby pixels within a GBT beam width around the center position to get the spectrum of each galaxy. The chosen aperture radius is $\theta_{\rm GBT}/2 = 1.22\frac{\lambda}{D_{\rm GBT}}\sim$ 4.5 arcmin, with slight variations depending on the frequency at which the \HI emission appears. After getting the pixels' combined spectra, we subtract the spectrum baseline with arPLS (asymmetrically reweighted Penalized Least Squares, \citealt{baek2015baseline}). An example of the sky map near the galaxy \HI-MaNGA 8260-12703 and its spectrum are shown in \reffg{fig:HIgal_example}. In the left figure, bright areas can be seen at the corresponding position of the bright source marked by the cross on the map averaged over $\sim$ 1387.1 - 1388.5 MHz. In the right plot, we can see clear \HI emission and the spectrum measured with CRAFTS (red line) matches well with \HI-MaNGA (blue line).

\begin{figure*}
    \centering
    \includegraphics[width=0.47\textwidth]{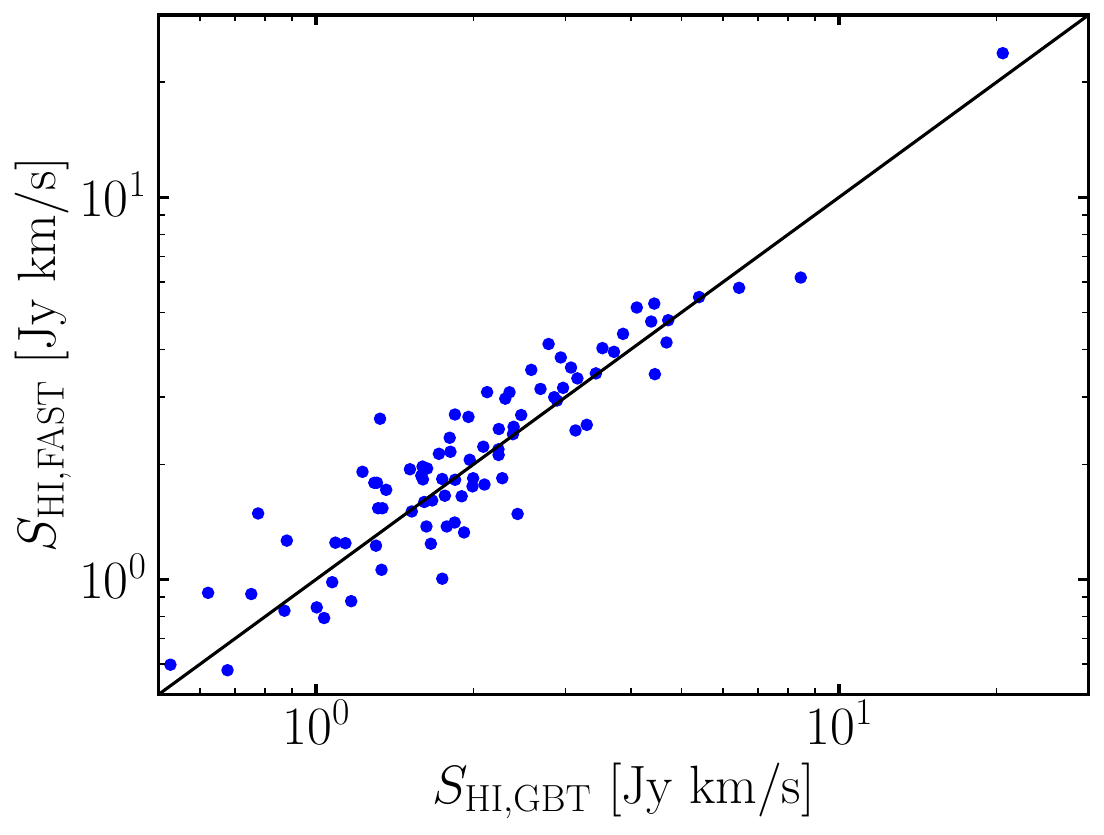}
    \hspace{3mm}
    \includegraphics[width=0.47\textwidth]{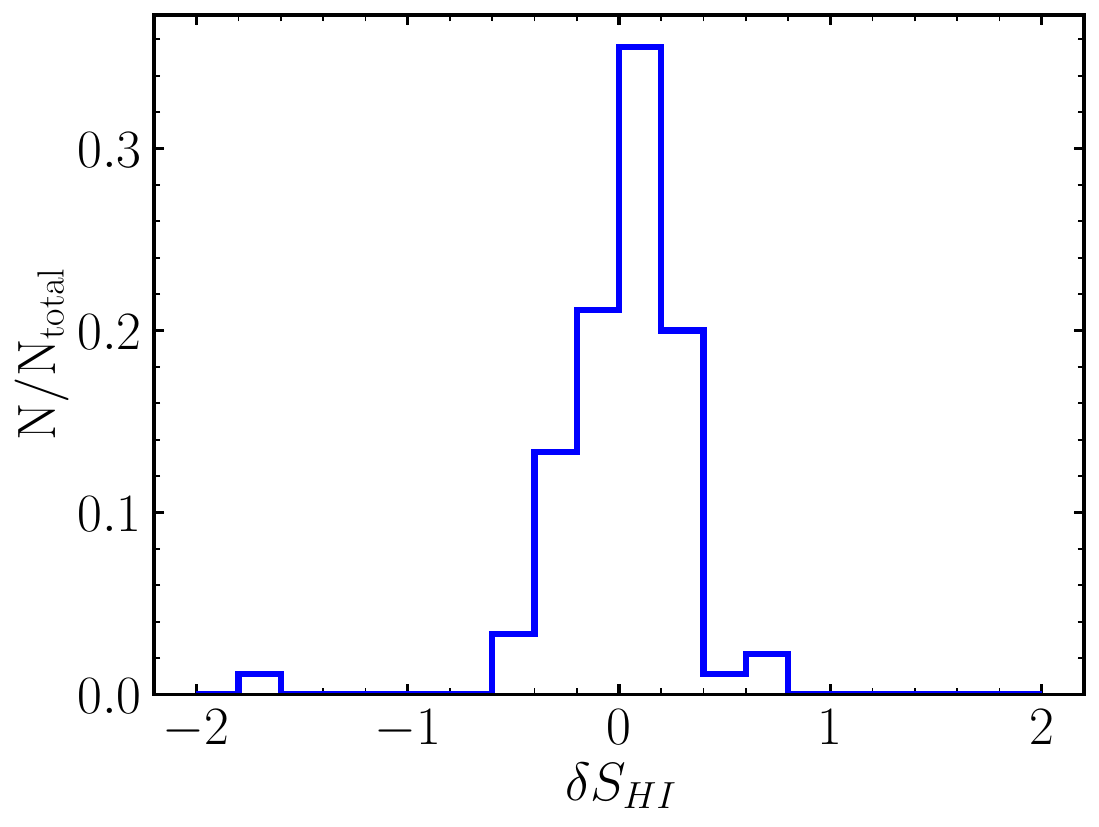}
    \caption{ Left: comparison of the flux of 90 sources measured on CRAFTS map and in GBT \HI-MaNGA catalog. The black line is the $y = x$ relation and each blue dot represents one source. Right: histogram of the relative error of the \HI integral flux. }
    \label{fig:HIgal_map}
\end{figure*}

The integral \HI flux of each galaxy is obtained by calculating the area between the frequency $\nu_c \pm 0.6 \cdot W_{20}$, where $\nu_c$ is the central frequency of the emission line, $W_{20}$ is the full width between the two channels with 20\% of the peak flux and both of them are given in \HI-MaNGA catalog. The width $1.2 W_{20}$ we use is an empirical value for the whole emission width used in \cite{2019MNRAS.488.3396M}. To avoid the difference between FAST and GBT results caused by different baseline subtraction methods and integral parameters, we use the baseline unsubtracted spectra obtained by GBT to re-calculate the integral flux, RMS (root-mean-square) and S/N for \HI-MaNGA with the same baseline subtraction algorithm, signal width, and frequency range. 

The measured flux, RMS, S/N, and relative errors of the 90 galaxies from CRAFTS and \HI-MaNGA are listed in \reftb{tb:HIgal_map}. We compare the measured integral flux of all galaxies in our sample in \reffg{fig:HIgal_map}. The overall relative error is $\sim$ 16.7\%, which is derived by the method described in \refsc{subsubsec:tod_continuum}. The spectra and flux of most galaxies measured with CRAFTS are consistent with \HI-MaNGA, apart from a very few outliers. Many reasons can cause  outliers. For example, although we tried to use the GBT beam size to choose our integral pixels, the photometric apertures for the two observations are not exactly the same. In addition, for some \HI emission which coincide with RFI-flagged frequency channels, it is hard to recover the signal, leading to unpredictable large errors of their integral flux. To quantify the influence of RFI, we define a parameter $\xi_{\rm mask}$ to represent the ratio of RFI flagged data point for each signal, which is also listed in \reftb{tb:HIgal_map}. For galaxies with $\xi_{\rm mask} > 0.1$, we conservatively assume the measured flux is severely influenced by RFI and not reliable. Three galaxies in our sample are contaminated according to this criteria and are excluded when we calculate the overall relative error. \add{The errors in our measurement are also competitive compared with other studies of \HI galaxies. For example, \cite{2011AJ....142..170H} compared \HI line flux density measurements from ALFALFA and HIPASS for 1888 galaxies, reporting a scatter with $\sigma$ = 23.8\%; \cite{2022PASA...39...58W} compared the Wide-field ASKAP L-band Legacy All-sky Blind Survey (WALLABY) measurements with those from single-dish surveys (ALFALFA and HIPASS), finding the scatter of 34\%; \cite{2024SCPMA..6719511Z} examined the FAST all sky \HI survey (FASHI) which is also carried out by FAST, compared \HI fluxes from FAST and ALFALFA for 3620 galaxies and also found significant dispersion, though no specific numerical value was provided. Although these studies may differ in the observation parameters and the way to compute errors, they can serve as useful references, suggesting that our measurement uncertainty is relatively well-constrained.}

The RMS of CRAFTS spectra is a little bit larger than the \HI-MaNGA results, as the CRAFTS integration time per beam is much shorter than that of the GBT \HI-MaNGA observation(several minutes). Besides, there are also differences in the frequency resolution and the smoothing process between the CRFATS and \HI-MaNGA. Nevertheless, the comparable results given by FAST confirm its high performance for HI galaxy detection, and we can expect lower noise with an additional scan (which is in CRAFTS future plan). Because of the high sensitivity of the FAST map, fainter galaxies are expected to have better detections. Some \HI galaxy surveys with FAST are underway, e.g. FAST all sky \HI survey (FASHI, \citealt{2024SCPMA..6719511Z}), \HI Intensity Mapping and Galaxy Survey (\citealt{2023ApJ...954..139L}, Shu et al,  in preparation). \add{In future work, we will also conduct a more detailed analysis of \HI emission lines of galaxies with CRAFTS data and further improve our results through additional tests and optimization.}

\section{Summary}\label{sec:summary}

In this work, we describe the \HI data processing pipeline for the spectrum data from the Commensal Radio Astronomy FAST Survey (CRAFTS) project. According to the continuity of sky and the data quality, we select a patch of sky area of $\sim 270 \rm deg^2$ at RA from 12h to 17h, Dec from $40^{\circ}$ to $45^{\circ}$. Our data reduction pipeline consists of nine steps: pre-processing, bandpass calibration, RFI flagging, temporal drift calibration, absolute flux calibration, temporal baseline subtraction, flux correction, map-making, and standing wave removal. Many systematic issues are carefully studied. We investigate the pointing deviation and errors during the drift scan observation, and the effect of noise diode overflow and its correction. 

We compare the theoretical sensitivity and the real noise level of our data. For the time-ordered-data, the theoretical noise level is estimated to be $\sigma_{\rm TOD,theo} = 5 \sim 6 ~\rm mJy$, and we obtain an observed value of $\sigma_{\rm TOD,obs} = 6.2~ \rm mJy$ for the 1050-1150 MHz band; and $\sigma_{\rm TOD,obs} = 5.5~ \rm mJy$ for the 1300-1450 MHz. For the sky map, the expected and observed noise levels are $\sigma_{\rm map,theo} = 1.58~ \rm mJy$ and $\sigma_{\rm map,obs} = 1.60~ \rm mJy$ for the 1315-1415 MHz band. Our results are consistent with expectations considering acceptable errors of the parameters we use to estimate the ideal sensitivity.

Principal Components Analysis (PCA) is applied to our map at 1315-1415 MHz to test foreground removal. To improve the PCA results, we remove RFI-contaminated pixels and frequency channels and cut off the map edges to achieve approximately uniform sensitivity. Most foreground components associated with contamination are effectively mitigated after we subtract the first 30 PCA modes. The eigenvalues decrease rapidly and reach a plateau after 30 modes are removed, which gives us guidance on the choice of the number of foreground modes to subtract. 

With the processed time-ordered data and the map, we measure the flux of 447 continuum point sources near 1400 MHz. Compared with the NVSS catalog, our results yield a relative flux error of $\sim 8.3\%$ for TOD and $\sim 11.6\%$ for the map. The consistency between these two surveys verifies the validity of our data processing. By dividing the sample of sources into three sub-groups according to the position of the beam that scans the source, we find larger errors in sources measured by beams located at the outer ring of the 19-beam receiver, which indicates the beam pattern may deviate more from the 2D Gaussian shape than the central beam.

We also select 90 \HI emission galaxies from the \HI-MaNGA catalog and re-measure their flux with the CRAFTS map. While choosing the aperture size of the integral pixels equivalent to the GBT beam size, we get the well-matched \HI flux between these two observations, and the overall relative error is $\sim 16.7\%$, verifying the accuracy of our processed data. The comparable S/N of these sources measured in CRAFTS drift scan observation and GBT ON-OFF tracking observation proves the high detection ability of FAST. 

The processed CRAFTS data product we obtained provides a valuable foundation for future \HI cosmology research and has been validated for \revise{point sources and \HI galaxies} detection. With the increased observed sky area in the CRAFTS project and the deployment of wider band receivers in the near future, we plan to carry out further \HI cosmological studies with more FAST data. Searching for faint \HI galaxies would also be a valuable work thanks to the high sensitivity of FAST. For research about the large-scale structure of the \add{Universe} with an intensity mapping approach, \revise{we still need to conduct additional validation tests to refine the calibration strategies and precisely quantify their influence for future cosmological analysis. We will also improve the} foreground removal by investigating unknown structures in the data and trying other methods like FastICA (Fast Independent Component Analysis) and GMCA (Generalised Morphological Component Analysis) before estimating the power spectrum.


\section*{Acknowledgments}
This work made use of the data from FAST (Five-hundred-meter Aperture Spherical radio Telescope, \url{https://cstr.cn/31116.02.FAST}). FAST is a Chinese national mega-science facility, operated by National Astronomical Observatories, Chinese Academy of Sciences.
This work is supported by the National SKA Program of China (Nos. 2022SKA0110100 and 2022SKA0110101), the NSFC International (Regional) Cooperation and Exchange Project (No. 12361141814). 
WY acknowledges the financial support from the China Scholarship Council (CSC, No. 202204910346). 
LW is a UK Research and Innovation Future Leaders Fellow [grant MR/V026437/1]. 
SC is supported by a UK Research and Innovation Future Leaders Fellowship grant [MR/V026437/1]. 
YL acknowledges the support of the National Natural Science Foundation of China (No. 1247309).
DL is a new Cornerstone investigator and acknowledges the support from the National Natural Science Foundation of China (No. 11988101). 
The authors thank Weiwei Zhu, Bo Zhang, Aishrila Mazumder, Sourabh Paul for helpful discussion. 



\section*{Data Availability}

The data underlying this article will be shared on reasonable request to the corresponding authors.



\appendix

\setcounter{figure}{0}
\setcounter{table}{0}
\renewcommand{\thefigure}{A\arabic{figure}}
\renewcommand{\thetable}{A\arabic{table}}

\add{\section{Noise diode}\label{appx:TND}}


\begin{figure*}
    \centering
    \includegraphics[width=0.6\textwidth]{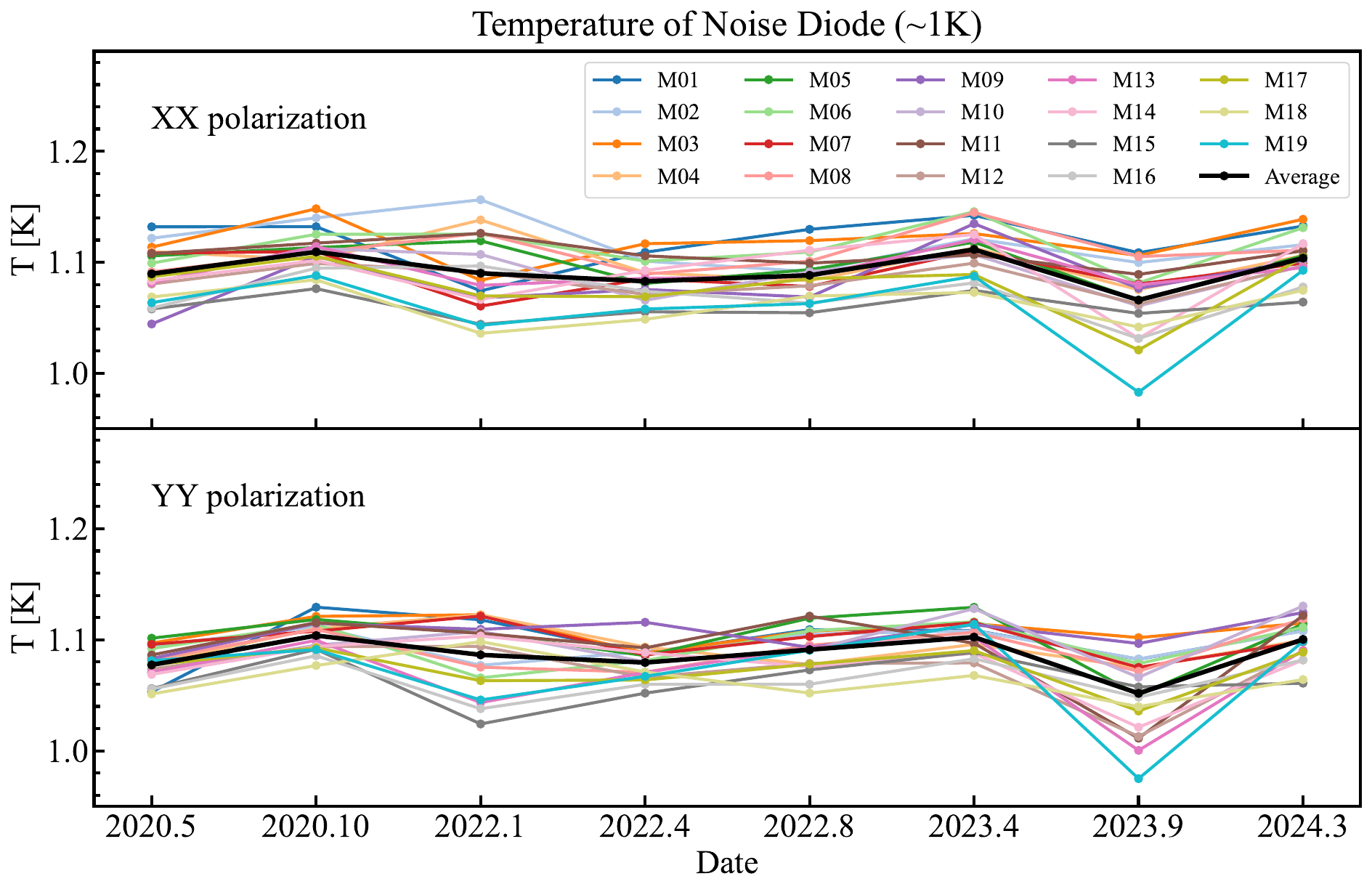}
    \caption{\add{The average noise diode temperature for each beam at each measurement. The x-axis represents the date of measurement in the format YYYY.M, e.g. ``2020.5'' means the measurement in May 2020. Different colored lines with dots show the temperature of different beams.} }
    \label{fig:TND_mean}
\end{figure*}

\add{The temperature of the noise diode are measured with the hot load method by FAST group. To assess the stability of the noise diode, we check its spectra from eight measurements conducted between 2020 and 2024. Here we focus on the low-level noise diode ($\sim$ 1K) because the high-level noise diode ($\sim$ 10K) is not used in CRAFTS observations. }

\add{
Through the comparison of eight measurements, we find the temperature spectral profiles of the low-level noise diode remained stable over the four-year period, despite slight variation in amplitude. The stability of the spectral shape enables us to correct the calibration error caused by noise diode variation during the flux correction process in \refsc{subsec:flux_corr}. We also plot the mean temperature for each feed in \reffg{fig:TND_mean} to see the trend of amplitude variation. Compared to the average temperature of all beams indicated by the black line, the temperatures of different beams deviate within $\sim$ 5\%, with no significant similarity in their variation trends. }

\section{Tests for RFI flagging}\label{appx:jackknife}

\begin{figure*}
    \centering
    \includegraphics[width=0.85\textwidth]{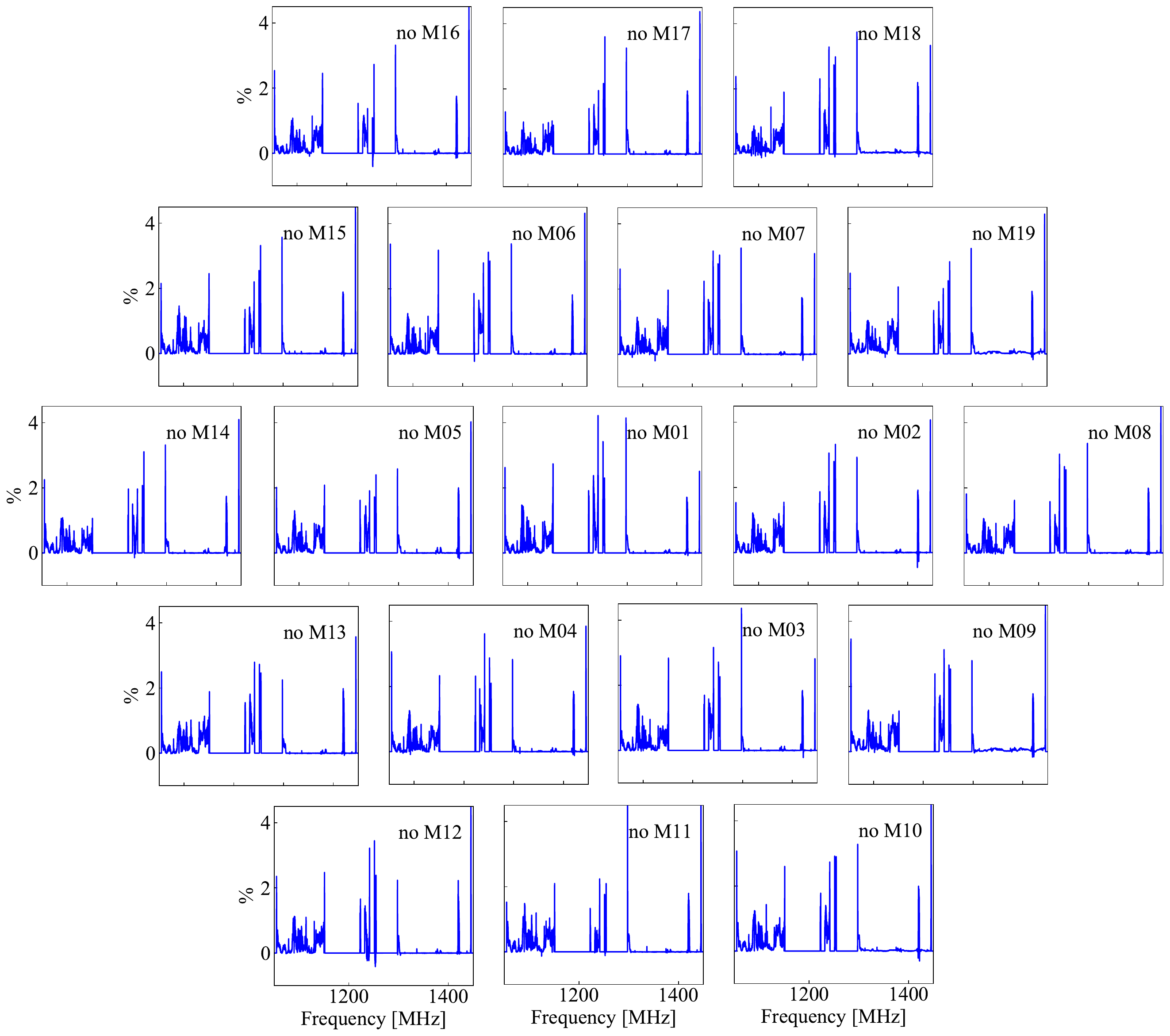}
    \caption{Percentage of different flagged points at each frequency channel while applying our RFI flagging algorithm on 19 beams averaged data and 18 beams averaged data. }
    \label{fig:RFIflag_jackknife}
\end{figure*}

\begin{figure*}
    \centering
    \includegraphics[width=0.45\textwidth]{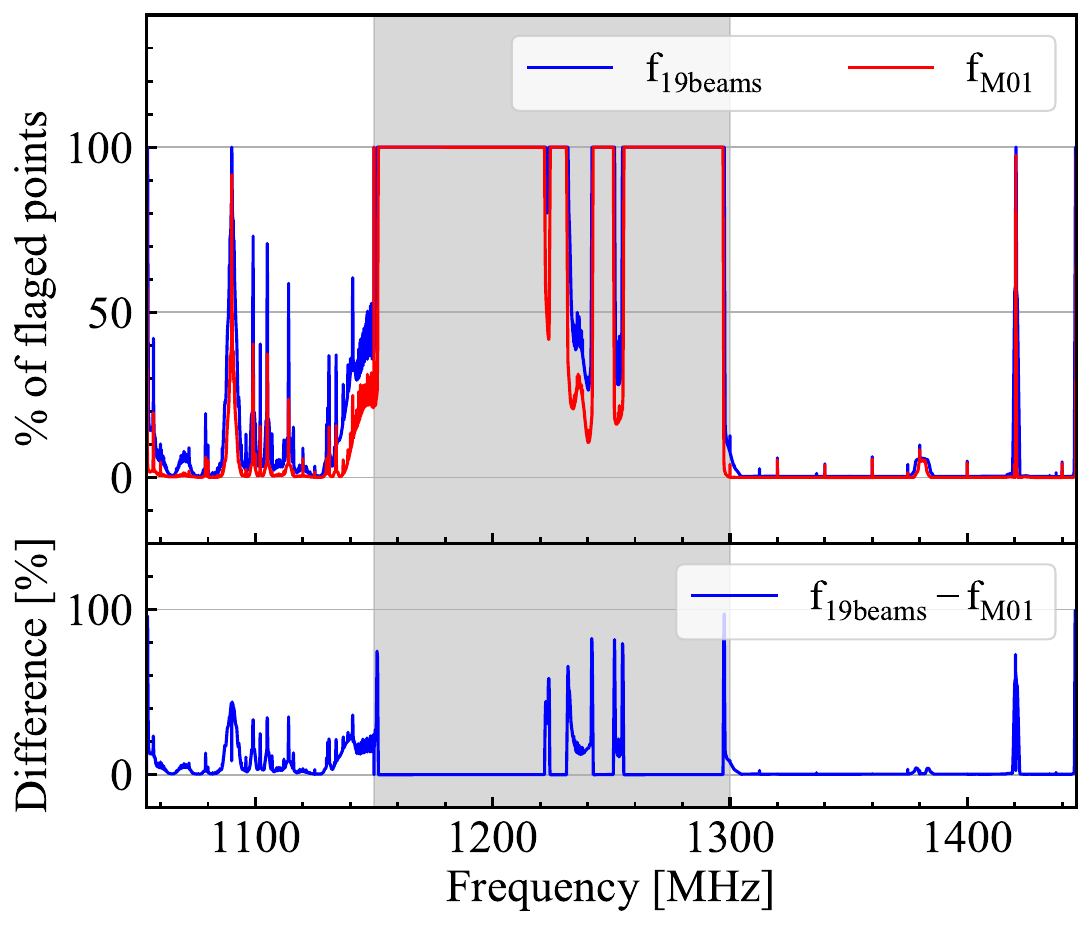}
    \includegraphics[width=0.45\textwidth]{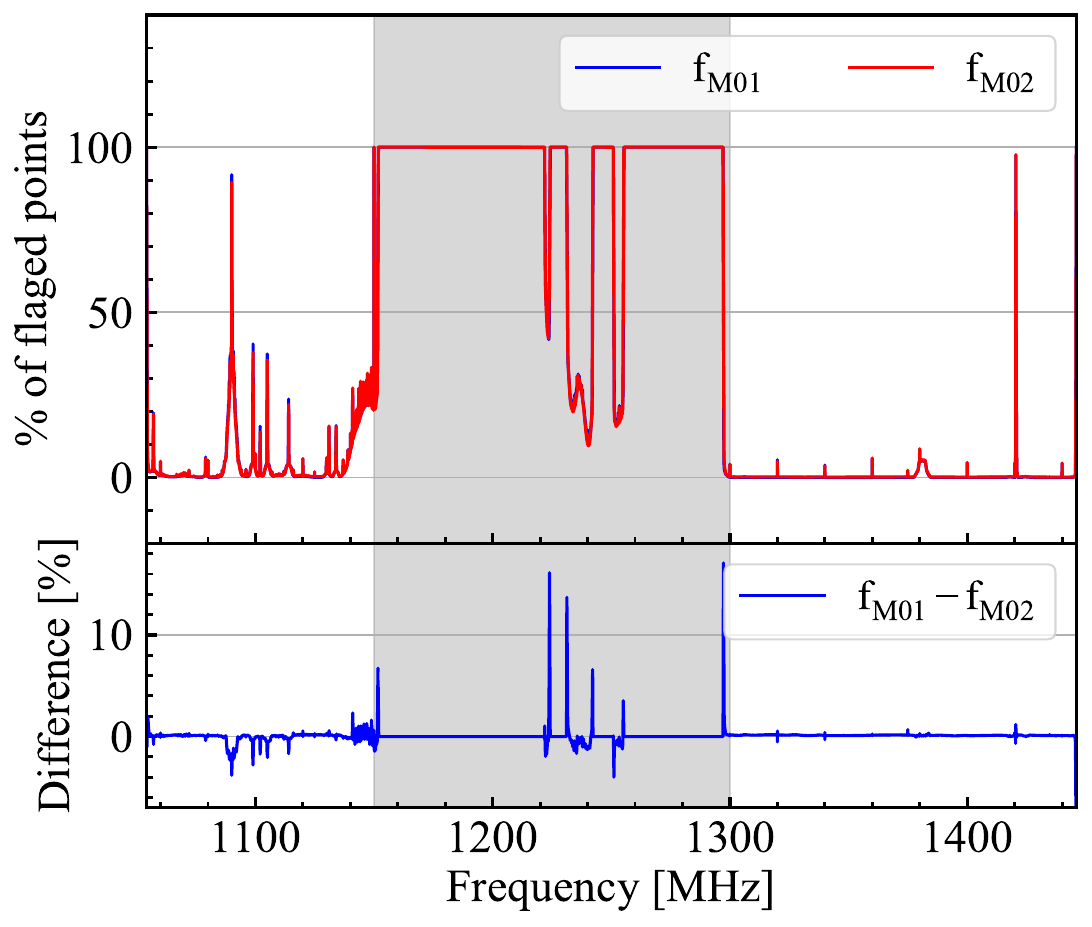}
    \caption{\add{Left: the upper panel shows comparison of the percentage of flagged time points at each frequency channels for the all-beam stacked RFI flagging (blue line) and the single-beam flagging on M01 data (red line). The lower panel shows the difference between the upper two lines. The gray shaded area is the severely RFI contaminated region 1150-1300MHz that is not included in our scientific analysis. Right: similar to the left figure but for the comparison between RFI flagging results with single-beam flagging method for M01 (blue line) and M02 (red line) respectively.} }
    \label{fig:RFIflag_perbeam}
\end{figure*}

\subsection{Jackknife tests}

In the RFI flagging part, we use the Jackknife test to check the influence of a single beam. We set 20 groups as the input data of our RFI flagging algorithm, one of them is the 19 beams averaged data as our fiducial group, and data in each of the other 19 groups is only 18 beams averaged for testing. In \reffg{fig:RFIflag_jackknife}, we plot the percentage of different flagged points between the fiducial masking results and the testing results at each frequency channel, i.e. 
\begin{equation}\label{eq:f_diff_flag}
    f_{{\rm diff}, i}(\nu) = \frac{\sum_t (\mathcal{M}_{\rm 19beam}(t,\nu) - \mathcal{M}_{{\rm 18 beam}, i}(t,\nu))}{N_t}\,,
\end{equation}
in which $\mathcal{M}$ represents the mask array with the value 1 at flagged points and 0 at unflagged points, $\mathcal{M}_{\rm 19beam}$ and $\mathcal{M}_{{\rm 18beam},i}$ are the mask array obtained by applying out RFI flagging algorithm on 19 beams averaged data or 18 beams averaged data (with $i$-th beam excluded) respectively, $N_t$ is the number of time points at each frequency channel. Each sub-figure of \reffg{fig:RFIflag_jackknife} represents the result with different excluded beams in $\mathcal{M}_{\rm 18beam}$. We can see the values of $f_{{\rm diff}, i}(\nu)$ are mostly between 0\% and 3\%, indicating that our RFI flagging process provides a stricter identification of RFIs with the 19 beams averaged input data than 18 beams averaged input data. The level of different flagging points is also consistent with the different theoretical noise levels of 19 beams or 18 beams averaged data, which is $\sim 2.7\%$ estimated using the radiometer equation in \refeq{eq:sensitivity}. The 19 sub-figures in \reffg{fig:RFIflag_jackknife} are very similar, which means a single beam would not significantly influence the structure of the flagging array and confirm the robustness of our RFI flagging process.

\subsection{Per-beam flagging tests}

\add{Instead of applying RFI flagging on individual beams separately, we performed this process on the stacked data across all beams. Our primary motivation for using all-beam stacked flagging is to reduce the noise level, making weak RFIs easier to identify while also avoiding excessive removal of bright sources. To confirm this, we conducted a test comparing per-beam flagging with all-beam stacked flagging. The results presented in the left panel of \reffg{fig:RFIflag_perbeam}, show that the all-beam stacked approach identifies more RFIs, particularly weak RFIs or those at the edges of strong RFIs. Furthermore, as shown in the right panel of \reffg{fig:RFIflag_perbeam}, the per-beam flagging results for different beams exhibit strong similarity, indicating that RFI features tend to be consistent across beams. Based on these findings, we maintain the use of all-beam stacked RFI flagging in our pipeline.}

\section{Bad data identification}\label{appx:bad_data}

\begin{figure*}
    \centering
    \includegraphics[width=0.85\textwidth]{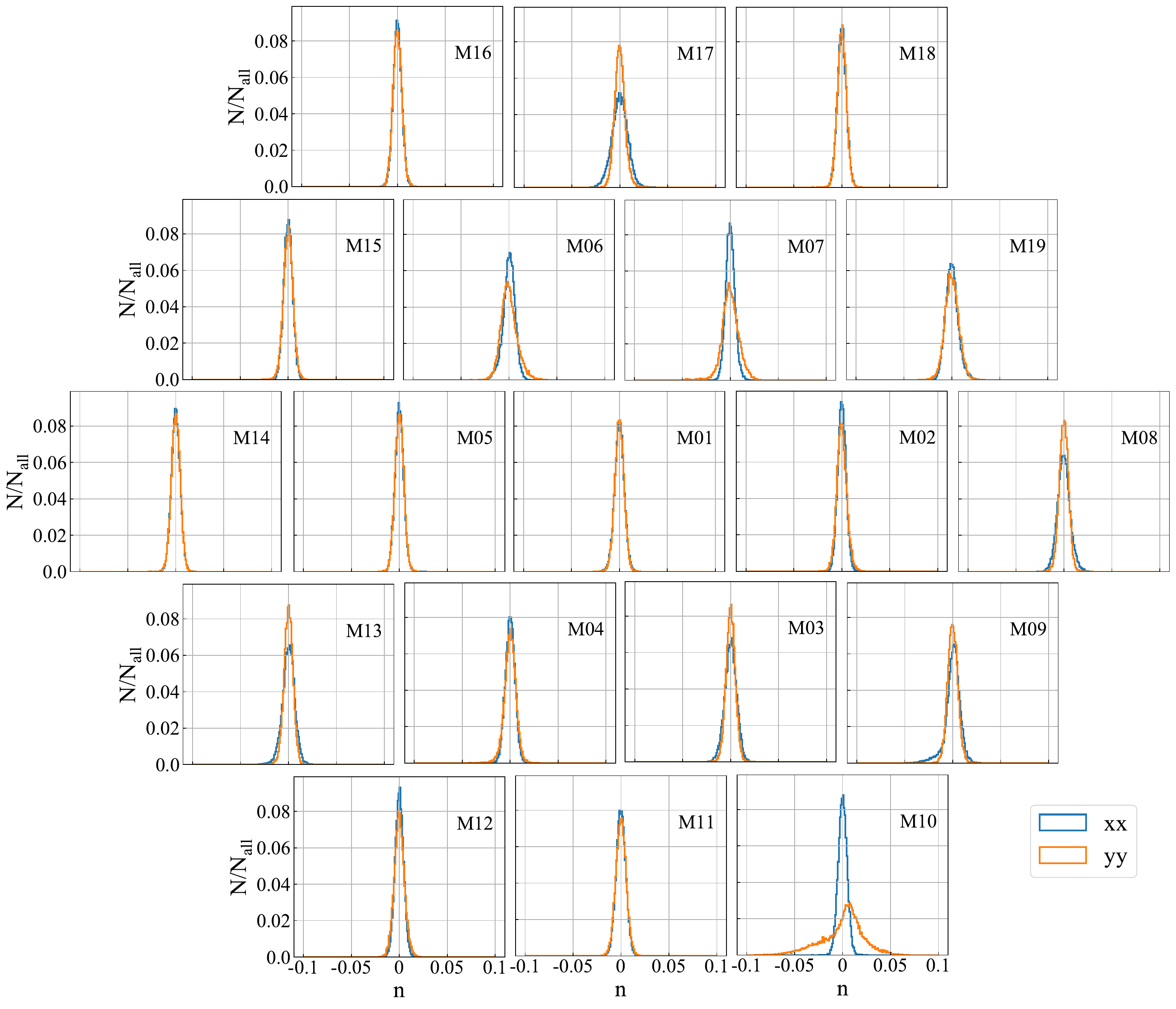}
    \caption{Histograms of the baseline subtracted $g(t)$. Each subplot represents data from one beam the blue lines are for XX polarization and the orange lines are for YY polarization. }
    \label{fig:gt_hist}
\end{figure*}

\begin{figure}
    \centering
    \includegraphics[width=0.49\textwidth]{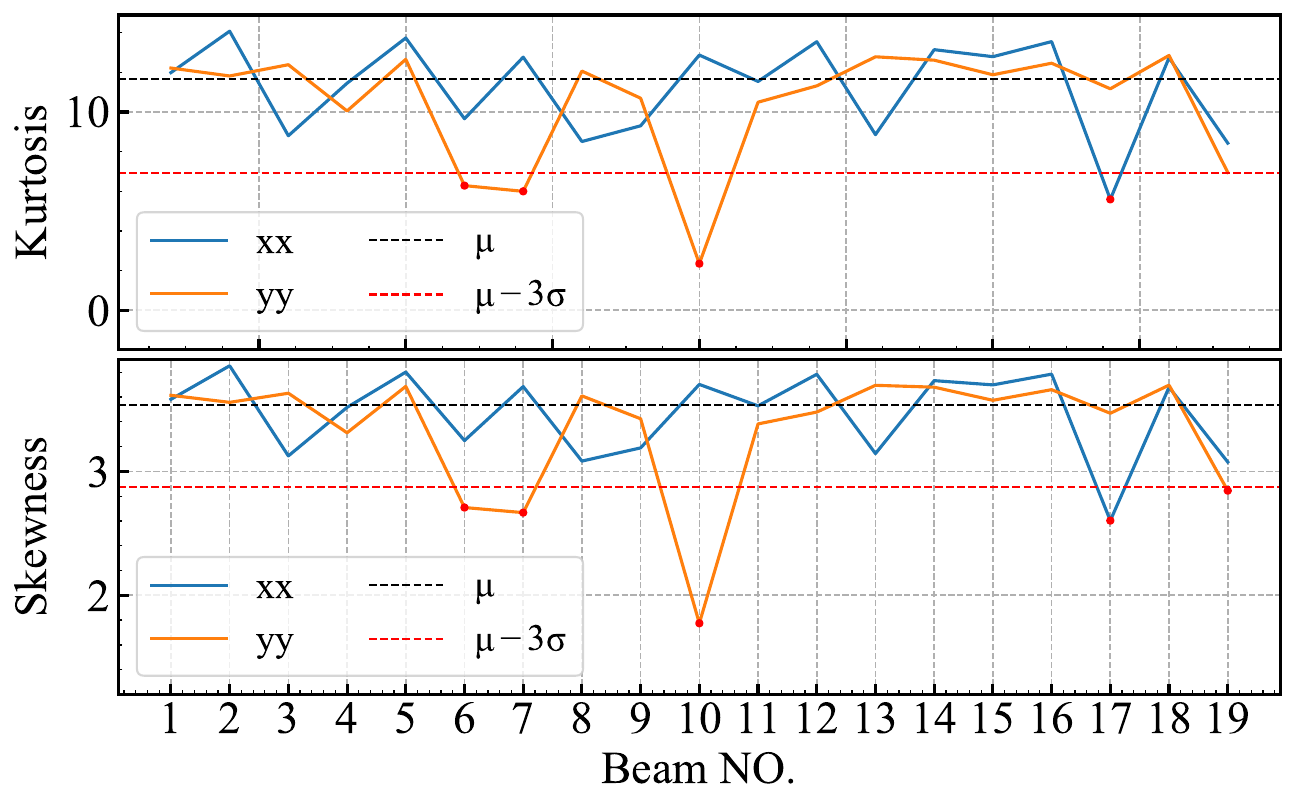}
    \caption{The Kurtosis values (upper panel) and Skewness values (lower panel) of all 19 beams and 2 polarizations. The blue lines show values from XX polarization data and the orange lines are for YY polarization. The black dashed lines represent the mean value $\mu$ of Kurtosis or Skewness of 19 beams and 2 polarizations and the red dashed lines show the threshold $\mu - 3\sigma$ as we set to identify bad data. }
    \label{fig:gt_kurt+skew}
\end{figure}

As mentioned in \refsc{subsec:gtcal}, to identify the bad data that should be excluded in later analysis, we calculate the Kurtosis and Skewness value of the temporal drift $g(t)$ for each beam to check to what extent the time-ordered-data deviates from Gaussian distribution. The Kurtosis (Kurt) and Skewness (Skew) value of dataset $X$ are defined as
\begin{equation}\label{eq:kurtosis}
    {\rm Kurt}(X) = {\rm E}\Big{[}\Big{(}\frac{X-\mu}{\sigma}\Big{)}^4\Big{]} \,,
\end{equation}
and
\begin{equation}\label{eq:skewness}
    {\rm Skew}(X) = {\rm E}\Big{[}\Big{(}\frac{X-\mu}{\sigma}\Big{)}^3\Big{]} \,,
\end{equation}
where $\mu$ is the mean value, $\sigma$ is the standard deviation and E represents the calculation of expected value. They can be calculated by functions in the Python package {\tt scipy.stats}.

Examples for the histograms of the baseline subtracted $g(t)$ for all 19 beams and 2 polarizations are shown in \reffg{fig:gt_hist}. We can easily notice the profile of the histograms for bad data (e.g. M10, YY polarization) are lower than normal ones and deviate from Gaussian shape. The kurtosis and skewness values for all 19 beams and 2 polarizations are shown in \reffg{fig:gt_kurt+skew}. For data with kurtosis or skewness value deviate over $3\sigma$ from the mean value of all 19 beams and polarizations, we record them as bad data (marked with red points in \reffg{fig:gt_kurt+skew}) and do not include them in our map.

\section{\HI sources catalog}

The catalog of 90 \HI sources that we measure in this work as described in \refsc{subsec:HIgal} are shown in \reftb{tb:HIgal_map}. The introduction of each column is listed below:

\begin{itemize}
    \item Column 1: Number of the galaxy measured in this work.
    \item Column 2: Name of the galaxy in \HI-MaNGA catalog.
    \item Column 3: Position including RA and Dec (J2000) of the galaxy in the unit of degree given in \HI-MaNGA catalog.
    \item Column 4: Heliocentric velocity of the \HI emission given in \HI-MaNGA catalog, $cz_{\sun}$ in $\rm km/s$.
    \item Column 5: Integrated \HI flux of the source re-measured by us with GBT spectra data in the unit of $\rm Jy \cdot km/s$.
    \item Column 6: Integrated \HI flux of the source in the unit of $\rm Jy \cdot km/s$ measured on CRAFTS map with the photometric aperture size equal to GBT beam size ($\sim 9'$).
    \item Column 7: Noise level of the spatially integrated spectra in the unit of mJy re-measured by us with GBT spectra data.
    \item Column 8: Noise level of the spatially integrated spectra in the unit of mJy measured with CRAFTS map.
    \item Column 9: S/N of the source re-measured by us with GBT spectra data.
    \item Column 10: S/N of the source measured with CRAFTS map.
    \item Column 11: Relative error between the two integrated flux values measured with GBT and CRAFTS.
    \item Column 12: The RFI flagging parameter to reflect the ratio of masked data points.
    
\end{itemize}

\begin{longtable}{ccccccccccccc}
\caption{Parameters of 90 \HI galaxies measured by CRAFTS and GBT \HI-MaNGA program.}\\
    \hline
     NO. & Name & RA, Dec & $v_{\rm \HI}$ & $F_{\rm \HI, GBT}$ & $F_{\rm \HI, FAST}$ & ${\sigma}_{\rm GBT}$ & ${\sigma}_{\rm FAST}$ & $S/N_{\rm GBT}$ & $S/N_{\rm FAST}$ & $\delta S$ & $\rm \xi_{mask}$ \\
    &  & ([$^{\circ}$], [$^{\circ}$]) & [${\rm km/s}$] & [$\Jy \cdot {\rm km/s}$] & [$\Jy \cdot {\rm km/s}$] & [mJy] & [mJy] &  &  & [\%] &  \\
     \hline
     \endfirsthead

    \hline
     NO. & Name & RA, Dec & $v_{\rm \HI}$ & $F_{\rm \HI, GBT}$ & $F_{\rm \HI, FAST}$ & ${\sigma}_{\rm GBT}$ & ${\sigma}_{\rm FAST}$ & $S/N_{\rm GBT}$ & $S/N_{\rm FAST}$ & $\delta S$ & $\rm \xi_{mask}$ \\
    &  & ([$^{\circ}$], [$^{\circ}$]) & [${\rm km/s}$] & [$\Jy \cdot {\rm km/s}$] & [$\Jy \cdot {\rm km/s}$] & [mJy] & [mJy] &  &  & [\%] &  \\
     \hline
     \endhead

     \hline
     \endfoot
     
    1 & 11750-12703 & (191.05, 41.67) & 5120.52 & 1.76 & 1.66 & 1.89 & 1.92 & 35.78 & 29.15 & -6.27 & 0 \\
    2 & 11750-12705 & (189.16, 40.27) & 7081.80 & 1.78 & 1.38 & 1.35 & 2.02 & 34.94 & 15.73 & -25.73 & 0 \\
    3 & 11750-6103 & (189.25, 40.35) & 7101.79 & 1.07 & 0.98 & 1.30 & 1.81 & 23.55 & 13.51 & -8.80 & 0 \\
    4 & 11755-12705 & (188.65, 42.44) & 5901.28 & 2.09 & 2.23 & 1.23 & 1.96 & 39.43 & 22.93 & 6.33 & 0 \\
    5 & 11755-9102 & (187.75, 42.53) & 11572.80 & 1.92 & 1.33 & 1.30 & 2.24 & 33.54 & 11.60 & -37.05 & 0 \\
    6 & 11941-12703 & (247.44, 41.78) & 9410.95 & 0.62 & 0.92 & 1.11 & 2.68 & 17.47 & 9.29 & 39.70 & 0 \\
    7 & 11941-9101 & (246.16, 41.02) & 8174.22 & 0.88 & 1.26 & 1.98 & 1.79 & 15.57 & 21.13 & 36.44 & 0 \\
    8 & 11945-6102 & (252.90, 40.02) & 8859.08 & 1.09 & 1.25 & 1.25 & 2.23 & 20.69 & 11.52 & 13.62 & 0 \\
    9 & 7443-1902 & (231.99, 42.97) & 5358.00 & 4.11 & 5.15 & 1.52 & 3.13 & 39.30 & 20.75 & 22.66 & 0.01 \\
    10 & 7443-6103 & (230.23, 42.78) & 5501.10 & 1.84 & 1.41 & 1.20 & 2.02 & 42.38 & 16.90 & -26.82 & 0 \\
    11 & 8259-12701 & (179.96, 43.74) & 5869.00 & 1.00 & 0.85 & 1.54 & 2.96 & 19.87 & 7.49 & -17.19 & 0 \\
    12 & 8259-12704 & (180.16, 44.41) & 7065.00 & 3.16 & 3.36 & 1.38 & 2.37 & 96.72 & 50.43 & 6.08 & 0 \\
    13 & 8260-12701 & (181.24, 43.33) & 7130.24 & 1.72 & 2.13 & 1.48 & 1.87 & 37.62 & 31.77 & 21.64 & 0 \\
    14 & 8260-12702 & (181.77, 42.98) & 7067.38 & 1.74 & 1.83 & 1.98 & 2.03 & 44.15 & 38.86 & 4.95 & 0 \\
    15 & 8260-12703 & (182.64, 43.44) & 7053.53 & 2.97 & 3.17 & 1.74 & 1.82 & 45.20 & 40.42 & 6.68 & 0 \\
    16 & 8260-12705 & (183.14, 41.62) & 7238.50 & 2.43 & 1.48 & 1.43 & 2.00 & 45.01 & 17.14 & -49.80 & 0 \\
    17 & 8260-6101 & (182.41, 42.01) & 6848.66 & 2.24 & 2.48 & 1.54 & 1.59 & 35.02 & 32.67 & 10.15 & 0 \\
    18 & 8260-6103 & (182.31, 44.09) & 11247.46 & 4.38 & 4.73 & 1.26 & 1.78 & 72.43 & 48.33 & 7.80 & 0 \\
    19 & 8260-9101 & (182.29, 44.00) & 11247.66 & 2.23 & 2.19 & 1.38 & 1.67 & 35.20 & 24.44 & -1.96 & 0 \\
    20 & 8261-6102 & (182.71, 44.51) & 6901.00 & 4.68 & 4.17 & 2.85 & 1.84 & 38.29 & 45.88 & -11.54 & 0 \\
    21 & 8262-3702 & (183.66, 43.54) & 7281.02 & 2.10 & 1.77 & 1.27 & 1.87 & 48.32 & 24.15 & -17.03 & 0 \\
    22 & 8262-9102 & (184.55, 44.17) & 7400.15 & 1.67 & 1.61 & 1.26 & 1.74 & 34.35 & 20.49 & -3.55 & 0 \\
    23 & 8263-12705 & (186.06, 44.94) & 14402.71 & 1.84 & 2.70 & 1.35 & 1.88 & 31.03 & 28.08 & 38.46 & 0 \\
    24 & 8313-12701 & (239.49, 41.79) & 10474.00 & 1.66 & 1.24 & 1.57 & 2.54 & 23.12 & 9.25 & -29.18 & 0 \\
    25 & 8313-12702 & (240.68, 41.20) & 9953.00 & 0.22 & 2.83 & 1.63 & 2.05 & 2.70 & 23.81 & 329.25 & 0 \\
    26 & 8313-12703 & (240.37, 42.39) & 10649.00 & 2.12 & 3.09 & 1.37 & 2.11 & 34.07 & 27.80 & 37.74 & 0 \\
    27 & 8313-12705 & (242.68, 41.15) & 9477.00 & 0.78 & 1.49 & 1.64 & 1.98 & 8.05 & 11.15 & 66.40 & 0 \\
    28 & 8313-6103 & (239.44, 41.71) & 7131.00 & 0.87 & 0.83 & 1.76 & 2.38 & 24.06 & 14.97 & -4.94 & 0 \\
    29 & 8313-9102 & (239.99, 41.48) & 9991.00 & 1.59 & 1.87 & 1.39 & 2.00 & 26.03 & 18.60 & 16.17 & 0 \\
    30 & 8314-9101 & (242.38, 40.06) & 5279.84 & 1.30 & 1.23 & 1.26 & 2.08 & 32.61 & 15.94 & -6.06 & 0 \\
    31 & 8317-6104 & (194.89, 42.76) & 7177.00 & 2.23 & 2.12 & 3.21 & 1.90 & 16.61 & 25.76 & -5.38 & 0 \\
    32 & 8317-9102 & (192.98, 44.09) & 7267.00 & 1.14 & 1.24 & 1.55 & 1.68 & 28.80 & 25.11 & 8.90 & 0 \\
    33 & 8318-9101 & (196.09, 45.06) & 8495.30 & 1.90 & 1.65 & 1.46 & 2.50 & 33.13 & 14.62 & -14.03 & 0 \\
    34 & 8327-12702 & (204.81, 44.25) & 10180.00 & 1.33 & 1.06 & 1.16 & 1.81 & 35.40 & 15.79 & -23.10 & 0 \\
    35 & 8328-3704 & (211.68, 44.43) & 3829.13 & 1.34 & 1.54 & 0.85 & 1.78 & 60.64 & 27.90 & 13.68 & 0 \\
    36 & 8329-12701 & (213.42, 43.87) & 10517.00 & 4.72 & 4.77 & 1.79 & 1.91 & 61.64 & 50.36 & 1.12 & 0 \\
    37 & 8329-12702 & (211.68, 44.43) & 3829.13 & 1.31 & 1.54 & 1.23 & 1.78 & 41.17 & 27.90 & 15.55 & 0 \\
    38 & 8330-12703 & (203.37, 40.53) & 8076.00 & 1.63 & 1.38 & 1.23 & 2.32 & 31.48 & 12.18 & -16.72 & 0 \\
    39 & 8330-9101 & (203.96, 40.27) & 10466.00 & 2.85 & 3.00 & 1.82 & 1.88 & 37.08 & 34.65 & 4.94 & 0 \\
    40 & 8332-12701 & (206.71, 42.65) & 8472.00 & 3.53 & 4.03 & 2.00 & 2.05 & 48.33 & 46.61 & 13.29 & 0 \\
    41 & 8332-1902 & (209.27, 41.85) & 2265.00 & 20.57 & 23.84 & 1.84 & 2.53 & 257.96 & 196.80 & 14.77 & 0 \\
    42 & 8333-12705 & (214.28, 42.68) & 7776.63 & 2.39 & 2.51 & 1.22 & 1.80 & 75.28 & 47.09 & 4.96 & 0 \\
    43 & 8333-1902 & (216.55, 42.71) & 5108.02 & 1.31 & 1.79 & 1.27 & 1.89 & 34.19 & 27.34 & 31.54 & 0 \\
    44 & 8335-12702 & (215.49, 40.20) & 5636.00 & 0.75 & 0.92 & 1.56 & 1.92 & 27.69 & 24.69 & 19.72 & 0 \\
    45 & 8335-12703 & (216.88, 40.96) & 5489.00 & 2.27 & 1.84 & 1.74 & 2.17 & 31.55 & 17.73 & -21.05 & 0 \\
    46 & 8335-6102 & (216.90, 40.41) & 5769.00 & 1.51 & 1.94 & 1.31 & 2.24 & 31.98 & 20.99 & 25.10 & 0 \\
    47 & 8335-9102 & (218.19, 40.72) & 5240.00 & 2.69 & 3.15 & 1.15 & 2.05 & 71.19 & 40.25 & 15.97 & 0 \\
    48 & 8550-9101 & (247.41, 40.24) & 8451.55 & 1.17 & 0.88 & 1.50 & 1.94 & 21.60 & 10.97 & -28.69 & 0 \\
    49 & 8551-12705 & (233.94, 44.83) & 8878.00 & 2.94 & 3.81 & 1.44 & 2.01 & 60.30 & 48.42 & 26.07 & 0 \\
    50 & 8552-12701 & (226.43, 44.40) & 8492.00 & 1.85 & 1.82 & 1.37 & 2.06 & 25.10 & 14.27 & -1.18 & 0.31 \\
    51 & 8552-12702 & (227.93, 43.97) & 8232.00 & 1.23 & 1.91 & 1.67 & 1.92 & 21.36 & 24.74 & 44.76 & 0 \\
    52 & 8552-12705 & (228.30, 44.06) & 5188.00 & 1.04 & 0.79 & 1.68 & 1.52 & 18.79 & 13.70 & -26.90 & 0 \\
    53 & 8552-3704 & (229.02, 43.16) & 5403.00 & 5.40 & 5.48 & 1.96 & 3.66 & 61.29 & 28.86 & 1.47 & 0 \\
    54 & 8552-6101 & (227.02, 42.82) & 5411.00 & 0.53 & 0.60 & 1.52 & 2.06 & 19.90 & 15.03 & 12.75 & 0 \\
    55 & 8588-6103 & (250.35, 40.22) & 9817.47 & 1.63 & 0.37 & 1.65 & 1.90 & 27.39 & 4.68 & -163.72 & 0 \\
    56 & 8600-12701 & (245.73, 41.52) & 8417.22 & 1.63 & 1.95 & 1.38 & 1.88 & 36.41 & 27.99 & 17.98 & 0 \\
    57 & 8600-12704 & (244.88, 41.66) & 8194.12 & 0.64 & 0.38 & 1.17 & 3.55 & 17.91 & 3.07 & -53.27 & 0.01 \\
    58 & 8600-3703 & (245.87, 41.71) & 8338.00 & 3.14 & 2.45 & 1.28 & 2.31 & 46.87 & 17.49 & -24.55 & 0 \\
    59 & 8600-3704 & (245.85, 41.65) & 8318.05 & 3.87 & 4.39 & 1.33 & 2.05 & 54.89 & 34.77 & 12.80 & 0 \\
    60 & 8603-6104 & (247.42, 40.69) & 9147.30 & 2.34 & 3.09 & 1.52 & 2.19 & 31.02 & 24.60 & 27.65 & 0 \\
    61 & 8604-9102 & (246.46, 40.35) & 8696.85 & 1.78 & -0.64 & 1.63 & 5.42 & 27.90 & -2.62 & -226.49 & 0.35 \\
    62 & 8978-12701 & (248.78, 40.99) & 9001.26 & 3.29 & 2.54 & 1.66 & 1.77 & 38.09 & 23.61 & -26.14 & 0 \\
    63 & 8978-12704 & (250.79, 42.19) & 8522.72 & 2.78 & 4.13 & 1.60 & 2.18 & 32.88 & 31.16 & 39.72 & 0 \\
    64 & 8978-1901 & (248.91, 42.46) & 9557.02 & 1.96 & 2.66 & 2.47 & 2.39 & 22.30 & 26.95 & 30.93 & 0 \\
    65 & 8978-1902 & (249.83, 42.18) & 8213.66 & 0.68 & 0.58 & 3.85 & 1.98 & 6.54 & 9.48 & -15.77 & 0 \\
    66 & 8979-6102 & (241.82, 41.40) & 10400.23 & 1.97 & 2.06 & 1.36 & 1.96 & 35.02 & 21.98 & 4.42 & 0 \\
    67 & 8979-6104 & (242.45, 42.33) & 11599.40 & 1.60 & 1.97 & 1.28 & 2.20 & 28.65 & 17.80 & 21.11 & 0 \\
    68 & 8980-12704 & (225.59, 41.92) & 4884.58 & 1.99 & 1.75 & 1.49 & 2.51 & 40.90 & 18.29 & -12.82 & 0 \\
    69 & 8980-12705 & (223.98, 42.16) & 8176.72 & 1.29 & 1.79 & 1.32 & 2.16 & 29.01 & 21.19 & 32.69 & 0 \\
    70 & 8980-3703 & (226.49, 42.23) & 5072.58 & 1.33 & 2.63 & 1.39 & 2.26 & 26.66 & 28.26 & 69.98 & 0 \\
    71 & 8980-3704 & (224.21, 41.60) & 4861.18 & 2.47 & 2.69 & 1.34 & 2.20 & 53.97 & 31.24 & 8.73 & 0 \\
    72 & 8988-6104 & (186.91, 40.16) & 11129.76 & 2.89 & 2.94 & 1.16 & 2.28 & 63.24 & 27.97 & 1.71 & 0 \\
    73 & 9026-6101 & (249.23, 44.38) & 9089.00 & 8.45 & 6.17 & 1.50 & 12.40 & 75.39 & 5.77 & -31.64 & 0.19 \\
    74 & 9029-12705 & (247.58, 41.10) & 8850.48 & 2.58 & 3.54 & 1.43 & 2.02 & 65.23 & 54.32 & 31.60 & 0 \\
    75 & 9035-1902 & (235.90, 43.84) & 11048.18 & 1.80 & 2.35 & 2.01 & 1.89 & 21.46 & 25.59 & 26.59 & 0 \\
    76 & 9036-9101 & (237.90, 44.23) & 11886.58 & 3.07 & 3.58 & 1.34 & 1.88 & 38.72 & 27.87 & 15.39 & 0 \\
    77 & 9036-9102 & (240.57, 42.92) & 7339.18 & 1.60 & 1.83 & 1.74 & 1.93 & 26.64 & 23.55 & 13.24 & 0 \\
    78 & 9037-12701 & (233.65, 43.04) & 6027.90 & 2.30 & 2.97 & 1.69 & 1.64 & 38.77 & 45.16 & 25.67 & 0 \\
    79 & 9037-12703 & (235.98, 44.31) & 10691.23 & 4.43 & 5.27 & 1.59 & 2.10 & 56.34 & 43.77 & 17.30 & 0 \\
    80 & 9869-9101 & (246.59, 40.91) & 8858.52 & 1.61 & 1.60 & 1.78 & 2.27 & 24.02 & 16.27 & -1.00 & 0 \\
    81 & 9871-12701 & (227.33, 42.68) & 5431.93 & 1.74 & 1.00 & 1.12 & 2.12 & 28.24 & 7.40 & -55.85 & 0 \\
    82 & 9871-12703 & (229.36, 42.96) & 5365.14 & 1.36 & 1.72 & 2.19 & 3.06 & 33.32 & 26.26 & 23.17 & 0 \\
    83 & 9871-12704 & (229.16, 42.95) & 5390.94 & 3.43 & 3.46 & 1.05 & 2.61 & 108.42 & 38.23 & 0.92 & 0 \\
    84 & 9871-1901 & (228.92, 43.18) & 5500.02 & 4.45 & 3.45 & 1.64 & 3.51 & 64.86 & 20.52 & -25.58 & 0 \\
    85 & 9871-1902 & (229.23, 43.11) & 5448.78 & 2.00 & 1.84 & 1.24 & 6.60 & 48.77 & 7.31 & -8.23 & 0 \\
    86 & 9871-3701 & (227.36, 42.60) & 5230.81 & 1.52 & 1.51 & 1.31 & 2.09 & 32.41 & 17.50 & -1.19 & 0 \\
    87 & 9871-3703 & (228.80, 43.16) & 5453.72 & 3.71 & 3.94 & 1.13 & 5.53 & 74.59 & 14.15 & 5.95 & 0 \\
    88 & 9871-6103 & (226.99, 41.90) & 5055.58 & 1.81 & 2.16 & 1.18 & 2.17 & 41.47 & 23.50 & 17.78 & 0 \\
    89 & 9871-6104 & (228.98, 43.17) & 5407.09 & 6.44 & 5.80 & 1.31 & 3.24 & 111.00 & 34.97 & -10.58 & 0 \\
    90 & 9871-9101 & (227.55, 42.61) & 5189.91 & 2.38 & 2.40 & 1.20 & 1.88 & 51.78 & 28.93 & 0.83 & 0 \\
    \label{tb:HIgal_map}
\end{longtable}



\bibliography{main}
\bibliographystyle{aasjournal}



\end{document}